\documentclass[draft]{article}
\usepackage{amsmath,amsfonts,latexsym, amssymb}

\usepackage{color}

\newcommand{\wt}{\widetilde}
\newcommand{\wh}{\widehat}

\newcommand{\e}{\varepsilon}
\newcommand{\eps}{\varepsilon}
\newcommand{\pt}{\partial}
\newcommand{\rd}{{\rm d}}
\newcommand{\bR}{{\mathbb R}}

\newcommand{\K}{n}

\newcommand{\Tr}{\mbox{Tr\,}}

\newcommand{\ba}{{\bf{a}}}

\newcommand{\bx}{{\bf{x}}}
\newcommand{\by}{{\bf{y}}}
\newcommand{\bu}{{\bf{u}}}
\newcommand{\bv}{{\bf{v}}}

\newcommand{\tbx}{\widetilde\bx}

\newcommand{\tby}{\widetilde\by}

\newcommand{\bla}{\mbox{\boldmath $\lambda$}}
\newcommand{\bnu}{\mbox{\boldmath $\nu$}}
\newcommand{\bbeta}{\mbox{\boldmath $\beta$}}

\newcommand{\al}{\alpha}

\newcommand{\be}{\begin{equation}}
\newcommand{\ee}{\end{equation}}

\newcommand{\la}{\lambda}
\newcommand{\Om}{{\Omega}}
\newcommand{\om}{{\omega}}

\newcommand{\cG}{{\mathcal G}}
\newcommand{\cY}{{\mathcal Y}}

\newcommand{\cM}{{\mathcal M}}

\newcommand{\fN}{{\frak N}}

\newcommand{\im}{{\text Im }}

\newcommand{\N}{{\mathbb N}}
\renewcommand{\P}{{\mathbb P}}
\newcommand{\bC}{{\mathbb C}}

\newtheorem{theorem}{Theorem}
\newtheorem{corollary}[theorem]{Corollary}
\newtheorem{lemma}[theorem]{Lemma}
\newtheorem{proposition}{Proposition}

\newtheorem{remark}{Remark}
\newtheorem{definition}{Definition}
\newcommand{\qed}{\hfill\fbox{}\par\vspace{0.3mm}}

\numberwithin{equation}{section}
\numberwithin{theorem}{section}
\numberwithin{definition}{section}
\numberwithin{proposition}{section}
\numberwithin{remark}{section}

\usepackage{geometry}     
\usepackage{graphicx}

\numberwithin{equation}{section}
\def\cal{}
\def\RR{{\mathbb R}}

\def\ZZ{{\mathbb Z}}
\def\EE{{\mathbb E}}\def\PP{{\mathbb P}}
\def\NN{{\mathbb N}}
\def\CC{{\mathbb C}}

\def\N{{\mathcal N}}
\def\cH{{\mathcal H}}

\def\W2{W^{1,2}({\cal O}(M))}

\def\O{\mathcal{ O}}

\def\1half{\frac{1}{2}}

\def\b{\beta}

\title{Universality of sine-kernel for Wigner matrices with
a small Gaussian perturbation}

\author{L\'aszl\'o Erd\H os${}^1$\thanks{Partially supported
by SFB-TR 12 Grant of the German Research Council}, Jos\'e A. 
Ram{\'\i}rez${}^2$,
Benjamin Schlein${}^3$\;
and Horng-Tzer Yau${}^4$\thanks{Partially supported
by NSF grants DMS-0602038, 0757425, 0804279} \\
\\
Institute of Mathematics, University of Munich, \\
Theresienstr. 39, D-80333 Munich, Germany \\ lerdos@math.lmu.de ${}^1$ \\ \\
Department of Mathematics, Universidad de Costa Rica\\
San Jose 2060, Costa Rica \\ alexander.ramirezgonzalez@ucr.ac.cr ${}^2$ \\ \\
Department of Pure Mathematics and Mathematical Statistics
\\  University of Cambridge \\
Wilberforce Rd, Cambridge CB3 0WB, UK \\b.schlein@dpmms.cam.ac.uk ${}^3$ \\ \\
Department of Mathematics, Harvard University\\
Cambridge MA 02138, USA \\  htyau@math.harvard.edu ${}^4$ \\ \\
\\}

\begin{document}

\date{March 30, 2010}

\maketitle

\begin{abstract}

We consider  $N\times N$ Hermitian random matrices with
independent identically distributed
entries (Wigner matrices). We assume that the distribution
of the entries have a Gaussian component with variance
$N^{-3/4+\beta}$ for some 
positive $\beta>0$. We prove that the local eigenvalue
statistics follows the universal Dyson sine kernel.

\end{abstract}


\medskip

{\it Running title:} Universality for Wigner matrices


\medskip

Submitted to EPJ on Jun 18, 2009


\bigskip
\section{Introduction}

Certain spectral statistics of  broad classes of $N\times N$ 
random matrix ensembles 
are believed to follow a universal behavior in the limit $N\to\infty$.
Wigner has observed
\cite{W} that the density of eigenvalues 
of large  symmetric or hermitian 
matrices $H$ with independent entries (up 
to the symmetry requirement) converges, as $N\to\infty$, to 
a universal density, the Wigner semicircle law. 
Dyson has observed that the local correlation statistics
of neighboring eigenvalues inside the bulk of the spectrum follows 
another universal
pattern, the Dyson sine-kernel in the $N\to\infty$ limit
\cite{Dy1}. Moreover,
any $k$-point correlation function can be
obtained as a determinant of the two point correlation
functions. The precise form of the universal two point function in the
bulk
seems to depend only on the symmetry class of the matrix ensemble
(a different  universal behavior emerges near the spectral edge
\cite{Sosh}).

Dyson has proved this fact for the {\it Gaussian Unitary Ensemble}
(GUE), where the matrix elements are independent,
identically distributed  complex Gaussian 
random variables (subject to the hermitian constraint). A 
characteristic feature of GUE is that the distribution is
invariant under unitary conjugation, $H \to U^*H U$ for 
any unitary matrix $U$.
Dyson found an explicit formula for the joint density function of the $N$
eigenvalues. The formula contains a characteristic
Vandermonde determinant and therefore it coincides with the Gibbs measure
of a
particle system interacting via a logarithmic potential
analogously to the two dimensional Coulomb gas.
Dyson also observed that the computation of two point function can
be reduced to asymptotics of Hermite polynomials. 

His approach
has later been substantially generalized to include a large class
of random matrix ensembles, but always with unitary (orthogonal, 
symplectic, etc.) invariance. For example, a general class of 
invariant ensembles can be given by the measure 
$Z^{-1}\exp(-\Tr V(H))\rd H$ on the space of hermitian matrices,
where $\rd H$ stands for the Lebesgue measure for all independent matrix
entries, $Z$ is the normalization and $V$ is a real function
with certain smoothness and growth properties. For example, the
GUE ensemble corresponds to $V(x)=x^2$.

The joint density function is explicit in all
these cases and the evaluation of the two point function
 can be reduced to certain asymptotic properties
of  orthogonal polynomials with respect to the weight
 function $\exp(-V(x))$ on the real line.
The sine kernel can thus be proved for a wide range of potentials
$V$. Since the references in this direction are enormous, 
 we can only refer the reader to the book by Deift \cite{D}
for the Riemann-Hilbert approach,
the paper by  Levin and Lubinsky
\cite{LL}  and references
 therein for approaches based on classical analysis of
orthogonal polynomials, or
the paper by Pastur and Shcherbina \cite{PS} for  a
  probabilistic/statistical physics approach.
The book by Anderson et al  \cite{AGZ1} or the book
by Metha \cite{M} also contain 
 extensive lists of  literatures.

Since the computation of the explicit formula of the joint
density relies on the unitary invariance, there have been very little
progress in understanding non-unitary invariant ensembles.
The most prominent example
is the {\it Wigner ensemble} or {\it Wigner matrices},
i.e.,  hermitian random matrices with i.i.d.
entries. Wigner matrices are not unitarily invariant unless
the single entry distribution is Gaussian, i.e. for the GUE case.
The disparity between our understanding of the Wigner ensembles
and the unitary invariant ensembles is startling.
Up until the very  recent work of \cite{ESY2}, there was 
no proof that the density
follows the semicircle law in small spectral windows
unless the number of eigenvalues in the window is at least $\sqrt N$.
This is entirely due to a serious lack of analytic tools 
for studying eigenvalues
once the mapping between eigenvalues and Coulomb gas ceases
to apply. At present,
there are only two rigorous  approaches to eigenvalue
distributions: the moment method
and Green function method. The moment method is restricted to
studying the spectrum
near the edges \cite{Sosh}; the precision of the Green function
method seems to be still very far from getting information on 
level spacing \cite{BK}.

Beyond the unitary ensembles, Johansson  \cite{J} proved  the sine-kernel for
a broader category of ensembles, i.e.,  for matrices  of the form $H+ s V$
where $H$  is a Wigner matrix,
$V$ is an independent GUE matrix
and $s$ is a positive constant of order one.
(Strictly speaking, in the original work \cite{J}, the 
range of the parameter  $s$ depends
on the energy $E$.  This restriction was later removed
by  Ben Arous and P\'ech\'e
\cite{BP}, who also extended this approach to Wishart ensembles).
 Alternatively formulated, if the matrix elements are
normalized to have variance one, then the distribution of the matrix
elements
of the ensemble  $H+ s V$ is given by $\nu\ast \cG_s$, where $\nu$ is
the distribution of the Wigner matrix elements and $\cG_s$
is the centered Gaussian
law with  variance $s^{2}$.
Johasson's work is based on the analysis of the  
explicit formula  for the joint eigenvalue distribution
of the matrix $H+ s V$  (see also \cite{BH}). 


Dyson has introduced a dynamical version of generating random matrices.
 He considered  a matrix-valued process 
$H + sV$ where $V$ is a matrix-valued Brownian motion. 
The distribution of the eigenvalues then evolves according to 
a process called Dyson's Brownian motions. For the convenience of analysis, 
we replace the Brownian motions by
an Ornstein-Uhlenbeck process so that the distribution of 
GUE is the invariant measure of this
modified process, which we still call Dyson's Brownian motion.
Dyson's Brownian motion thus can be viewed as a reversible
interacting particle system with a long range (logarithmic) interaction.
This process is well adapted for studying the 
evolution of the empirical measures
of the eigenvalues, see \cite{GZ}. 
The sine kernel, on the other hand, is a very detailed property 
which typically cannot be obtained from considerations of interacting
particle systems.  The Hamiltonian for GUE, however, is strictly convex
and thus the Dyson's Brownian motion satisfies the logarithmic
Sobolev inequality (LSI).  It was noted in the derivation of the 
Navier-Stokes equations \cite{EMY, QY} that the combination of
the Guo-Papanicolaou-Varadhan \cite{GPV} approach and LSI  provides 
very detailed estimates on the dynamics.

The key  observation of the present paper is that this method can
also be used to estimate
the approach to local equilibria so precisely that, after
combining it with existing techniques from orthogonal polynomials, 
the Dyson sine kernel emerges. In pursuing this approach, we face 
two major obstacles: 1. Good  estimate of the initial entropy, 
2. Good understanding of the
structure of local equilibria. It turns out that the initial 
entropy can be estimated using 
the explicitly formula for the transition kernel of the Dyson's 
Brownian motion (see \cite{BH} and \cite{J}) provided strong
inputs on the local semicircle law  \cite{ESY2} and level repulsion
\cite{ESY3} are
available. 

The structure of local equilibria, however, is much harder to analyze.
Typically, the local equilibrium measures are finite volume Gibbs
measures with short range interaction and the boundary effects can be
easily
dealt with in the high temperature phase. In the GUE case, the logarithmic
potential
does not even decay at large distance and the equilibrium measure
can depend critically on the boundary conditions. The theory of
orthogonal polynomials provides explicit formulae for the correlation
functions of this highly correlated Gibbs measure. These formulae
can be effectively analyzed if the external potential (or logarithm of
the weight function in the terminology of the orthogonal polynomials)
is very well understood. 
Fortunately, we have proved the local
semicircle law up to scales of order
$1/N$ and the level repulsion, which can be used to  control the
boundary effects.
By invoking the theorem of Levin and Lubinsky \cite{LL} and
the method of Pastur and Shcherbina \cite{PS} we are led
to the sine kernel.

It is easy to see  that
adding a Gaussian component of size much smaller than $N^{-1}$ to the
original Wigner matrix
would not move the eigenvalues  sufficiently to change the local
statistics. Our  requirement that the Gaussian component is at least of
size $N^{-3/4}$
comes from technical estimates to control the initial global entropy
and it does not have any intrinsic meaning.
The case that the variance is of order $N^{-1}$, however,  is an
intrinsic
barrier which is difficult to cross.  Nevertheless, we believe 
that our method may offer a possible strategy
to prove the universality of sine kernel for general Wigner matrices.

\bigskip

After this manuscript had been completed, we  found a different
approach to prove the Dyson sine kernel \cite{ERSY}, partly based
on a contour integral representation for the two-point correlation
function \cite{BH, J}. Shortly after our manuscripts were completed,
we learned that our main result was also obtained by Tao and Vu in
\cite{TV} with a different method under no regularity conditions on the initial
distribution $\nu$ provided the third moment of $\nu$ vanishes.

Although the results in this paper  are weaker  than those in 
\cite{ERSY} and \cite{TV}, we believe that
the method presented here has certain independent  interest.
Unlike \cite{ERSY} and \cite{TV}, this approach does not
use the contour integral representation of the two point
correlation function. Hence,  it may potentially have  a broader applicability
to  other  matrix ensembles for which such representation is not available.

\medskip

{\it Acknowledgements.} We would like to thank the referees 
 for suggesting several improvements of the presentation.

\section{Main theorem and conditions}

Fix $N\in\NN$ and we consider a Hermitian matrix ensemble 
of $N\times N$ matrices $H=(h_{\ell k})$ 
with the normalization 
\be
   h_{\ell k} = N^{-1/2} z_{\ell k}, \qquad z_{\ell k}=   x_{\ell k}
  +i y_{\ell k},
\label{scaling}
\ee
where $x_{\ell k}, y_{\ell k}$ for $\ell<k$ are independent,
identically distributed random variables
with distribution $\nu=\nu^{(N)}$ that has
zero expectation and  variance $\frac{1}{2}$. The diagonal
elements are real, i.e. $y_{\ell\ell}=0$ and
and $x_{\ell \ell}$ are also i.i.d., independent
from the off-diagonal ones with distribution $\wt \nu=\wt \nu^{(N)}$
that has zero expectation and  variance one. The superscript
indicating the $N$-dependence of $\nu$, $\wt \nu$ will be omitted.

We assume that the probability measures $\nu$
and $\wt\nu$ have a small Gaussian component of variance
$N^{-3/4+\beta}$ where $\beta>0$ is some
fixed positive number.
More precisely,  we assume 
there exist probability measures $\nu_0$ and $\wt\nu_0$ 
with zero expectation and variance $\frac{1}{2}$
and $1$, respectively, such that
\be
   \nu = \nu_s\ast G_{s/\sqrt{2}}, \quad  \wt\nu = \wt\nu_s\ast
G_{s}, 
\label{muN}
\ee
where $G_s(x) = (2\pi s)^{-1} \exp (-x^2/2s)$ is the Gaussian
law with variance $s^2$ and $\nu_s$, $\wt\nu_s$ are
the rescaling of the laws $\nu_0$, $\wt\nu_0$ to ensure
that $\nu$ and $\wt\nu$ have variance $1/2$ and 1; i.e, 
explicitly
$$
  \nu_s(\rd x) =  (1-s^2)^{-1/2} \nu_0(\rd x (1-s^2)^{-1/2}),
\qquad \wt\nu_s(\rd x) =  (1-s^2)^{-1/2} \wt\nu_0(\rd x (1-s^2)^{-1/2}).
$$

This requirement is equivalent to considering random matrices
of the form
\be
       H = (1-s^2)^{1/2} \wh H + s V,
\label{HN}
\ee
where $\wh H$ is a Wigner matrix with single entry distribution $\nu_0$
and $\wt \nu_0$, and $V$ is a GUE matrix whose
elements are centered Gaussian random variables
with variance $1/N$.

Furthermore, we assume that $\nu$ is absolutely continuous with
positive density functions $h(x)>0$,
i.e. we can write it as
$\rd\nu(x) = h(x) \rd x =\exp(-g(x))\rd x$
with some real function $g$.
We assume the following conditions:
\begin{itemize}
\item{} The measure $\rd \nu$ satisfies the logarithmic Sobolev
inequality,  i.e. there exists a constant $S$ such
that
\be
  \int_\RR u \log u \; \rd\nu \le S \int_\RR |\nabla
\sqrt{u}|^2\rd\nu
\label{Sobol}
\ee
holds for any density function $u>0$ with $\int u\, \rd\nu=1$.

\item{} The Fourier transform
of the functions $h$ and
$h(\Delta g)$
satisfy the decay estimates
\be
  |\wh h(t,s)|\leq \frac{1}{\left[1+\om(t^2+s^2)\right]^9},
\qquad |\wh{h\Delta g}(t,s)|
\leq \frac{1}{\left[1+\wt\om(t^2+s^2)\right]^9}
\label{cond2}
\ee
with some constants $\om, \wt\om> 0$.
\item{} 
There exists a $\delta_0>0$ such that for the 
distribution of the diagonal elements
\be
D_0:=\int_\RR \exp{\big[\delta_0 x^2\big]}\rd \wt\nu(x)  <\infty \; .
\label{cond1}
\ee
\end{itemize}
Although the conditions are stated  directly for the measures $\nu$
and $\wt\nu$, 
it is easy to see that it is sufficient to assume that $\nu_0$ satisfies
\eqref{Sobol} and \eqref{cond2} and $\wt\nu_0$ satisfies \eqref{cond1}.
We remark that \eqref{Sobol} implies that \eqref{cond1}  holds for $\nu$
instead of $\wt\nu$ as well (see \cite{L}).


The eigenvalues of $H$ are denoted by $\la_1, \la_2, \ldots \la_N$.
The law of the matrix ensemble induces a 
probability measure on the set of eigenvalues whose
density function will be denoted by
$p(\la_1, \la_2, \ldots , \la_N)$. The eigenvalues are considered
unordered
for the moment and thus $p$ is a symmetric function. For any
$k=1,2,\ldots, N$,
let 
$$ 
  p^{(k)}(\la_1, \la_2,\ldots \la_k):= 
 \int_{\bR^{N-k}} p(\la_1, \la_2, \ldots , \la_N)\rd\la_{k+1}\ldots \rd
\la_N
$$
be the $k$-point correlation function of the eigenvalues. The $k=1$ point
correlation function (density) is denoted by $\varrho(\lambda):= p^{(1)}(\la)$.
With our normalization convention, the density $\varrho(\la)$
is supported in $ [- 2,   2]$ and in the $N\to\infty$ limit
it converges to the Wigner semicircle law given by the density
\be
 \varrho_{sc}(x)= \frac{1}{2\pi} \sqrt{4-x^2}\, 1_{[-2,2]}(x).
\label{def:sc}
\ee

The main result of this paper is the following theorem:

\begin{theorem} \label{mainthm} Fix arbitrary positive
constants $\beta>0$ and $\kappa>0$.
Consider the Wigner
matrix ensemble with a  Gaussian convolution of variance 
$s^2=N^{-3/4+\beta}$ given by 
\eqref{HN} and assume \eqref{Sobol}--\eqref{cond1}.
 Let $p^{(2)}$ be the two point correlation
function of the eigenvalues of this ensemble.
Let $|E_0|<2-\kappa$ and
\be
  O(a,b) = g(a-b) h\big( \frac{a+b}{2}\big)
\label{def:O}
\ee
with $g,h$   smooth and compactly supported functions 
such that $h\ge 0$ and $\int h=1$.
Then we have
\be
\begin{split}\label{eq:mainthm}
\lim_{\delta\to 0}
 \lim_{N\to \infty}  \frac{1}{2\delta}  \int_{E_0-\delta}^{E_0+\delta}  
\rd E \int\!\!\int  \rd a \rd b & \, O(a,b)  
 \frac{1}{\rho_{sc}^2(E)} \;p^{(2)}\Big(E+ \frac{ a } { \rho_{sc}(E) N}, E+
\frac{ b} 
{ \rho_{sc}(E) N }
  \Big) \\
&  = \int_\RR g(u) \left [ 1-  \Big(  \frac{\sin \pi u}{\pi u}\Big)^2 
\right ]  \rd u.
\end{split}
\ee
\end{theorem}

 The factor $g$ in the  observable  \eqref{def:O} 
tests the eigenvalue differences. The factor $h$, that disappears in the
right hand side of \eqref{eq:mainthm}, is only a normalization factor.
Thus the  special form of observable  \eqref{def:O}  directly exhibits
the fact that the local statistics is translation invariant.

{\it Conventions.} All integrations with unspecified domains are on $\RR$.
We will use the letters $C$ and $c$ to denote general constants whose
precise values are irrelevant and they may change from line to line. 
These constants may depend on the constants in \eqref{Sobol}--\eqref{cond1}.

\subsection{Outline of the proof}

Our approach has three main ingredients. In the first step, we use the entropy method
from hydrodynamical limits to establish a local equilibrium of
the eigenvalues in a window of size $N^{-1+\e}$ (with some small $\e>0$),
i.e. window
that typically contains $n=N^\e$ eigenvalues.
This local equilibrium is subject to an external
potential generated by all other eigenvalues. In the second step we then 
prove that the density of this equilibrium measure is locally constant
by using methods from orthogonal polynomials. Finally, in the third step, we employ
a recent result \cite{LL} to deduce the
sine-kernel. We now describe each step in more details.

\medskip

{\it Step 1.}

\medskip

We generate the Wigner matrix
with a small Gaussian component by running a matrix-valued
Ornstein-Uhlenbeck process \eqref{matrixOU}
for a short time of order $t\sim N^{-\zeta}$, $\zeta>0$.
This generates a stochastic process for the eigenvalues which
can be described as Ornstein-Uhlenbeck processes for the individual
eigenvalues with a strong interaction \eqref{sde}. 

This process is
 the celebrated Dyson's Brownian motion (DBM) \cite{Dy} and the equilibrium
measure
is the GUE distribution of eigenvalues.
The transition kernel can be computed explicitly  \eqref{gkernel}
and it contains the determinantal structure
of the joint probability density of the GUE eigenvalues that
is responsible for the sine-kernel. This kernel was analyzed
by  Johansson \cite{J} assuming that the time $t$ is of order one, 
which is the same order as the relaxation  time to equilibrium
for the Dyson's Brownian motions.
 The sine-kernel,
however, is a local statistics, and {\it local} equilibrium can
be reached within a much shorter time scale. To implement this 
idea, we first control the global entropy on time scale $N^{-1}$
by $N^{1+\al}$, with $\al >1/4$ 
(Section \ref{sec:entropybound}).

More precisely, recall that the entropy of $f\mu$ with
respect to a probability measure $\mu$ is given by 
$$
S(f)=S_\mu(f): = S(f\mu|\mu) = \int f(\log f)\rd \mu.
$$
In our application, the measure $\mu$ is the Gibbs measure for the equilibrium distribution 
of the (ordered)
 eigenvalues of the GUE, given by the Hamiltonian
\be
\cH(\bla) =  
N \left [ \sum_{i=1}^N \frac{\lambda_{i}^{2}}{2} -  \frac{2}{N} \sum_{i<
j} 
\log |\lambda_{j} - \lambda_{i}| \right ].
\label{Hamm}
\ee
If $f_t$ denotes the joint probability density of the eigenvalues
 at the time $t$ with respect to $\mu$, then 
the evolution of $f_t$ is given by the equation 
\be \pt_t f_t =Lf_t,
\label{evolu}
\ee
 where 
the generator  $L$  is defined via the Dirichlet form 
\[
D(g) = \int g (-L) g \rd \mu =  \frac 1 {2 N} \sum_{j=1}^N \int  (\nabla_{\la_j} g)^2 \rd \mu. 
\]
The evolution
of the entropy  is given by the equation 
$$
\partial_{t} S(f_t) = -  D(\sqrt {f_t}).
$$
The key initial entropy estimate is the inequality that 
\be
S_\mu(f_{s}):=S(f_s\mu|\mu) \le  C_\al N^{1+\al}, \quad s = 1/N
\label{Sest}
\ee
for any $\al>\frac{1}{4}$ and for sufficiently large $N$.
The proof of this estimate  uses the explicit formula for the transition kernel
of \eqref{evolu}
and several inputs from our previous papers \cite{ESY1, ESY2, ESY3}
on the local semicircle law and on the level repulsion
for general Wigner matrices. We need to strengthen some of these inputs;
the new result will be presented in
Section \ref{sec:goodglobal} with proofs deferred to Appendix \ref{sec:B}, 
Appendix \ref{sec:C} and Appendix \ref{sec:grid}.

It is natural to think of each eigenvalue as a particle and we will use the language of
 interacting particle systems. 
We remark that the entropy per particle  is typically  of order one in the
 interacting particle systems. 
But in our setting, due to the factor $N$ in front of the Hamiltonian \eqref{Hamm},
the typical size of entropy per particle is of order $N$. Thus for a system bearing
little relation to the equilibrium measure $\mu$, 
we expect the total entropy to be $O(N^2)$. So the bound  \eqref{Sest}  already
 contains nontrivial information. 
However, we believe that one should be able to improve this bound to $\alpha \sim 0$ and 
the additional $\al>1/4$ power in \eqref{Sest}
is only for technical reasons.
This is the main reason why our final result holds only for a Gaussian
convolution with variance larger than  $ N^{-3/4}$.  
The additional $N^\al$ factor originates from Lemma \ref{lm:detappr} 
where we approximate the Vandermonde determinant appearing in the
transition kernel by estimating the fluctuations around the local
semicircle law. We will explain the origin of $\al>1/4$
in the beginning of Appendix \ref{sec:det} where the proof of Lemma \ref{lm:detappr}
is given.

{F}rom the initial entropy estimate,  it follows that the time
integration of the Dirichlet form is bounded by 
the initial entropy. For the DBM, due to convexity of the Hamiltonian of the equilibrium measure $\mu$, 
the Dirichlet form is actually decreasing. Thus  
for  $t =\tau N^{-1}$ with some $\tau \ge 2$  we have
\[
D(\sqrt f_t) \le 2S(f_{N^{-1}}) t^{-1} \leq CN^{2+\al}\tau^{-1}.
\]
The last estimate says that the Dirichlet form per particle is bounded by  $N^{1+\al}\tau^{-1}$.
 So if we take an interval 
of $n$ particles (with coordinates given by $\bx = (x_1, \ldots, x_n)$), then on average 
the total Dirichlet form of these particles is bounded by $n N^{1+\al}\tau^{-1}$.
We will choose $n=N^\e$ with some very small $\e>0$. 
As always in the hydrodynamical limit approach, we  consider the probability law 
of these $n$ particles 
given that all other particles (denoted by $\by$) are fixed.  Denote by  $\mu_\by (\rd \bx) $ 
the equilibrium measure 
of $\bx$ given that the coordinates of the other $N-n$ particles $\by$ are fixed. 
Let $f_{\by, t}$ be  the conditional density 
of $f_t$ w.r.t. $\mu_\by (\rd \bx) $ with $\by$ given.  
The Hamiltonian of the measure $\mu_\by (\rd \bx) $  is given by 
\[
\cH_{\by} (\bx) =  
N \left [ \sum_{i=1}^n \frac{1}{2} x_{i}^{2}
-  \frac{2}{N} \sum_{1\le i< j\le n} 
\log |x_{j} - x_{i}|  -    \frac{2}{N} \sum_{k}\sum_{ i=1 }^n 
\log |x_{i} - y_{k}|   \right ]
\]
and it satisfies the convexity estimate  
\[
\mbox{Hess}  \, \cH_{\by}(\bx)  
\ge  \sum_{k}   |x- y_{k}|^{-2}.
\]
If $\by$  are regularly distributed, we have the convexity bound 
\[
\mbox{Hess}  \, \cH_{\by}(\bx) 
 \ge \frac{cN^2}{n^2}. 
\]
This implies  the logarithmic Sobolev inequality 
\[
S_{\mu_\by}(f_{\by} ) \le C{\K}^2
   N^{-1}D_{\by} (\sqrt {f_{\by} } )
\le  C{\K}^{6} N^{\al}\tau^{-1}, \quad 
\]
where in the last estimate some additional $\K$-factors were
needed to convert the local Dirichlet form estimate per particle
on average to an estimate that holds for a typical particle.
Thus  we obtain
\[
 \left [ \int |f_{\by}-1| \rd \mu_\by \right ]^{2} \le 
S_{\mu_\by}(f_{\by} ) 
\le  C{\K}^{6} N^{\al}\tau^{-1} \le n^{-4} \ll 1,
\]
provided we choose  $t= N^{-1}\tau = N^{\beta-1}$ with $\beta\ge 10\e + \al$ (Section \ref{sec:loceq}).
The last inequality asserts that the two measures $f_\by \mu_\by$ and $\mu_\by$ 
are almost  the same and thus we only need to establish the sine kernel 
for the measure $\mu_\by$. 
At this point, we remark that this argument is valid only if $\by$ is regularly distributed 
in a certain sense which  we will call {\it good configurations} (Definition \ref{defi:good}).
Precise estimates on the local  semicircle law can be
used to show that most external configurations are
good. Although the rigorous treatment of the good configurations and estimates on the 
bad configurations occupy a large part of this paper,  it is of technical nature
and we deferred the proofs of several steps to the appendices.

\medskip

{\it Step 2.}

\medskip

In Sections \ref{sec:derop}, \ref{sec:smeared} and
\ref{sec:regdens}, we refine the precision on the local
density and prove that the density is
essentially constant pointwise. Direct probabilistic arguments to
establish 
the local semicircle law in \cite{ESY3} rely
on the law of large numbers and they give information
on the density on scales of much larger than $N^{-1}$,
i.e. on scales that contain many eigenvalues.
The local equilibrium is reached in a window of size $n/N$
and within this window, we can conclude that the local semicircle
law holds on scales of size $n^\gamma/N$ with an arbitrary small
$\gamma>0$. However, this still does not control
the density  {\it pointwise}. 
To get this information,
we need to use orthogonal polynomials.

 The density 
in  local equilibrium  can be expressed
in terms of sum of squares of orthogonal polynomials $p_1(x), p_2(x), \ldots$
with respect to the weight function $\exp{(-n U_\by(x))}$ generated by the
external configuration $\by$ (see Section \ref{sec:derop} for precise definitions).
 To get a pointwise bound from the appropriate
bound on average, we need only to control the derivative of the density, that,
in particular, can be expressed in terms of derivatives
of the orthogonal polynomials $p_k$. 
Using integration by parts
and orthogonality properties of $p_k$,
it is possible to control the $L^2$ norm of $p_k'$
in terms of the $L^2$ norm of $p_k(x) U_\by'(x) $.
Although the derivative of the potential is singular, $\| p_k U_\by'\|_2$
 can be estimated by a Schwarz inequality
at the expense of treating  higher $L^p$ norms of $p_k$
(Lemma \ref{9.1}).
 In this content,
we will exploit the fact that we are dealing with polynomials 
by using the Nikolskii
inequality which estimates higher $L^p$ norms in terms of lower ones at the expense of
a constant depending on the degree.
To avoid a very large constant   in
the Nikolskii inequality, in Section \ref{sec:cutoff} we first cutoff the 
external potential and thus we reduce  the degree of the weight function.

We remark that our approach of using orthogonal polynomials to
control the density pointwise was motivated by the work of Pastur
and Shcherbina \cite{PS}, 
where they proved sine-kernel for
unitary invariant matrix ensembles with a three times differentiable
potential function on the real line. In our case, however,
the potential is determined
by the external points and it is logarithmically divergent
near the edges of the window.

\medskip

{\it Step 3.}

\medskip

Finally, in Section \ref{sec:proofmain},
we complete the proof of the sine-kernel by applying
the main theorem of \cite{LL}. This result establishes
the sine-kernel for orthogonal polynomials with respect to
an $n$-dependent sequence of weight functions under general conditions.
The most serious condition to verify is that the density is
essentially constant pointwise -- the main result we have
achieved in the Step 2 above. We also need to identify the support
of the equilibrium measure which will be done in Appendix \ref{sec:supp}.

We remark that, alternatively, it is possible to complete the third step
along the lines of the  argument of \cite{PS} without
using  \cite{LL}.
Using explicit formulae from orthogonal polynomials
and the pointwise control on the density and on its derivative,
it is possible to prove that the local two-point correlation function
$p^{(2)}_n(x,y)$
is translation invariant as $n\to\infty$. After having established 
the translation invariance of $p^{(2)}$, it is easy to derive an equation
for
its Fourier transform and obtain the sine-kernel as the unique
solution of this equation. We will not pursue this alternative
direction in this paper.

\section{Dyson's Brownian motion}

\subsection{Ornstein-Uhlenbeck process}

We can generate our matrix $H$ \eqref{HN} from a  stochastic process
with initial condition $\wh H$.
Consider the following matrix valued stochastic differential equation
\be
  \rd H_t =\frac{1}{\sqrt{N}} \rd \bbeta_t - \frac{1}{2}H_t\rd t
\label{matrixOU}
\ee
where $\bbeta_t$ is a hermitian matrix-valued stochastic
process whose diagonal matrix elements
are standard real Brownian motions and whose off-diagonal matrix elements
are standard complex Brownian motions. 

For completeness we describe this matrix valued Ornstein-Uhlenbeck process
more precisely. The rescaled matrix elements $z_{ij} = N^{1/2}h_{ij}$
evolve according to the complex Ornstein-Uhlenbeck
process
\be
  \rd z_{ij}= \rd \beta_{ij} - \frac{1}{2} z_{ij}\rd t, \qquad
i,j=1,2,\ldots N.
\label{zij}
\ee
For $i\neq j$, $\beta = \beta_{ij}$ is a complex Brownian
motion with variance one. The  real and imaginary parts of $z=x+iy$
satisfy
$$
  \rd x = \frac{1}{\sqrt{2}}\rd \beta_x - \frac{1}{2} x\rd t,\qquad
  \rd y = \frac{1}{\sqrt{2}}\rd \beta_y - \frac{1}{2} y\rd t
$$
with $\beta = \frac{1}{\sqrt{2}} (\beta_x + i\beta_y)$ and
where $\beta_x,\beta_y$ are independent standard real Brownian
motions.
For the diagonal elements $i=j$ in \eqref{zij}, 
$\beta_{ii}$ is a standard real Brownian motion with variance 1.

To ensure $z_{ij} = \bar z_{ji}$, for $i<j$ 
we choose $\beta_{ij}$  to be independent complex Brownian motion with 
$\EE \, |\beta_{ij}|^2=1$, we set $\beta_{ji}:= \bar
\beta_{ij}$ and we let $\beta_{ii}$ to be a real
Brownian motion with $\EE \, \beta_{ii}^2 =1$. Then
\be
   (\rd z_{ik})(\rd z_{\ell j}) =  (\rd \beta_{ik})(\rd \bar
\beta_{j\ell})
  = \delta_{ij}\delta_{k\ell}\rd t .
\label{diff}
\ee
We  note that $\rd \Tr H^2 =0$, thus
\be
  \Tr H^2 = N
\label{trace}
\ee
remains constant for all time.

\medskip

If the initial condition of \eqref{matrixOU} 
is distributed according to the law of $\wh H$, then the solution
of \eqref{matrixOU} is clearly
$$
   H_t = e^{-t/2}\wh H + (1-e^{-t})^{1/2} V
$$
where $V$ is a standard GUE matrix (with matrix elements
having variance $1/N$) that is independent of $\wh H$.
With the choice of $t$ satisfying $(1-e^{-t}) = s^2 = N^{-3/4+\beta}$,
i.e.
$t=-\log (1- N^{-3/4+\beta}) \approx N^{-3/4+\beta}$,
we see that $H$ given in \eqref{HN} 
has the same law as $H_t$. 

\subsection{Joint probability distribution of the eigenvalues}

We will now analyze the
eigenvalue distribution of $H_t$. 
Let $\bla(t)= (\lambda_1(t), \lambda_2(t), \ldots, \lambda_N(t))\in 
\RR^N$ denote the eigenvalues of $H_t$.
As $t\to\infty$, the Ornstein-Uhlenbeck process \eqref{matrixOU} converges
to the standard GUE. The joint distribution of the GUE
eigenvalues is given by the following measure
$\wt\mu$ on $\RR^N$ 
\be\label{H}
\wt\mu=\wt\mu(\rd\bla)= \frac{e^{-\cH(\bla)}}{Z}\rd\bla,\qquad \cH(\bla) =  
N \left [ \sum_{i=1}^N \frac{\lambda_{i}^{2}}{2} -  \frac{2}{N} \sum_{i<
j} 
\log |\lambda_{j} - \lambda_{i}| \right ].
\ee
The measure $\wt\mu $ has a density with respect to Lebesgue measure given by
\be
\wt u(\bla) = \frac{ N^{N^2/2}}{(2\pi)^{N/2} \prod_{j=1}^N j!}
\exp\left[ -\frac{N}{2}\sum_{j=1}^N \lambda_j^2\right] \, 
\Delta_N(\bla)^2,  \qquad
  \wt\mu(\rd\bla) =\wt u(\bla) \rd \bla,
\label{def:tildeu}
\ee
where $\Delta_N(\bla) = \prod_{i<j} (\lambda_i-\lambda_j)$.
This is the joint probability distribution of the eigenvalues
of the standard GUE ensemble normalized in such a way that the matrix
elements 
have variance $1/N$ (see, e.g. \cite{M}).  
With this normalization convention, the bulk of the one point function
(density) is supported in $ [- 2,   2]$ and in the $N\to\infty$ limit
it converges to the Wigner semicircle law \eqref{def:sc}.

For any finite time $t<\infty$ we will represent the joint
probability density of the eigenvalues of $H_t$ as 
$f_t(\bla)\wt u(\bla)$, with $\lim_{t\to\infty} f_t (\bla) =1$. In particular,
we write the joint distribution  of the eigenvalues of the initial
Wigner matrix $\wh H$ as $f_0(\bla)\wt\mu(\rd\bla)=f_0(\bla)\wt u(\bla)\rd\bla$.

\subsection{The generator of Dyson's Brownian motion}

The Ornstein-Uhlenbeck  process \eqref{matrixOU} induces a
stochastic process for the eigenvalues. 


Let $L$ be the generator given by 
\be
L=   \sum_{i=1}^N \frac{1}{2N}\partial_{i}^{2}  +\sum_{i=1}^N
\Bigg(- \frac{1}{2} \lambda_{i} +  \frac{1}{N}\sum_{j\ne i} 
\frac{1}{\lambda_i - \lambda_j}\Bigg) \partial_{i}
\label{L}
\ee
acting on $L^2(\wt\mu)$
and let
\be
D(f) = -\int  f L f  \rd \wt\mu =  \sum_{j=1}^N \frac{1}{2N}
\int (\partial_j f)^2 \rd \wt\mu
\label{def:dir}
\ee
be the corresponding Dirichlet form, where
$\partial_j=\partial_{\lambda_j}$.
Clearly $\wt\mu$ is an invariant measure
for the dynamics generated by $L$.

Let the distribution of the eigenvalues of the
Wigner ensemble be given by $f_0 (\bla)\wt\mu(\rd \bla)$.
We will evolve this 
distribution by the dynamics given by $L$:
\be\label{dy}
\partial_{t} f_t =  L f_t
\ee
The corresponding stochastic differential equation
for the eigenvalues $\bla(t)$ is now given by  (see, e.g.
Section 12.1 of \cite{AGZ})
\be\label{sde}
 \rd  \lambda_i  =    \frac{\rd B_i}{\sqrt{N}} +  \left [ -  \frac{1}{2}
\lambda_i+  \frac{1}{N}\sum_{j\ne i} 
\frac{1}{\lambda_i - \lambda_j}  \right ]  \rd t, \qquad 1\leq i\leq N,
\ee
where $\{ B_i\; : \; 1\leq i\leq N\}$ is a collection of
independent Brownian motions and with 
initial condition $\bla(0)$ that is distributed according to 
the probability density $f_0(\bla)\wt\mu(\rd\bla)$.

We remark that $\wt u(\bla)$ and $f_t(\bla)$ are symmetric functions
of the variables $\lambda_j$
and $\wt u$ vanishes whenever two points coincide. By the level repulsion
we also know that $f_0(\bla)\wt u(\bla)$ vanishes whenever
$\lambda_j=\lambda_k$
for some $j\neq k$.  We can label the eigenvalues
according to their ordering,
$\lambda_1 <\lambda_2 < \ldots < \lambda_N$, i.e. one can consider
the configuration space
\be
\label{def:Xi}
 \Xi^{(N)}:=\Big\{ \bla= (\lambda_1, \lambda_2, \ldots ,\lambda_N)\; : \;
 \lambda_1 <\lambda_2 < \ldots < \lambda_N\Big\} \subset \RR^N.
\ee
instead of the whole $\RR^N$.
 With an initial point in $\Xi^{(N)}$, the 
equation  \eqref{sde} has a unique solution and
the trajectories do not cross each other, i.e. the ordering
of eigenvalues is preserved under the time evolution and thus the
dynamics generated by $L$ can be restricted to $\Xi^{(N)}$; see, e.g. 
Section 12.1 of \cite{AGZ}. The main reason is that near
a coalescence point $\lambda_i=\lambda_j$, $i> j$, the generator is
$$
  \frac{1}{N} \Big[ \frac{1}{2}\partial_{\lambda_i}^2 + \frac{1}{2}
   \partial_{\lambda_j}^2 
  + \frac{1}{\lambda_i-\lambda_j}(\partial_{\lambda_j}-
\partial_{\lambda_i}) \Big]
 =  \frac{1}{2N} \Big[ \frac{1}{2}\partial_{a}^2 +
\frac{1}{2}\partial_{b}^2 
  + \frac{1}{b}\partial_{b}\Big]
$$
with $a=\frac{1}{2}(\lambda_i+\lambda_j)$, $b=
\frac{1}{2}(\lambda_i-\lambda_j)$. 
The constant $1$ in front of the drift term is critical for the Bessel
process $\frac{1}{2}\partial^2_b + \frac{1}{b}\pt_b$ 
not to reach the boundary point $b=0$.

Note that the symmetric density function $\wt u(\bla)$ defined on $\RR^N$
can be  restricted to $\Xi^{(N)}$ as 
\be
 u(\bla)=N! \; \wt u(\bla){\bf 1}(\bla\in
\Xi^{(N)}).
\label{def:u}
\ee
 The density function of the ordered eigenvalues is thus
$f_t(\bla) u(\bla)$ on $\Xi^{(N)}$. 
Throughout this paper, with the exception of Section
\ref{sec:entropybound}, we work on the 
space $\Xi^{(N)}$, i.e., the equilibrium measure $\mu(\rd\bla) = u(\bla)\rd\bla$ with
density  $u(\bla)$ and the density function
 $f_t(\bla)$ will be considered restricted to  $\Xi^{(N)}$.

\section{Good global configurations}\label{sec:goodglobal}

Several estimates in this paper will
rely on the fact that the number of eigenvalues
$\N_I$ in intervals $I$ with length much larger than $1/N$ is given by
the semicircle law \cite{ESY3}. 
In this section we define the set of good global configurations,
i.e. the event that the semicircle law holds on all subintervals
in addition to a few other typical properties.

Let 
\be
  \om(\rd x) = \frac{1}{N} \sum_{j=1}^N \delta(x-\lambda_j)
\label{def:om}
\ee
be the empirical density of the eigenvalues.
For an interval $I= [a,b]$ we introduce the notation 
$$
 \N_I  =\N[a; b] = N\int_a^b \om(\rd x)
$$
for the number of eigenvalues in $I$. For the interval $[E-\eta/2,
E+\eta/2]$
of length $\eta$ and centered at $E$ we will also use the notation
$$
  \N_\eta(E):= \N[E-\eta/2; E+\eta/2]\, .
$$
Let
\be
   \omega_{\eta} (x): = (\theta_{\eta} \ast \omega ) (x), \qquad
\mbox{with}\quad 
   \theta_{\eta} (x) = \frac{1}{\pi}\frac{\eta} {x^2 + \eta^2}
\label{def:ometa}
\ee
be the empirical density smoothed out on scale $\eta$. Furthermore, 
let 
$$ 
 m(z) =\frac{1}{N}\sum_{j=1}^N \frac{1}{\lambda_j-z} = \int_\RR
\frac{\om(\rd x)}{x-z}
$$
be the Stieltjes transform of the empirical eigenvalue
distribution and 
\be
 m_{sc}(z) = \int_\RR \frac{\varrho_{sc}(x)}{x-z}\rd x = -\frac{z}{2} +
\sqrt{
 \frac{z^2}{4}-1}
\label{def:msc}
\ee
be the Stieljes transform of the semicircle law. 
The  square root here is defined as the analytic extension
(away from the branch cut $[-2,2]$)
of the positive square root on large positive numbers.
Clearly $\om_y(x) =\pi^{-1} \mbox{Im}\; m(x+iy)$ for $y>0$.

We will need an improved version of Theorem 4.1 from
\cite{ESY3} that is also applicable near the spectral edges.
The proof of the following theorem is given in Appendix \ref{sec:B}.

\begin{theorem}\label{thm:semi}  Assume that the Wigner
matrix ensemble satisfies conditions \eqref{Sobol}--\eqref{cond1}
and assume that $y$ is such that $(\log N)^4/N \leq |y| \leq 1$. 

(i) For any $q\ge 1$ we have
\be
  \EE\, |m(x+iy)|^q\le \, C_q 
\label{eq:mup}
\ee
\be 
\EE \; [\om_y(x)]^q\le \,  C_q
\label{eq:omup}
\ee
where $C_q$ is independent of $x$ and $y$.

\medskip

(ii) Assume that $|x| \leq K$ for some $K >0$. 
Then there exists $c>0$ such that
\be
\P \left( \left| m (x+iy) - m_{\text{sc}} (x+iy) \right| \geq \delta
\right) \leq 
C\, e^{-c \delta \sqrt{N|y| \, |2-|x||}} 
\label{Pm}
\ee
for all $\delta >0$ small enough and all $N$ large enough (independently
of $\delta$). 
Consequently, we have
\be\label{eq:E|m|}
\EE \, | m(x+iy)- \EE \, m(x+iy) |^q  \leq \,  \frac{C_q}{(N|y| |2
-|x||)^{q/2}}
 +C_q{\bf 1} \big( N|y| |2 -|x|| \le (\log N)^4\big)
\ee
with some $q$-dependent constant $C_q$. Moreover,
\begin{equation}\label{eq:Em-msc-2} 
| \EE \, m(x+iy) - m_{\text{sc}} (x+iy) |  \leq \frac{C}{N|y|^{3/2} |2
-|x||^{1/2}}
\end{equation}
for all $N$ large enough (independently of $x,y$).

\medskip

(iii) Assuming $|x|\le K$ and that $\sqrt{N|y| |2-|x||} \geq (\log N)^2$
we also have 
\begin{equation}\label{eq:Em-msc-1}
| \EE \, m(x+iy) - m_{\text{sc}} (x+iy) |  \leq   \frac{C}{N |y|
|2-|x||^{3/2}} \, .
\end{equation}
\end{theorem}

As a corollary to Theorem \ref{thm:semi}, the semicircle law 
for the density of states holds locally on very short scales. 
 The next proposition can be proved, starting from Theorem 
\ref{thm:semi}, exactly as Eq. (4.3) was shown in \cite{ESY1}.
\begin{proposition}\label{prop:shortsc} Assuming \eqref{Sobol}--\eqref{cond1},
for any sufficiently small $\delta$ and for any
$\eta^*$ with 
$$
C\delta^{-2}(\log N)^4/N\le\eta^*\le
C^{-1} \min\{ \kappa, \delta\sqrt{\kappa} \}
$$
(with a sufficiently large constant $C$)
 we have 
\be
 \PP \Big\{ \sup_{E\in [-2+\kappa, 2-\kappa]}
 \Big| \frac{\N_{\eta^*}(E)}{2N\eta^*} - \varrho_{sc}(E)\Big|
 \ge\delta\Big\}\leq Ce^{-c\delta^2\sqrt{N\eta^*\kappa}}.
\label{NI}
\ee
\end{proposition}
We also need an estimate directly on the number of eigenvalues in
a certain interval, but this will be needed only away from the spectral
edge. The following two results estimate the deviation of the 
normalized empirical counting function
$\frac{1}{N}\N[-\infty, E] = \frac{1}{N}\#\{ \lambda_j\le E\}$  and its
expectation
\be
  \fN (E) : = \frac{1}{N}\, \EE\, \N[-\infty, E]
\label{def:NE}
\ee
from the distribution function of
the semicircle law, defined as
\be
 \fN_{sc}(E) := \int_{-\infty}^E \varrho_{sc}(x) \rd x.
\label{def:Nsc}
\ee
\begin{proposition}\label{prop:distr}
Assume that the Wigner
matrix ensemble satisfies conditions  \eqref{Sobol}--\eqref{cond1}.
Let $\kappa >0$ be fixed. For any $0<\delta <1$ and $|E|\le 2-\kappa$,  we
have
\be
  \PP \Big\{  \Big| \frac{\N[-\infty, E]}{N} - \fN_{sc}(E)\Big| \ge \delta
 \Big\} \le C \; e^{- c\delta\sqrt{N}}
\label{NNE}
\ee
with $\kappa$-dependent constants.
Moreover, there exists a constant $C >0$ such that 
\be
  \int_{-\infty}^\infty |\fN (E) - \fN_{sc} (E)| \rd E  \leq \frac{C}{N^{6/7}}
\, .
\label{NNint}
\ee
\end{proposition}
The proof of this proposition will be given in Appendix \ref{sec:C}.
\bigskip

Next we define the good global configurations; the idea is that good 
global configurations are configurations for which the semicircle
 law holds up to scales of the order $(\log N)^4/N$ (and so that
 some more technical conditions are also satisfied). By 
Proposition \ref{prop:shortsc} and Proposition \ref{prop:distr}, 
we will see that set of these configurations have, asymptotically, a full 
measure. As a consequence, we will be able to neglect all configurations 
that are not good.

\medskip

Let
\be
  n:= 2[N^\e/2]+1, \qquad \eta^*_m= 2^m n^\gamma N^{-1},\quad
\delta_m = 2^{-m/4}\K^{-\gamma/6} 
\label{nchoice}
\ee
with some small constants $0<\e, \gamma\le \frac{1}{10}$
 and $m=0,1,2,\ldots, \log N$. 
Here $[x]$ denotes the integer part of $x\in \bR$.
Note that within this range of $m$'s,  $C\delta_m^{-2}(\log N)^4/N
\le \eta^*_m\le\kappa^{3/4}\delta_m^{1/2}$
is satisfied if $\e, \gamma$ are sufficiently small.
Let
\be
  \Omega^{(m)}:=\Big\{ \sup_{E\in [-2+\kappa/2, 2-\kappa/2]}
 \Big| \frac{\N_{\eta^*_m}(E)}{N\eta^*_m} - \varrho_{sc}(E)\Big| 
 \leq \frac{1}{(N\eta^*_m)^{1/4}}n^{\gamma/12}
 \Big\} 
\label{omm}
\ee
then we have
\be
  \PP (\Omega^{(m)}) \ge 1- Ce^{-c\K^{\gamma/6}}
\label{pom}
\ee
with respect to any Wigner ensemble. This gives rise to the following
definition.

\begin{definition}\label{defi:good} Let $\eta^*_m=2^m n^{\gamma}N^{-1}$
with some small constant $\gamma>0$, $m=0,1,2,\ldots \log N$,
and let $K$ be a
fixed big constant. The event
\be
 \Omega : =\bigcap_{m=0}^{\log N} \Omega^{(m)}
 \cap \Big\{
\big|\frac{\N[-\infty, 0]}{N/2} -1| \le n^{-\gamma/6}\Big\}
\label{Omstar}
\cap \Big\{ \sup_E \N_{\eta^*_0}(E)  \le KN\eta^*_0\Big\}\cap
  \Big\{ \N(-K, K)=N\Big\}
\ee
will be called the set of {\bf good global configurations}. 
\end{definition}

\begin{lemma}
The probability of good global configurations satisfies
\be
 \PP (\Omega) \ge 1- Ce^{-c\K^{\gamma/6}}
\label{NI1}
\ee
with respect to any Wigner ensemble
satisfying the conditions \eqref{Sobol} and \eqref{cond2}
\end{lemma}

{\it Proof.}  The probability of $\Omega^{(m)}$ was estimated
in \eqref{pom}. The probability of the
second event in \eqref{Omstar} can be estimated by \eqref{NNE} from
Proposition \ref{prop:distr} and from $\fN_{sc}(0)=1/2$.
The third event is treated by the large deviation estimate on $\N_I$
for any interval $I$ with length $|I|\ge (\log N)^2/N$
(see Theorem 4.6 from \cite{ESY3}; note that there is a small error in the statement of this theorem, since the conditions $y \geq (\log N)/N$ and $|I| \geq (\log N)/N$ should actually be replaced by the stronger assumptions $y \geq (\log N)^2/N$ and $|I| \geq (\log N)^2 /N$ which are used in its proof):
\be
 \PP \{ \N_I \ge KN|I|\}\le e^{-c \sqrt{KN|I|}}.
\label{NILDE}
\ee
The fourth event is a large deviation of the largest
eigenvalue, see, e.g. Lemma 7.4. in \cite{ESY1}. $\Box$

\bigskip

In case of good configurations, the
location of the eigenvalues
are close to their equilibrium localition given 
by the semicircle law. The following lemma contains 
the precise statement and it will be proven in Appendix
\ref{sec:grid}.

\begin{lemma} \label{lm:grid}
Let $\lambda_1<\lambda_2 <\ldots < \lambda_N$ denote the
eigenvalues in increasing order and let $\kappa>0$. Then on the set $\Omega$
and if $N\ge N_0(\kappa)$, 
it holds that
\be 
     |\lambda_a -\fN_{sc}^{-1}(aN^{-1})|\le C\kappa^{-1/2}n^{-\gamma/6}
\label{y}
\ee
for any $N\kappa^{3/2} \le a \le N(1-\kappa^{3/2})$
(recall the definition of $\fN_{sc}$ from \eqref{def:Nsc}), and
\be
 \Big| N\varrho_{sc}(\lambda_a)(\lambda_b -\lambda_a )
-(b-a) \Big| \le C\kappa^{-1/2} \big[ n^\gamma|b-a|^{3/4}+ N^{-1}|b-a|^2\big]
\label{yy}
\ee
for any $N\kappa^{3/2} \le a < b\le N(1-\kappa^{3/2})$
and $ |b-a| \le CN n^{-\gamma/6}$.
\end{lemma}

\subsection{Bound on the level repulsion and potential 
for good configurations}
\label{sec:levrep}

\begin{lemma}\label{lm:rep} 
On the set $\Omega$ and with the choice $n$ given in
\eqref{nchoice}, we have 
\be
\frac{1}{N} \EE  \sum_{\ell = N\kappa^{3/2}}^{(1-\kappa^{3/2})N} 
\sum_{j \ne \ell}
\frac{{\bf 1}_\Omega}{[N(\lambda_j-\lambda_\ell)]^2} \le Cn^{2\gamma}.
\label{eq:rep}
\ee
and
\be
 \frac{1}{N} \EE
\sum_{\ell = N\kappa^{3/2}}^{(1-\kappa^{3/2})N} \sum_{j \ne \ell}
 \frac{{\bf 1}_\Omega}{N(\lambda_\ell-\lambda_j)} \leq   C\K^{2\gamma} 
\label{above} 
\ee
with respect to any Wigner ensemble satisfying
the conditions \eqref{Sobol} and \eqref{cond2}
\end{lemma}

{\it Proof.} First we partition the interval $[-2+\kappa, 2-\kappa
]$ into subintervals
\be
     I_r = \big[ n^{\gamma} N^{-1}(r-\frac{1}{2}), n^{\gamma} N^{-1}
  (r+\frac{1}{2})\big], \qquad  r\in \ZZ, \;\;
|r|\le  r_1:=(2-\kappa)N n^{-\gamma},
\label{largepart}
\ee
that have already been used in the proof of Lemma \ref{lm:grid}.
On the set $\Omega$ we have the bound
\be
   \N(I_r) \le KN|I_r| \le C\K^\gamma
\label{apr}
\ee
on the number of eigenvalues in each interval $I_r$.
Moreover, the constraint 
 $N\kappa^{3/2}
 \le \ell \le N(1-\kappa^{3/2})$ implies, by \eqref{y},
that $|\lambda_\ell|\le 2 -\kappa$ for sufficiently small $\kappa$,
thus $\lambda_\ell \in I_r$ with $|r|\le r_1$.

We estimate \eqref{eq:rep} as follows:
\be
\begin{split}
A:= & \frac{1}{N} \EE {\bf 1}_\Omega  \sum_{j< \ell}^* 
\frac{1}{[N(\lambda_j-\lambda_\ell)]^2}  \cr
= &\frac{1}{N} \EE{\bf 1}_\Omega 
 \sum_{j < \ell}
\sum_{k\in\ZZ}  \sum_{|r|\leq  r_1}
\frac{{\bf 1}(\lambda_\ell \in I_r){\bf 1}( 2^{k}
\leq N|\lambda_j -\lambda_\ell|\leq 
2^{k+1})}{[N(\lambda_j-\lambda_\ell)]^2} 
\cr
 \leq & \frac{1}{N} \EE{\bf 1}_\Omega  \sum_{|r|\leq  r_1}\sum_{j <
\ell}
\sum_{k\in\ZZ}
2^{-2k} {\bf 1} \Big\{ \lambda_\ell \in I_r, 
\; 2^{k}
\leq N|\lambda_j -\lambda_\ell|\leq 
2^{k+1}\Big\}
\end{split}
\ee
where the star in the 
first summation indicates a restriction to $N\kappa^{3/2}\le j<\ell\le
(1-\kappa^{3/2})N$.
By \eqref{apr}, for any fixed $r$, 
the summation over $\ell$ with $\lambda_\ell\in I_r$
contains at most $C\K^\gamma$ elements. 
The summation over $j$ contains at most
$Cn^\gamma$ elements if $k<0$, since  $\lambda_\ell \in I_r$
and $|\lambda_j- \lambda_\ell|\le 2^{k+1}N^{-1}\le N^{-1}$
imply that $\lambda_j\in I_r \cup I_{r+1}$.
If $k\ge0$, then the $j$-summation has at most
$C (2^{k}+n^\gamma)$ elements since in this case
$\lambda_j \in \bigcup \{ I_{s}\; : \; |s-r|\le C 
\cdot 2^{k}n^{-\gamma}+1\}$.
Thus  we can continue the above estimate as
\be
\begin{split}
A\leq &
 \frac{C\K^{2\gamma}}{N}
\sum_{k<0} \sum_{|r|\leq  r_1} 2^{-2k}    \PP \Big\{ 
\exists I\subset I_{r-1}\cup I_r\cup I_{r+1}\; : \; |I| \leq
2^{k+1} N^{-1}, \; \N_I \ge 2\Big\} \cr
& +  \frac{C\K^{\gamma}}{N}
\sum_{k\ge 0} \sum_{|r|\leq  r_1} 2^{-2k} (n^\gamma + 2^k)  .
\end{split}
\ee
The second sum is bounded by $Cn^{3\gamma}$. In the first sum,
we use the level repulsion estimate by decomposing
$I_{r-1}\cup I_r\cup I_{r+1} = \bigcup_m J_m$
into intervals of length $2^{k+2}N^{-1}$ that overlap
at least by $2^{k+1}N^{-1}$, more precisely
$$
J_m = \Big[ n^\gamma N^{-1}(r-1-\frac{1}{2}) + 2^{k+1}N^{-1}(m-1),
  n^\gamma N^{-1}(r-1-\frac{1}{2}) + 2^{k+1}N^{-1}(m+1)\Big],
$$
where $m=1,2, \ldots , 3 \K^\gamma \cdot 2^{-k-1}$. Then
$$
   \PP \Big\{ 
\exists I\subset I_{r-1}\cup I_r\cup I_{r+1}\; : \; |I| \leq
2^{k+1} N^{-1}, \; \N_I \ge 2\Big\} \leq
 \sum_{m=1}^{3n^\gamma \cdot 2^{-k-1}} \PP \big\{ \N_{J_m}\ge 2\big\} 
$$
Using 
the level repulsion estimate given in  Theorem 3.4 of \cite{ESY3} 
(here the condition \eqref{cond2} is used)
 and the fact that $J_m\subset I_{r-1}
\cup I_r\cup I_{r+1}\subset [-2+\kappa,
2-\kappa]$ since $|r|\le r_1$, we have
$$
 \PP \big\{ \N_{J_m}\ge 2\big\} \leq C (N|J_m|)^4
$$
and thus
$$
A\leq 
\frac{C\K^{3\gamma}}{N}
\sum_{k=-\infty}^{-1} \sum_{|r|\leq  r_1} 2^{-2k} 2^{-k-1} (2^{k+2} )^{4}
\leq C\K^{2\gamma}.
$$
and this completes the proof of  \eqref{eq:rep}.

For the proof of \eqref{above}, 
we note that  it is sufficient
to bound the event when $N|\la_j-\la_\ell|\ge 1$
after using \eqref{eq:rep}. Inserting the
partition \eqref{largepart}, we get
\be
\begin{split}\label{lalal}
\frac{1}{N} \EE {\bf 1}_\Omega  \sum_{j <\ell}^*
\frac{{\bf 1}(  N|\lambda_\ell
-\lambda_j|\ge1)}{N(\lambda_\ell-\lambda_j)}
 & = \frac{1}{N} \sum_{|r|, |s|\leq  r_0}
\EE {\bf 1}_\Omega \sum_{j <\ell}
\frac{{\bf 1}(\lambda_j\in I_r, \lambda_\ell\in I_s)
{\bf 1}(  N|\lambda_\ell -\lambda_j|\ge1)}{N(\lambda_\ell-\lambda_j)}
\cr
&  \leq \frac{C}{N} \sum_{|r|, |s|\leq  r_0}
\EE {\bf 1}_\Omega 
\frac{ \N_{I_r}\N_{I_s}}{n^\gamma [|s-r|-1]_++ 1}
\cr
& \leq \frac{C\K^{2\gamma}}{N} \sum_{|r|, |s|\leq  r_0}
\frac{1}{n^\gamma [|s-r|-1]_++ 1} \cr & \le C\K^\gamma\log N .
\nonumber
\end{split}
\ee
Recalling the choice of $n$ completes the 
proof of Lemma \ref{lm:rep}.  $\Box$


\section{Global entropy}\label{sec:glob}

\subsection{Evolution of the entropy}

Recall the definition of the entropy of $f\mu$ with
respect to $\mu$
$$
S_\mu(f): = S(f\mu|\mu) = \int f(\log f) \rd \mu
$$
and let $f_t$ solve \eqref{dy}. Then the evolution
of the entropy  is given by the equation 
$$
\partial_{t} S(f_t) = -  D(\sqrt {f_t}) 
$$
and thus using that $S(f_t)>0$ we have
\be\label{3}
\int_{s}^{t} D(\sqrt {f_u}) \rd u \le S(f_s).
\ee
For dynamics with energy $\cH$ and the convexity condition 
\be\label{convexity} 
\mbox{Hess} (\cH) =\nabla^2\cH \ge \Lambda
\ee
for some constant $\Lambda$,
the following Bakry-Emery inequality \cite{BE}  holds: 
$$
\partial_{t}D(\sqrt {f_t})  \leq  - \frac{\Lambda}{N} D(\sqrt {f_t})
$$
(notice the additional $N$ factor due to the $N^{-1}$ in front 
of the second order term in the generator $L$, see \eqref{L}).
This implies the  logarithmic Sobolev inequality that 
for any probability density $g$, with respect to $\mu$, 
\be\label{lsi}
D(\sqrt g) = -\int  \sqrt g L \sqrt g  \rd \mu \ge \frac{\Lambda}{N} S(g)
\ee
In this case, the Dirichlet form is a decreasing function 
in time and we thus have  
for any $t>s$ that
\be
D(\sqrt f_t) \le \frac { S(f_s)} {t-s}
\label{DS}
\ee

In our setting,  we have 
\be\label{4}
 \mbox{Hess} (\cH) = \frac{\partial^2 \cH}{\partial \lambda_i \pt \lambda_j}
 = \delta_{ij} \left(N +  \sum_{k \neq j} \frac{2}{(\lambda_j - \lambda_k)^2}
 \right) - \delta_{i\ne j} \frac{2}{(\lambda_i - \lambda_j)^2} 
\ge N \,\cdot \mbox{Id}
\ee
as a matrix inequality away from the singularities
(see remark below how to treat the singular set).
Thus we have
\be
\partial_{t}D(\sqrt f_t) \le - D(\sqrt f_t)
\label{eq:BE}
\ee
and by \eqref{lsi} 
\be\label{3.1}
\partial_{t} S(f_t) \le  -  S(f_t) 
\ee
This tells us that $S(f_t)$ in \eqref{dy} 
is exponential decaying as long as $t \gg 1$. 
But for any time $t \sim 1$ fixed, the entropy 
is still  the same order as the initial one. 
Note that  $t \sim 1$ is the case considered in  Johasson's work \cite{J}.   

\begin{remark}\label{BErem} {\rm 
The proof of \eqref{4} and the application of
the Bakry-Emery condition in \eqref{eq:BE} requires further
justification. Typically,
Bakry-Emery condition is applied for Hamiltonians $\cH$ defined
on spaces without boundary. Although the Hamiltonian $\cH$ \eqref{H} is
defined
on $\bR^N$, it is however convex
only away from any coalescence points $\lambda_i=\lambda_j$
for some $i\ne j$; the Hessian of the logarithmic terms 
has a Dirac delta singularity with the wrong (negative)
sign whenever two particles overlap. In accordance with the
convention that we work on the space $\Xi^{(N)}$ throughout the paper,
we have to consider $\cH$
restricted to $\Xi^{(N)}$, where it is convex, i.e. \eqref{4} holds,
but we have to verify that
the  Bakry-Emery result still applies. We review the proof of Bakry and
Emery
and prove that the contribution of the boundary term is zero.

Recall that the invariant measure $\exp(-\cH)\rd \bla$ 
and the dynamics $L = \frac{1}{2N} [\Delta - (\nabla \cH) \nabla]$ are
 restricted to $\Xi=\Xi^{(N)}$.
With $h=\sqrt{f}$ we have
$$
  \pt_t h^2 = Lh^2 = 2hLh + \frac{1}{N}(\nabla h)^2,
\quad\mbox{i.e.}\quad
 \pt_t h = Lh + \frac{1}{2N}h^{-1} (\nabla h)^2.
$$
Computing $\partial_t D(\sqrt{f_t})$, we have
\be
\begin{split}
  \pt_t \frac{1}{2N}\int_{\Xi} (\nabla h)^2 e^{-\cH} \rd \bla 
  & =\frac{1}{N} \int_\Xi \nabla h \nabla \Big( Lh+ \frac{1}{2N}
 h^{-1} (\nabla h)^2\Big)e^{-\cH} \rd\bla \cr
  & = \frac{1}{N}
  \int_\Xi \Big[ \nabla h L \nabla h -\frac{1}{2N} \nabla h (\nabla^2
\cH)\nabla h
+   \frac{1}{2N} (\nabla h)\nabla[h^{-1} (\nabla h)^2]\Big]e^{-\cH}
\rd\bla
\cr
& = \frac{1}{2N^2}\int_\Xi\Big[
-\nabla h (\nabla^2 \cH)\nabla h - \sum_{i,j}\big(\partial_{ij}^2 h -
\frac{\pt_i h
\pt_j h}{h}\Big)^2\Big] e^{-\cH} \rd\bla
\cr
&\le -D(\sqrt{f_t})
\end{split}
\ee
assuming that the boundary term
\be
   \int_{\pt \Xi} \pt_i h\; \pt_{ij}^2 h \; e^{-\cH}=0
\label{bdry}
\ee
in the integration by parts vanishes.

To see \eqref{bdry},
consider a  segment $\lambda_i=\lambda_{i+1}$ of the boundary $\pt \Xi$.
{F}rom the explicit representation \eqref{ft}, \eqref{gkernel}
in the next section,  we will see that $f_t \ge 0$ is a  
meromorphic function in each
variable in the domain $\Xi$ for any $t>0$. It can be represented as
by $(\lambda_{i+1}-\lambda_{i})^{\beta} F(\bla)$ with some $\beta\in\ZZ$,
where
$F$ is analytic and
$0< F<\infty$ near $\lambda_i=\lambda_{i+1}$. 
 Since $f_t\ge0$,
we obtain that the exponent $\beta$ is non-negative and even.
Therefore $f_t^{1/2}$ behaves as $|\la_{i+1}-\la_i|^{\beta/2}$
with a non-negative integer exponent $\beta/2$ near 
$\lambda_i=\lambda_{i+1}$.
It then follows that $\pt_i \sqrt{f}\; \pt_{ij}^2 \sqrt{f} e^{-\cH}$ 
vanishes at the boundary
due to the factor $(\lambda_{i+1}-\lambda_i)^2$
in $e^{-H}$, i.e. the integral \eqref{bdry} indeed vanishes.

}
\end{remark}

\subsection{Bound on the entropy}\label{sec:entropybound}

\begin{lemma}\label{lm:entropy} Let $s= N^{-1}$. 
For any $\al>\frac{1}{4}$ we have
\be\label {C1}
S_\mu(f_s):=S(f_s\mu|\mu) \le  C N^{1+\al}
\ee
with $C$ depending on $\al$.
\end{lemma}

{\it Proof.} 
In the proof  we consider the probability density $u(\bla)$
and the equilibrium measure $\mu$ extended
to  $\RR^N$ (see \eqref{def:u}), i.e. the eigenvalues 
are not ordered.
Clearly $S(f_s\mu|\mu) = S(f_s\wt\mu|\wt\mu)$ and we estimate
the relative entropy of the extended measures.

Given the density $f_0(\bla)\wt \mu(\rd\bla)$ of the eigenvalues
of the Wigner matrix
as an initial distribution, the eigenvalue density $ f_s(\bla)$ for the
matrix evolved under the Dyson's Brownian motion is given by
\be
      f_s(\bla) \wt u(\bla)= \int_{\RR^N} g_s(\bla,\bnu) \, 
f_0(\bnu) \wt u(\bnu)\rd \bnu
\label{ft}
\ee
with a kernel 
\be
g_s(\bla,\bnu) = \frac{N^{N/2}}{(2\pi)^{N/2}c^{N(N-1)/2}(1-c^2)^{N/2}}
\frac{\Delta_N(\bla)}{\Delta_N(\bnu)} 
\det\left( \exp\left[ 
\frac{-N(c\lambda_j-\nu_k)^2}{2(1-c^2)}\right]\right)_{j,k},
\label{gkernel}
\ee
where $c=c(s) = e^{-s/2}$ for brevity. The derivation of \eqref{gkernel}
follows very similarly to Johansson's presentation
of the Harish-Chandra/Itzykson-Zuber formula (see Proposition 1.1
of \cite{J}) with the difference that in our case the matrix elements
move by the Ornstein-Uhlenbeck process  \eqref{matrixOU} instead of
the Brownian motion.


 In particular,  formula \eqref{gkernel} implies 
that $f_s$ is an analytic function for any $s>0$ since
$$ 
 f_s(\bla) = \frac{h_s(\bla)}{\Delta_N(\bla)} \int_{\RR^N} 
\det\left( \exp\left[ 
\frac{-N(c\lambda_j-\nu_k)^2}{2(1-c^2)}\right]\right)_{j,k} 
 f_0(\bnu) \frac{\wt u(\bnu)}{\Delta_N(\bnu)}\rd \bnu
$$
with an explicit analytic function $h_s(\bla)$. Since the determinant
is analytic in $\bla$, we see that $f_s(\bla)$ is meromorphic in each
variables and
the only possible poles of $ f_s(\bla)$ come from the factors
$(\lambda_i-\lambda_j)^{-1}$
in $\Delta_N(\bla)$ near the coalescence points. But $ f_s(\bla)$ is a
non-negative function, so it cannot have a singularity of order $-1$,
thus these singular factors cancel out from a factor
$(\lambda_i-\lambda_j)$
from the integral. Alternatively, using the Laplace expansion the
determinant, one can explicitly
see that each 2 by 2 subdeterminant from the $i$-th and $j$-th columns
carry a factor $\pm (\lambda_i-\lambda_j)$.

Then, by Jensen inequality from \eqref{ft} and from the
fact that $f_0(\bnu) \wt{u}(\bnu)$ 
 is a probability density, we have
\[
S_{\wt\mu}(f_s ) = \int_{\RR^N}  f_s ( \log f_s) \rd \wt \mu
\leq  \iint_{\RR^N\times\RR^N}
\log\left(\frac{g_s(\bla,\bnu)}{\wt{u} (\bla)}\right) 
g_s(\bla,\bnu)\,  f_0(\bnu) \wt u(\bnu)\rd \bla\, \rd \bnu .
\]
Expanding this last expression we find, 
after an exact cancellation of the term $(N/2)\log(2\pi)$, 
\begin{eqnarray*}
S_{\wt\mu}(f_s) & \leq & \iint_{\RR^N\times\RR^N}
\left\{ \frac{N}{2}\log{N} - \frac{N(N-1)}{2}\log{c}
- \frac{N}{2}\log(1-c^2) + \log\Delta_N(\bla) -\log\Delta_N(\bnu) \right.
\\ 
&& + \log\det\left( \exp\left[  \frac{-N(c\lambda_j-\nu_k)^2}{2(1-c^2)}
\right]\right)_{j,k} 
- \frac{N^2}{2}\log{N} \\
&& \left.  + \frac{N}{2}\sum_{i=1}^N \lambda_i^2 -2\log\Delta_N(\bla) + 
\sum_{j=1}^N \log{j!} \right\} g_s(\bla,\bnu)  f_0(\bnu)\wt u(\bnu)\rd
\bla\rd \bnu.
\end{eqnarray*}
Since $s = N^{-1}$, we have 
$\log{c}=-1/2N$ and $ \log(1-c^2) =-\log N+O(N^{-1}) $.
Hence 
\begin{eqnarray}
S_{\wt\mu}(f_s) & \leq & \iint_{\RR^N\times\RR^N} \left\{ C N \log{N}
+ \log\Delta_N(\bla) -\log\Delta_N(\bnu) + 
\log\det\left( \exp\left[ 
\frac{-N(c\lambda_j-\nu_k)^2}{2(1-c^2)}\right]\right)_{j,k} 
\right. 
\nonumber
\\ && \left.  
- \frac{N^2}{2}\log{N}  + \frac{N}{2}\sum_{i=1}^N \lambda_i^2
-2\log\Delta_N(\bla)
+ \sum_{i=1}^N \log{j!} \right\} g_s(\bla,\bnu) f_0(\bnu)\wt u(\bnu)
\rd \bla\rd \bnu  . \label{123}
\end{eqnarray}
For the determinant term, we use that each entry is at most one, thus
\[
\log\det\left( \exp\left[ 
\frac{-N(c\lambda_j-\nu_k)^2}{2(1-c^2)}\right]\right)_{j,k} 
\leq \log N!.
\]
The last term in \eqref{123}
can be estimated using Stirling's formula and Riemann integration
\begin{equation}\begin{split}
\sum_{j=1}^N \log{j!} \leq \; & \sum_{j=1}^N  
\left( \log\left(\frac{j}{e}\right)^j 
+ C \log(2\pi j) \right)  \\
\leq \; & \int_1^{N+1} \rd x \, x \log x - \sum_{j=1}^N j + C N \log{N}
 \\ \leq \; &\frac{N^2 \log N}{2} -\frac{3}{4} N^2 + C N \log N
\end{split}\end{equation}
thus the $\frac{1}{2}N^2\log{N}$ terms cancel.
For the $N^2$ terms we need the following approximation
\begin{lemma}\label{lm:scappr} 
With respect to any Wigner ensemble whose single-site
distribution satisfies \eqref{Sobol}--\eqref{cond1}  and for
any $\al>1/4$ we have
\be
  \EE\Big[ \frac{N}{2}\sum_{i=1}^N \lambda_i^2 -2\log\Delta_N(\bla)\Big]
  = \frac{3}{4}N^2 
+O(N^{1+\al}),
\ee
where the constant in the error term depends on $\al$ and on the
constants in \eqref{Sobol}--\eqref{cond1}.
\end{lemma}
Note that \eqref{cond1}, \eqref{cond2} hold
for both the initial Wigner ensemble with density $f_0$
and for the evolved one with density $f_t$.
These conditions
ensure that  Theorem 3.5 of \cite{ESY3} is applicable.

\medskip

{\it Proof of Lemma \ref{lm:scappr}.} The quadratic term can be
computed explicitly using \eqref{trace}:
\be
   \frac{N}{2} \EE \sum_{i=1}^N \lambda_i^2 = \frac{N}{2} \EE \Tr H^2
= \frac{1}{2}N^2  = \frac{N^2}{2} \int x^2 \varrho_{sc}(x) \, 
\rd x ,
\label{quadr}
\ee
The second (determinant) term
will be approximated in the following
lemma whose proof is postponed to Appendix \ref{sec:det}.
\begin{lemma}\label{lm:detappr} 
 With respect to any Wigner ensemble whose single-site
distribution satisfies \eqref{Sobol}--\eqref{cond1}  and for
any $\al>1/4$ we have
\be
\EE \log\Delta_N(\bla) = 
\frac{N^2}{2} 
\int\!\int \log|x-y|\, \varrho_{sc}(x)\varrho_{sc}(y)\,\rd x\,\rd y
+O(N^{1+\al}).
\label{loglog1}
\ee
\end{lemma}

Finally, explicit calculation then shows that 
$$
\frac{1}{2} \int x^2 \varrho_{sc}(x) \, \rd x - 
\int\!\int \log|x-y|\, \varrho_{sc}(x)\varrho_{sc}(y)\,\rd x\,\rd y
 =\frac{3}{4},
$$
and this proves Lemma \ref{lm:scappr}. $\Box$

\bigskip

Hence, continuing the estimate \eqref{123}, we have the bound
\begin{eqnarray}
S_{\wt \mu}( f_s) & \leq &  C N^{1+\al} + \iint_{\RR^N\times\RR^N}
\left\{ \log\Delta_N(\bla)
-\log\Delta_N(\bnu)\right\} g_s(\bla,\bnu) f_0(\bnu)\wt u(\bnu)
\rd\bnu\rd\bla \nonumber \\
& \le &  C N^{1+\al} + \frac{N}{4} 
 \EE\sum_{j=1}^N [\la_j^2(s) -\la_j^2(0)] =  C N^{1+\al},
\label{entrop1}
\end{eqnarray}
where we used Lemma \ref{lm:scappr} both for the initial Wigner
measure and for the evolved one and finally we used that
the $\EE \, \Tr H^2$ is preserved, see \eqref{trace}.
This completes the proof of \eqref{C1}. $\Box$

\section{Local equilibrium}\label{sec:loceq}

\subsection{External and internal points}

Choose $t =\tau N^{-1}$ with some $\tau \ge 2$.
Thus  from \eqref{DS} with $s=N^{-1}$, we have
\be
D(\sqrt f_t) \le 2S(f_{N^{-1}}) t^{-1} \leq CN^{2+\al}\tau^{-1}
\label{Dbound}
\ee
by using \eqref{C1}.
Recall that the eigenvalues are ordered,
$\lambda_1 < \lambda_2  < \ldots < \lambda_N$.
Let $L\leq N-\K$  ($n$ was defined in \eqref{nchoice}) and define
$$ 
  \Pi_L (\bla) : = \{ \lambda_{L+1}, \lambda_{L+2},
\ldots  \lambda_{L+\K}\}
$$
and  
$$
  \Pi_L^c (\bla): =\{ \lambda_1, \lambda_2, \ldots \lambda_N\}
\setminus \Pi_L(\bla)
$$
its complement. For convenience, we will relabel the elements of $\Pi_L$
as
$\bx=\{ x_1, x_2, \ldots x_\K\}$ in increasing order.
The elements of $\Pi_L^c$ will be denoted by 
$$
  \Pi_L^c (\bla): = \by=
 (y_{-L}, y_{-L+1}, \ldots y_{-1}, y_1,
 y_2, \ldots y_{N-L-\K}) \in \Xi^{(N-\K)},
$$
again in increasing order ($\Xi$ was defined in \eqref{def:Xi}). We set
\be
  J_L : = \{ -L, -L+1, \ldots , -1, 1, 2, \ldots N-L-n\}
\label{JL}
\ee
to be the index set of the $y$'s.
We will refer to the $y$'s as {\bf external
points} and to the $x_j$'s as {\bf internal points}.
Note that the indices are 
chosen such that for any $j$
we have $y_k < x_j$ for $k<0$ and $y_k> x_j$ for $k>0$.
In particular, for any fixed $L$, we can split any $\by\in \Xi^{(N-\K)}$
as $\by=(\by_-, \by_+)$ where 
$$
   \by_-:=   (y_{-L}, y_{-L+1}, \ldots y_{-1}),
  \quad \by_+:= (y_1,  y_2, \ldots y_{N-L-\K}) 
$$
The set $\Xi^{(N-\K)}$ with a splitting mark after the $L$-th 
coordinate will be denoted by  $\Xi^{(N-\K)}_L$ and we use
the
$\by\in  \Xi^{(N-\K)} \Longleftrightarrow (\by_-, \by_+)\in\Xi^{(N-\K)}_L$
one-to-one correspondance.

For a fixed $L$ we will often consider the expectation of functions
$O(\by)$
on  $\Xi^{(N-\K)}$ with respect to $\mu$ or $f\mu$; 
this will always mean the marginal probability:
\begin{equation}
   \EE_\mu O :=  \int O(\by) u(\by_-, x_1, x_2, \ldots x_n, \by_+)
  \rd \by \rd \bx, \qquad \by = (\by_-, \by_+).
\end{equation}
\begin{equation}\label{eq:EEfO}
   \EE_f O :=  \int O(\by) (fu)(\by_-, x_1, x_2, \ldots x_n, \by_+)
  \rd \by \rd \bx.
  \end{equation}

For a fixed $L\le N-\K$
and $\by\in \Xi^{(N-\K)}$  let
\begin{equation}\label{eq:fLde}
f_{\by}^L (\bx)= f_{\by} (\bx)=  f_t(\by, \bx) \left [ 
\int f_t(\by, \bx) \mu_\by (\rd\bx) \right ]^{-1} 
\end{equation}
be the conditional density of $\bx$ given $\by$ with respect to the
conditional
equilibrium measure
\begin{equation}\label{eq:muyde}
 \quad 
\mu_{\by}^L (\rd\bx)=\mu_{\by} (\rd\bx)  = u_\by(\bx) \rd \bx, \qquad
u_\by(\bx):=  u(\by, \bx) \left [ \int u(\by, \bx) \rd \bx \right ]^{-1}
\end{equation}
Here $f_\by^L$ also 
depends on time $t$, but we will omit this dependence in the notation.
Note that for any fixed $\by\in \Xi^{(N-\K)}$, any value $x_j$
lies in the interval $I_\by:= [y_{-1}, y_1]$, i.e.
the functions $u_\by(\bx)$ and $f_\by(\bx)$
are supported on the set
$$
  \Xi_\by^{(\K)}: = \Big\{ \bx= (x_1, x_2, \ldots , x_{\K})\; : \;
  y_{-1} < x_1 < x_2 <\ldots < x_\K < y_1\Big\}\subset I_\by^\K.
$$

Now we localize the good set $\Omega$ 
introduced in Definition \ref{defi:good}.
For any fixed $L$ and $\by=(\by_-, \by_+)\in \Xi^{(N-n)}_L$ we define
$$
  \Omega_\by : =\{ \Pi_L(\bla) \; : \; \bla \in \Omega, 
\Pi_L^c(\bla) = \by \}
  =\{ \bx= (x_1, x_2, \ldots, x_n)\; : \;
 (\by_-, \bx,  \by_+) \in \Omega\}.
$$ 
Set
\be
 \Omega_1 =\Omega_1(L):=\big\{ \by \in \Xi^{(N-n)}_L \; : \;
\PP_{f_\by}(\Omega_\by) \ge
1- Ce^{-n^{\gamma/12}} \big\}. 
\label{def:om1}
\ee
Since
$$
  \PP(\Omega) = \PP_f \PP_{f_\by} (\Omega_\by) ,
$$
from \eqref{NI1} 
we have
\be
 \PP_f (\Omega_1) \ge 1- Ce^{-n^{\gamma/12}}.
\label{om1}
\ee
Here  $\PP_f (\Omega_1)$ is a short-hand notation for
the marginal expectation, i.e.
$$
  \PP_f (\Omega_1) :=\PP_f \big[ (\Pi_L^c)^{-1}(\Omega_1)\big],
$$
but we will neglect this distinction.

Note that $\by\in \Omega_1$ also implies, for large $N$, that 
there exists an $\bx\in I_\by^n$ such that $(\by_-,\bx, \by_+)\in
\Omega$. This ensures  that those  properties
of $\bla \in\Omega$ that are determined only by $\by$'s, 
will be inherited to the $\by$'s. E.g. $\by\in \Omega_1$
will guarantee that the local density of $\by$'s is
close to the semicircle law on each interval away from $I_\by$.
More precisely, note that
for any interval $I=[E-\eta_m^*/2, E+\eta_m^*/2]$
of length $\eta_m^* = 2^m n^\gamma N^{-1}$ and center $E$, $|E|\le
2-\kappa$,
that is disjoint from $I_\by$, we have,  by (\ref{omm}),
\be
   \by \in \Omega_1, \quad I\cap I_\by =\emptyset
 \quad \Longrightarrow \quad
\Big| \frac{\N(I)}{N|I|} - \varrho_{sc}(E)\Big| \le \frac{1}{
(N|I|)^{1/4}}
  n^{\gamma/12} \, .
\label{om3prop}
\ee
Moreover, for any interval $I$ with  $|I|\ge n^\gamma N^{-1}$ we have, 
 by (\ref{Omstar}),
\be
   \by \in \Omega_1, \quad I\cap I_\by =\emptyset
 \quad \Longrightarrow \quad
\N(I)\le K N|I|.
\label{om3prop1}
\ee

For any $L$ with $N\kappa^{3/2}\le L \le N(1-\kappa^{3/2})$,
let $E_L = \fN_{sc}^{-1}(LN^{-1})$, i.e.
\be
   N\int_{-2}^{E_L} \varrho_{sc}(\lambda)\rd \lambda = L.
\label{ELdef}
\ee
Then we have
\be
 -2+C\kappa\leq      E_L\leq 2- C\kappa, \qquad
 \varrho_{sc}(E_L) \ge c\kappa^{1/2}
\label{ELbound}
\ee
Using \eqref{y} and \eqref{yy} from Lemma \ref{lm:grid} on the set
$\Omega$
(see \eqref{Omstar}), we
for any $\by\in\Omega_1(L)$ we have
\be
   | y_{-1} - \fN_{sc}^{-1}(LN^{-1})| \le Cn^{-\gamma/6}, \qquad
\Big| |I_\by| - \frac{n}{N\varrho_{sc}(E_L)}\Big| 
  \leq   CN^{-1}n^{\gamma+3/4} 
\label{Il1}
\ee
in particular
\be
 |y_{-1}|, |y_1|\le 2-\kappa/2 \quad \mbox{and} 
\quad |I_\by|\leq  \frac{Cn}{N} 
\label{Ilength}
\ee
with $C=C(\kappa)$.

Let
\be
 \Omega_2= \Omega_2(L)= \Big\{ (\by_-, \by_+)\in \Xi^{(N-n)}_L, 
\; : \; |I_\by| \le KnN^{-1}\Big\}
\label{def:om2}
\ee
with some large constant $K$. 
On the set $\Omega$ we have $|I_\by|\le Kn/N$ (see \eqref{Ilength}),
thus  $\Pi_L^c(\Omega)\subset \Omega_2(L)$, i.e.
\be
 \PP_f (\Omega_2) \ge 1- Ce^{-n^{\gamma/6}}.
\label{om2}
\ee

\subsection{Localization of the Dirichlet form}

For any $L\le N-\K$
and any $\by\in \Xi^{(N-\K)}_L$,  we define the Dirichlet form 
$$
  D_{L,\by} (f) : = \int \frac{1}{2N}\big(\nabla_\bx f)^2
\rd\mu_\by^L(\bx)
$$
for functions $f=f(\bx)$ defined on $\Xi^{(n)}_\by$.
Hence from \eqref{Dbound} we have the inequality 
\be\label{ldb}
\frac{1}{N(1-2\kappa^{3/2})} \sum_{L=N\kappa^{3/2}}^{N(1-\kappa^{3/2})}
 \EE_{f_t} D_{L,\by} (\sqrt {f_{\by} (\bx)} ) 
\le C {\K} N^{-1} D( \sqrt f_t) 
\le  CN^{1+\al}\K\tau^{-1}
\ee
where the expectation $\EE_{f_t}$ is defined similarly to 
(\ref{eq:EEfO}), with $f$ replaced by $f_t$. In the first
 inequality in (\ref{ldb}), we used the fact that, by (\ref{eq:fLde})
 and (\ref{eq:muyde}),
\[  \begin{split} \EE_{f_t} D_{L,\by} &(\sqrt {f_{\by} (\bx)} ) 
 \\ = \; & \int \rd \bx \rd \by \;  f_t (\by,\bx) u(\by,\bx) \,
 D_{L,\by} (\sqrt {f_{\by} (\bx)} ) \\ = \; &\frac{1}{8N} 
\int \rd \bx \rd \by \,  f_t (\by,\bx) u(\by,\bx) 
\left[ \int \rd \bx'  \,  \frac{|\nabla_{x'} f_t (\by,\bx')|^2}{f_t (\by,\bx')}
 \, \frac{1}{\int \rd \bx' f_t (\by,\bx') u (\by, \bx')} u(\by,\bx') 
\right] \\ = \; &\frac{1}{8N} \sum_{j=1}^n 
\int \rd \bx \rd \by  \,  \frac{|\nabla_{x_j}
 f_t (\by,\bx)|^2}{f_t (\by,\bx)} \, u(\by,\bx) 
\end{split}\]
and therefore, when we sum over all $L \in \{ N \kappa^{3/2} , \dots ,
 N(1-\kappa^{3/2}) \}$ as on the l.h.s. of (\ref{ldb}), every
local Dirichlet form  is summed over at most $n$ times,
so we get the total Dirichlet form with a multiplicity at most $n$.

We define the set
\be
  \cG_1 =\Big\{ N\kappa^{3/2} \le L \le N(1-\kappa^{3/2}) \; :
\;  \EE_{f_t} D_{L,\by} (\sqrt {f_{\by} (\bx)} )
  \leq CN^{1+\al}\K^2\tau^{-1}\Big\},
  \label{def:cG}
\ee
then the above inequality guarantees that for the cardinality of $\cG_1$,
\be
  \frac{ |\cG_1|}{N(1-2\kappa^{3/2})} \ge 1- \frac{C}{n}.
\label{fractioncG1}
\ee
For $L\in \cG_1$, we define
\be
 \Omega_3=\Omega_3(L): =  \Big\{(\by_-, \by_+)\in \Xi_L^{(N-n)}\; : \;
  D_{L,\by} (\sqrt {f_{\by} (\bx)} )
  \leq CN^{1+\al}n^4\tau^{-1} \Big\},
\label{def:om4}
\ee
then 
\be
 \PP_f (\Omega_3^c) \le Cn^{-2}.
\label{om4}
\ee

\subsection{Local entropy bound}

Suppose that $L\in \cG_1$ and fix it.
For any $\by\in \Xi^{(N-\K)}_L$ denote by 
\be
\cH_{\by} (\bx) =  
N \left [ \sum_{i=1}^n \frac{1}{2} x_{i}^{2}
-  \frac{2}{N} \sum_{1\le i< j\le n} 
\log |x_{j} - x_{i}|  -    \frac{2}{N} \sum_{k\in J_L}\sum_{ i=1 }^n 
\log |x_{i} - y_{k}|   \right ]
\ee
Note that
\be\label{C2}
\mbox{Hess}  \, \cH_{\by}(\bx)  
\ge \inf_{ x\in I_\by}\sum_{k\in J_L}   |x- y_{k}|^{-2} 
\ee
for any $\bx\in I_\by^\K$ as a matrix inequality. 
On the set $\by\in \Omega_2(L)$ we have
$$
\inf_{ x\in I_\by}\sum_{k\in J_L}   |x- y_{k}|^{-2}  \ge 
\frac{1}{|y_1-y_{-1}|^2} \ge \frac{cN^2}{n^2}, \qquad \by \in \Omega_2(L).
$$
We can apply the logarithmic Sobolev inequality \eqref{lsi} 
to the local measure $\mu_\by$, taking into account Remark \ref{BErem}.
Thus  we have 
\be\label{ent1}
S_{\mu_\by^L}(f_{\by} ) \le c^{-1}{\K}^2
   N^{-1}D_{L,\by} (\sqrt {f_{\by} (\bx)} )
\le  C{\K}^{6} N^{\al}\tau^{-1} \quad 
\mbox{for any $\by\in\Omega_2(L)\cap\Omega_3(L)$, $L\in \cG_1$}.
\ee
Using the inequality
\be
\sqrt {S(f)} \ge C \int |f-1| \rd \mu
\label{entrop}
\ee
for $\mu=\mu_\by$ and $f=f_\by$,
we have also have
\be\label{L1}
\left [ \int |f_{\by}-1| \rd \mu_\by \right ]^{2} \le  
C{\K}^{6} N^{\al}\tau^{-1} \quad \mbox{for any
$\by\in\Omega_2(L)\cap\Omega_3(L)$,
$L\in \cG_1$}
\ee
We will choose $t= N^{-1}\tau $ with $\tau = N^{\beta}$ such that
\be
    C{\K}^{6} N^{\al}\tau^{-1} \le n^{-4}
\label{L1choice}
\ee
i.e. $\beta\ge 10\e + \al$.

\subsection{Good external configurations}\label{sec:ext}

\begin{definition}
The set of {\bf good $L$-indices} is defined
by
\be
\begin{split}
 \cG : =&  \Big\{ L \in \cG_1\; :
\; \EE_f  \sum_{j\ne L}
 \frac{{\bf 1}_\Omega
 }{[N(\lambda_j-\lambda_L)]^2} \leq   C\K^{3\gamma}, \;
 \EE_f \sum_{j\ne L+n+1}
 \frac{{\bf 1}_\Omega}{[N(\lambda_j-\lambda_{L+n+1})]^2} \leq  
C\K^{3\gamma}
 \Big\}
\cr
&\cap\Big\{ L \in \cG_1\; : \;
\EE_f \sum_{j\ne L}
 \frac{{\bf 1}_\Omega }{N|\lambda_j-\lambda_L|} \leq   C\K^{3\gamma},
\quad
 \EE_f 
\sum_{j\ne L+n+1}
 \frac{{\bf 1}_\Omega }{N|\lambda_{L+\K+1}-\lambda_j|}  \leq  
C\K^{3\gamma}
\Big\}.
\label{def:cg}
\end{split}
\ee
\end{definition}
Lemma \ref{lm:rep}
together with  \eqref{fractioncG1}
imply that
\be
  \frac{ |\cG|}{N(1-2\kappa^{3/2})} \ge 1-\frac{1}{\K^\gamma}.
\label{fractioncG}
\ee

Notice that for any fixed $L$ we can write
$$
 \EE_f  \sum_{j=L+1}^{L+\K}
 \frac{{\bf 1}_\Omega }{N(\lambda_j-\lambda_L)}
 = \EE_f \EE_{f_\by} \sum_{j=1}^\K
 \frac{{\bf 1}_{\Omega_\by} }{N(x_j-y_{-1})}
$$
$$
\EE_f  \sum_{j= L+1}^{L+n}
 \frac{ {\bf 1}_\Omega 
 }{[N(\lambda_j-\lambda_L)]^2} 
= \EE_f \EE_{f_\by} \sum_{j=1}^n
 \frac{{\bf 1}_{\Omega_\by}
 }{[N(x_j- y_{-1})]^2} 
$$
and we also have 
$$
 \EE_f  \sum_{ j\ne L+1, \ldots L+n}
 \frac{{\bf 1}_\Omega }{N|\lambda_j-\lambda_L|}
 = \EE_f  \sum_{j\in J_L, j\ne -1}
 \frac{1 }{N|y_j-y_{-1}|} \PP_{f_\by} (\Omega_\by) 
 \ge \frac{1}{2}\EE_f\sum_{j\in J_L, j\ne -1}
 \frac{ {\bf 1}(\by\in\Omega_1) }{N|y_j-y_{-1}|},
$$
and similar formulae hold when $\la_L$ is replaced with $\la_{L+n+1}$
and $y_{-1}$ with $y_1$. 

We also want to ensure that the density on scale 
$\eta :=\eta_0^* = n^\gamma N^{-1}$
is close to the semicircle law. Let
$$
   \O_E (x) =  {\bf 1}(|x-E|\le \eta/2)
$$
be the  characteristic function of
the interval $[E-\eta/2, E+\eta/2]$. 
Consider $\Omega^{(0)}$ defined in \eqref{omm}, then
$\Omega\subset\Omega^{(0)}$ and \eqref{NI1} imply that
$$
    \EE_f {\bf 1}_\Omega \sup_{|E|\le 2-\kappa/2}\Big| \frac{1}{N\eta}
  \sum_{i=1}^N \O_E(\lambda_i) -\varrho_{sc}(E)\Big|
 \le (N\eta)^{-1/4} n^{\gamma/12} = n^{-\gamma/6}
$$
Fix $L\in \cG$, consider $\by \in \Xi^{(N-n)}_L$ and define
$$
  I^*_\by: = [y_{-1} + \eta/2, y_1-\eta/2]
$$
so that if $E \in I^*_\by$ then
$[E-\eta/2, E+\eta/2]\subset I_\by$. Moreover, on
the set $\Omega$ we know that $I_\by \subset [-2+\kappa/2, 2-\kappa/2]$
(see \eqref{Ilength}).
Therefore
\be
  \EE_f \EE_{f_\by} {\bf 1}_{\Omega_\by} 
\sup_{E\in I^*_\by }\Big| n^{-\gamma}
  \sum_{i=1}^n \O_E(x_i) -\varrho_{sc}(E)\Big| \le 
 \EE_f {\bf 1}_\Omega \sup_{|E|\le 2-\kappa/2}\Big| \frac{1}{N\eta}
  \sum_{i=1}^N \O_E(\lambda_i) -\varrho_{sc}(E)\Big|\le n^{-\gamma/6}.
\label{gg1}
\ee

This gives rise to the following definition:

\begin{definition}\label{def:good}
Let $L\in \cG$. The 
{\bf set of good external points} is given by
\be
\begin{split}
   \cY_L: = &  \Omega_1\cap \Omega_2\cap \Omega_3\cap
 \Bigg\{  \by = (\by_-, \by_+)\in \Xi^{(N-\K)}_L : \;
\sum_\pm \sum_{k\in J_L\atop k \ne \pm 1}
  \frac{1}{|N(y_{\pm 1}-y_k)|}\le  C\K^{3\gamma} , \cr
&   \EE_{f_\by} \sum_\pm
\sum_{j=1}^\K
 \frac{{\bf 1}_{\Omega_\by} }{N|x_j-y_{\pm 1}|} \leq   C\K^{4\gamma}, \;
\EE_{f_\by}  \sum_\pm\sum_{j=1}^n
 \frac{ {\bf 1}_{\Omega_\by} 
 }{[N(x_j- y_{\pm 1})]^2} \le C\K^{4\gamma}, \cr
 & \EE_{f_\by} {\bf 1}_{\Omega_\by} 
\sup_{E\in I^*_\by }\Big| n^{-\gamma}
  \sum_{i=1}^n {\bf 1}\Big(N|x_i-E| \le \frac{1}{2} n^\gamma\Big)
 -\varrho_{sc}(E)\Big| \le n^{-\gamma/12}
 \Bigg\} 
\label{def:goody}
\end{split}
\ee
\end{definition}
It follows from \eqref{om1}, \eqref{om2},  \eqref{om4},
\eqref{def:cg} 
and \eqref{gg1}
that
\be
  \PP_f \big( \cY_L \big)\ge 1-  C\K^{-\gamma/12}.
\label{cyprob}
\ee

\subsection{Bounds in equilibrium}
\label{sec:potdeneq}

In this section we translate the bounds in the second and third
lines of \eqref{def:goody}
into similar bounds with respect to equilibrium
using that the control on the local Dirichlet form 
also controls the local entropy for the good indices:

\begin{lemma}\label{lm:equipot}
Let $A>0$ be arbitrary and $\by\in\cY_L$. If $\tau \ge \K^{4A+8}N^\alpha$,
i.e. $\beta \ge (4A+8)\e+\al$,
then
for $p=1,2$ we have
\be
\EE_{\mu_\by}  \sum_\pm \sum_{j=1}^\K
 \frac{  {\bf 1}( N|x_j-y_{\pm 1}| 
\ge \K^{-A})}{[N|x_j-y_{\pm 1}|]^p} 
\leq C\K^{4\gamma}
\label{poteq}
\ee
Moreover, we also have
\be
\EE_{\mu_\by} 
\sup_{E\in I^*_\by }\Big| n^{-\gamma}
  \sum_{i=1}^n \O_E(x_i) -\varrho_{sc}(E)\Big| 
\le Cn^{-\gamma/12}.
\label{gg2}
\ee
\end{lemma}

{\it Proof.} Let $\O:\bR^n\to \bR$ be any observable and $\Omega_\by$ 
be any event.
Then for any fixed $\by\in \Xi^{(N-\K)}$ we have
$$
  \big|  \EE_{f_\by} {\bf 1}_{\Omega_\by}
\O -  \EE_{\mu_\by}  {\bf 1}_{\Omega_\by}
\O \big|^2 =
\Big[ \int {\bf 1}_{\Omega_\by}\O (f_\by-1)\rd \mu_\by \Big]^2
  \leq \| O \|_\infty^2 \Big[\int |f_\by-1|\rd\mu_\by \Big]^2
 \leq C \| O \|_\infty^2 S_{\mu_\by}(f_\by )
$$
by the entropy inequality \eqref{entrop}. If $L\in \cG$ and
$\by\in \Omega_2(L)$, then we have by \eqref{L1} that
\be
   \EE_{\mu_\by} {\bf 1}_{\Omega_\by}\O \leq  \EE_{f_\by} 
{\bf 1}_{\Omega_\by}\O +
C \| \O \|_\infty \Big( \K^6 N^\alpha
 \tau^{-1}\Big)^{1/2}.
\label{entrr}
\ee
For a given $\by\in \cY_L$, we set the observable
$$
 \O(\bx) =\sum_\pm\sum_{j=1}^\K 
  \frac{ {\bf 1}( N(x_j-y_{\pm 1}) \ge \K^{-A})}{[N|x_j-y_{\pm 1}|]^p}.
$$
with $\|\O\|_\infty \le Cn^{Ap+1}\le cn^{2A+1}$.
Then, for
$\tau \ge \K^{4A+8}N^\alpha$ 
we obtain from \eqref{def:goody} and
  \eqref{entrr} that
$$
  \EE_{\mu_\by}  \sum_\pm \sum_{j=1}^\K
 \frac{{\bf 1}_{\Omega_\by}  {\bf 1}( N|x_j-y_{\pm 1}|
\ge \K^{-A})}{[N|x_j-y_{\pm 1}|]^p} \leq   C\K^{4\gamma} + 
C\K^{2A+4} N^{\alpha/2}\tau^{-1/2} \leq C\K^{4\gamma}.
$$
On the complement set $\Om_\by^c$ we just use the crude supremum
bound together with the bound on $\PP_{f_\by} (\Omega_\by^c)$
in the definition of $\Omega_1$ \eqref{def:om1}:
$$
  \EE_{\mu_\by}  \sum_\pm \sum_{j=1}^\K
 \frac{{\bf 1}_{\Omega_\by^c}  {\bf 1}( N|x_j-y_{\pm 1}|
\ge \K^{-A})}{[N|x_j-y_{\pm 1}|]^p} \leq C n^{4A+1} e^{-n^{\gamma/12}} 
\le Cn^{4\gamma}.
$$
Combining the last two estimates
proves \eqref{poteq}.

The proof of \eqref{gg2} is analogous, here we use that 
the corresponding observable has an $L^\infty$ bound
$$
\Big| n^{-\gamma}
  \sum_{i=1}^n \O_E(x_i) -\varrho_{sc}(E)\Big| \le n^{1-\gamma}.
$$
This completes the proof of Lemma \ref{lm:equipot}.
$\Box$

\section{Cutoff Estimates}\label{sec:cutoff}

In this section, we cutoff the interaction 
with the far away particles.  We fix a good index $L\in \cG$
and a good external point configuration $\by\in \cY_L$.
Consider the measure  $\mu_\by= e^{-\cH_\by}/Z_\by$ with 
\be
\cH_{\by} (\bx) =  
N \left [ \sum_{i=1}^{\K} x_{i}^{2}/2 -  2 N^{-1} \sum_{1\leq i< j\leq n}
\log |x_{j} - x_{i}|  -    2 N^{-1} \sum_{k, i } 
\log |x_{i} - y_{k}|   \right ]
\label{def:Hy}
\ee
The measure $\mu_\by$ is supported on the  interval $I_\by=(y_{-1}, y_1)$.

For any fixed $\by$,  decompose 
\be
\cH_\by = \cH_1+ \cH_2,     \quad \cH_2 (\bx)=   \sum_{i=1}^\K  
V_{2}(x_i),
\ee
where 
\be
V_{2} (x) = \frac{N}{2} x^{2}  -    2  \sum_{ |k|\ge n^B } 
\log |x - y_{k}|
\ee 
and
\be
  \cH_1(\bx)=
-  2  \sum_{1\leq i< j\leq n} 
\log |x_{j} - x_{i}|  -     \sum_{i=1}^n V_1(x_i) 
\label{def:H1y}
\ee
with 
$$  
  V_1(x)  =  -    2  \sum_{ |k|< n^B } 
\log |x - y_{k}|
$$
where $B$ is a large positive number with $B\e< 1/2$.
We define the measure
\be
  \mu_\by^{(1)}(\rd \bx) := \frac{e^{- \cH_1(\bx)}\rd \bx}{Z_1}.
\label{mu1}
\ee
\begin{lemma}\label{lm:mumu} Let $L\in \cG$ and $\by\in \cY_L$.
For  $B\ge20$, we have
\be
  \sup_{\bx\in I_\by^n} \Big| 
 \frac{\rd \mu^{(1)}_\by}{\rd \mu_\by}(\bx)-1\Big|
 \leq C n^{-B/9 +2}
\label{mumu}
\ee
\end{lemma}
This lemma will imply
that one can cutoff all $y_k$'s in the potential
with $|k|\ge n^B$.

\bigskip

{\it Proof.} Let 
$$
 \delta V_2:= \max_{x\in I_\by} V_2 - \min_{x\in I_\by} V_2,
$$
then, by \eqref{def:om2} and $\by\in \cY_L$, we have 
$$
   \delta V_2\leq |I_\by| \| V'_2\|_\infty \le CnN^{-1} \| V'_2\|_\infty
$$
In Lemma \ref{lm:v2} we will give an upper bound on $\| V'_2\|_\infty$,
and then we have, for $B\ge 20$, that
$$
  \delta V_2 \leq Cn^{-B/9+1} .
$$
Since 
$$
 \Big| 
 \frac{\rd \mu^{(1)}_\by}{\rd \mu_\by}(\bx)-1\Big|
=\Big| e^{-\sum_{i=1}^n \big[ V_2(x_i)- \min V_2\big]}-1\Big| \leq 
 Cn\delta V_2 \leq Cn^{-B/9+2},
$$
we obtain \eqref{mumu}. $\Box$

\begin{lemma}\label{lm:v2} 
For $B\ge20$
 and for any $L\in\cG_1$, $\by\in \cY_L$  we have
\be\label{A1} 
\sup_{x\in I_\by}|V_{2}'(x)|  = \sup_{x\in I_\by}
\left |  -2 \sum_{ |k|\ge n^B } 
\frac 1 {x-y_{k}} + Nx \right | 
\le  CN \K^{\gamma/12-B/8}.
\ee
\end{lemma}

{\it Proof.} Recall that $ \by\in \cY_L\subset \Omega_1$ implies
that the density of the $y$'s is close the semicircle law in
the sense of \eqref{om3prop}. 
Let 
\be
  d:= \frac{ n^B}{N\varrho_{sc}( y_{-1})}.
\label{def:d}
\ee
Since $\by\in \Omega_1$, we know that
$|y_{-1}|, |y_1| \le 2-\kappa/2$ (see \eqref{Ilength}), thus
$\varrho_{sc}(y_{-1})\ge c>0$. Taking the imaginary part of \eqref{def:msc}
for $|z|\le 2$ and renaming the variables,
we have the identity
$$
  x  = 2\int_\bR \frac{\varrho_{sc}(y)}{x-y} \rd y.
$$
Furthermore, with $\bar y= \frac{1}{2}(y_{-1}+y_1)$ we have
$$
  \Big|\int_{|y-\bar y|\le d} \frac{\varrho_{sc}(y)}{x-y} \rd y\Big|
  \le Cd
$$
since $\bar y$ is away from the spectral
edge thus $\varrho_{sc}$ is continuously differentiable
on the interval of integration $[\bar y - d, \bar y +d]$. 
Thus 
$$
 \Big| Nx - 2N\int_{|y-\bar y|\ge d} \frac{\varrho_{sc}(y)}{x-y} \rd
y\Big|
\le CNd \le Cn^B
$$
therefore to prove \eqref{A1} it is sufficient to show
that
\be
\sup_{x\in I_\by} \Big| \frac{1}{N}\sum_{ |k|\ge n^B } 
\frac 1 {x-y_{k}} - 
\int_{|y-\bar y|\ge d} \frac{\varrho_{sc}(y)}{x-y} \rd y\Big|\le
Cn^{\gamma/12
-B/8}
\label{suf}
\ee
We will consider only $k\ge n^B$ and compare the sum with
the integral on the regime $y\ge \bar y +d$, the sum for $k\le -n^B$
is similar.

Define dyadic intervals
$$
  I_m = [\bar y + 2^{m} d, \bar y
+ 2^{m+1} d], \qquad m=0,1,2, \ldots , \log N
$$
Since
$\by\in \cY_L\subset \Omega_1$, i.e. $\max |y_k|\le K$,
there will be no $y_k$ above the last interval $I_{\log N}$.
We subdivide each $I_m$ into $n^{B/2}$ equal
disjoint subintervals of length $2^m d n^{-B/2}$
$$
  I_m =\bigcup_{\ell=1}^{n^{B/2}} I_{m,\ell}, \qquad
 I_{m,\ell} = [y^*_{m,\ell-1}, y_{m, \ell}^*]
\quad\mbox{with}\quad
y_{m,\ell}^* : =y_1 + 2^m d(1+\ell n^{-B/2}) .
$$
For $\by\in \cY_L\subset\Omega_1$, the estimate \eqref{yy} holds
for $y_1$ and $y_{n^B}$, i.e.
$$
  \Big| N\varrho_{sc}(y_1)( y_{n^B}- y_1) - (n^B-1)\Big| \le Cn^{\gamma+3B/4}
 \le Cn^{4B/5}
$$
if $B\ge 20$, 
which means that
\be
  |y_{n^B} - (y_1+d)| \le \frac{C n^{4B/5}}{N} + 
 \frac{n^B}{N} \Big|\frac{1}{\varrho_{sc}(y_{-1})} -
\frac{1}{\varrho_{sc}(y_1)}\Big| \le \frac{C n^{4B/5}}{N}
 +   \frac{C n^{B+1}}{N^2} \le \frac{C n^{4B/5}}{N}
\label{me1}
\ee
(using $B\e<1/2$, $n^B\le N^{1/2}$),
i.e.
\be
  |y_{n^B} - (\bar y+d)| \le \frac{Cn^{4B/5}}{N}
\label{me}
\ee
by using the definition of $d$ from \eqref{def:d},
the fact that $\varrho_{sc}(y_{\pm 1})$ is separated away from
zero and that $|I_\by|\le CnN^{-1}$ from \eqref{Ilength}.

Therefore
we can estimate
\be
\begin{split}
\label{calib}
\Big| \frac{1}{N} \sum_{ k\ge n^B } 
\frac 1 {x-y_{k}}  -& \frac{1}{N}\sum_{m= 0}^{\log
N}\sum_{\ell=1}^{n^{B/2}} 
\sum_{j\in J_L} \frac{ {\bf 1}(y_j\in I_{m,\ell})}{x-y_j}\Big| \cr
& \le
 \frac{1}{N} \sum_{ j\in J_L }\frac{{\bf 1}( j< n^B, y_j\ge \bar 
y+d)}{|x-y_j|}
 +  \frac{1}{N} \sum_{ j\in J_L }
\frac{{\bf 1}( j\ge n^B, y_j< \bar y +d)}{|x-y_j|} \cr
& \le Cn^{1-B/5}.
\end{split}
\ee
To see the last estimate, we notice that 
in the first summand we have $\bar y+d\le y_j \le y_{n^B} \le 
\bar y+d+ Cn^{4B/5} N^{-1}$ by \eqref{me}, i.e. all these $y_j$'s
lie in an interval of length $Cn^{4B/5}N^{-1}$, so their
number is bounded by $Cn^{4B/5}$ by \eqref{om3prop1}.
Thus the first term in the right hand side of 
\eqref{calib} is bounded by $Cn^{4B/5} N^{-1}d^{-1} \le Cn^{1-B/5}$;
the estimate of the second term is similar.

Using that
$$
  \Big| \max_{y\in I_{m,\ell}} \frac 1 {x-y} -
  \min_{y\in I_{m,\ell}} \frac 1 {x-y}\Big| \le  |I_{m,\ell}|
 \max_{x\in I_\by} \max_{y\in I_{m,\ell}} \frac{1}{(x-y)^2}
\le C\frac{2^m d n^{-B/2}}{(2^m d)^2} \le\frac{C}{2^m d n^{B/2}}
$$
we have
\be
\begin{split}\label{2e}
  \Big|  \frac{1}{N}\sum_{m=0}^{\log N}\sum_{\ell=1}^{n^{B/2}} 
\sum_{j\in J_L} \frac{ {\bf 1}(y_j\in I_{m,\ell})}{x-y_j}
 -\frac{1}{N}\sum_{m= 0}^{\log N}\sum_{\ell=1}^{n^{B/2}} 
\frac{\N (I_{m,\ell})}{x-y^*_{m,\ell}} \Big| 
& \le \frac{C}{N}\sum_{m=0}^{\log N}\sum_{\ell=1}^{n^{B/2}} 
\frac{\N (I_{m,\ell})}{ 2^md n^{B/2}}  \cr
& \le C\sum_{m=0}^{\log N}\sum_{\ell=1}^{n^{B/2}} 
\frac{1}{ n^{B}}\cr & \le Cn^{-B/2}\log N \le Cn^{-B/4}.
\end{split}
\ee
In the second line we used that $\N(I_{m,\ell})\le KN|I_{m,\ell}|$
by \eqref{om3prop1}
since $\by\in \Omega_1$ and $I_{m,\ell}\cap I_\by=\emptyset$.

We use that for $\by\in \Omega_1$ we can apply
\eqref{om3prop} for $I=I_{m,\ell}$ and we get
\be
\begin{split} \label{3e}
\Big| \frac{1}{N}\sum_{m\ge 0}\sum_{\ell=1}^{n^{B/2}} 
\frac{\N (I_{m,\ell})}{x-y^*_{m,\ell}}  
-  \frac{1}{N}\sum_{m= 0}^{\log N}\sum_{\ell=1}^{n^{B/2}} 
\frac{ N|I_{m,\ell}|\varrho_{sc}(y_{m,\ell}^*)}{x-y^*_{m,\ell}} 
\Big|  &\le  \frac{Cn^{\gamma/12}}{N}\sum_{m =0}^{\log N}
\sum_{\ell=1}^{n^{B/2}} 
\frac{ (N|I_{m,\ell}|)^{3/4} }{|x-y^*_{m,\ell}|} 
\cr
&\le \frac{Cn^{\gamma/12}}{N}\sum_{m= 0}^{\log N}\sum_{\ell=1}^{n^{B/2}} 
\frac{ ( 2^mn^{B/2})^{3/4} }{2^m n^B N^{-1}}  \cr
&\le Cn^{\gamma/12 - B/8},
\end{split}
\ee
where we used that $|I_{m,\ell}|= 2^md n^{-B/2} \le C\cdot
2^mn^{B/2}N^{-1}$
(see \eqref{def:d}) and that $|x-y^*_{m,\ell}|\ge 2^{m-1} d \ge c \cdot
2^m n^BN^{-1}$.

Finally, the second term on the left hand side of \eqref{3e} is
a Riemann sum of the integral in \eqref{suf} with an error
\be
 \Bigg|\sum_{m= 0}^{\log N}\sum_{\ell=1}^{n^{B/2}} 
\frac{ |I_{m,\ell}|\varrho_{sc}(y_{m,\ell}^*)}{x-y^*_{m,\ell}} 
-\int_{|y-\bar y|\ge d} \frac{\varrho_{sc}(y)}{x-y} \rd y \;\Bigg|\le
 \sum_{m= 0}^{\log N}\sum_{\ell=1}^{n^{B/2}} C\Big(\frac{
N}{2^mn^B}\Big)^2
 |I_{m,\ell}|^2 \le Cn^{-B/2}\log N,
\label{4e}
\ee
since on each interval $I_{m,\ell}$ we could estimate the derivative of
the integrand as
$$
   \sup_{y\in I_{m,\ell}} \Big| \frac{\rd}{\rd y} \frac{\varrho_{sc}(y)}
  {x-y}\Big|\le C\Big(\frac{ N}{2^mn^B}\Big)^2.
$$
Combining \eqref{calib}, \eqref{2e}, \eqref{3e} and \eqref{4e},
we have proved \eqref{suf} which completes the proof of
Lemma \ref{lm:v2}. $\Box$

\section{Derivative  Estimate of Orthogonal Polynomials}\label{sec:derop}

In the next few sections, we will prove the boundedness and small distance
regularity of the density.   Our proof 
follows  the approach of \cite{PS} (cf:  Lemma 3.3 and 3.4 in \cite{PS}), 
but the estimates are done in
a  different way due to the singularity of the potential.  
For the rest of this paper, it is convenient to  rescale the local
equilibrium 
measure to  the interval $[-1,1]$ as we now explain.

Suppose $L \in \cG$ and $\by\in \cY_L$. We change variables
by introducing the transformation
$$
T: I_\by\to [-1,1], \qquad
\wt w =T(w): = \frac{2( w - \bar y)}{|I_\by|}, \quad 
 \mbox{with}\quad  \bar y: = 
  \frac{y_{-1} + y_1}{2}
$$
and its inverse 
$$
    w = T^{-1}(\wt w)= \bar y + \frac{\wt w |I_\by|}{2},
$$
then $T(I_\by)=[-1,1]$.
Let $\wt\mu_{\tby}$ be  the measure $\mu^{(1)}_\by$
(see \eqref{mu1}) rescaled to the interval 
$[-1,1]$, i.e.,
\be
  \wt\mu_{\tby} (\rd \tbx): = \frac{1}{\wt Z_{n,\tby}}
 \exp{\Big[ - n\sum_{i=1}^n 
 U_{\tby}(\wt x_i) + 2\sum_{1\leq i<j\leq n} \log |\wt x_i-\wt x_j|\Big]}
\rd\tbx
\label{mutby}
\ee
on $[-1,1]^n$
with
\be
  U_{\tby} (\wt x):= -\frac{2}{n} \sum_{|k|< n^B} 
\log |\wt x- \wt y_k|.
\label{U}
\ee
The $\ell$-point correlation functions of $\mu_\by$ and $\wt\mu_{\tby}$
are related by 
\be
   p^{(\ell)}_n (x_1, x_2, \ldots x_n)=  p^{(\ell)}_n \Big(\bar y +
   \frac{\wt x_1 |I_\by|}{2}, \ldots \bar y +
   \frac{\wt x_n|I_\by|}{2}\Big)= \Big(\frac{2}{|I_\by|}\Big)^\ell
 \wt p^{(\ell)}_n( \wt x_1, \wt x_2, \ldots 
\wt x_n).
\label{corr}
\ee

Let $p_j(\la)$, $j=0, 1, \ldots$ 
denote the real orthonormal polynomials on $[-1,1]$ 
corresponding to the weight function $e^{-nU_{\tby}(\la)}$, i.e.
$\mbox{deg} \;  p_j = j$ and
$$ 
  \int_{-1}^1 p_j (\la) p_k(\la)
e^{-nU_{\tby} (\la)}\rd \la = \delta_{j k}
$$
and define 
\be
  \psi_j(\la) : = p_j(\la) e^{-nU_{\tby} (\la)/2}
\label{psi}
\ee
to be orthonormal functions with respect to the Lebesgue measure on
$[-1,1]$.  Everything
depends on $\by$, but $\by$ is fixed in this section and
we will omit this dependence from the notation.

We define the $n$-th reproducing kernel
\be
    K_n (\la,\nu ) = \sum_{j =0}^{n-1} \psi_j (\la) \psi_j (\nu)
\label{def:Kn}
\ee
that satisfies 
\be
   K_n (\la,\nu) = \int_{-1}^1 K_n(\la,\zeta)K_n(\zeta,\nu)\rd \zeta.
\label{repro}
\ee
The density  is given by
\be
  \wt \varrho_n(\lambda) = \wt p^{(1)}_n(\la) = n^{-1}
K_n(\lambda,\lambda)
\label{def:dens}
\ee 
and the general $\ell$-point correlation function is given by
\be
 \wt p^{(\ell)}_n(\la_1, \la_2, \ldots, \la_\ell) =
 \frac{(n-\ell)!}{n!} \mbox{det}\{ K_n(\lambda_j, \la_k)\}_{j,k=1}^\ell
\label{determ}
\ee
following the standard identities in orthogonal polynomials.
For the rest of the paper we drop the tilde  and all variables will
denote the rescaled ones, i.e. all $x$ variables will be on the 
interval $[-1,1]$.
All integrals in this section are understood on $[-1,1]$.

The basic ingredients of the approach \cite{PS} can be described as
follows: 
Suppose that the following two  properties hold
for the normalized function $\psi=\psi_j$, $j=n-1, n$,
and for some fixed $\kappa>0$
\be\label{21}
\int_{|x|\le 1 -\kappa/2}  |\psi'(x)|^{2}  \rd x 
\le C n^{2 + \bar\e} 
\ee
\be\label{22}
n^{\delta} \int_{|x-x_0|\le n^{-\delta}}  \psi^{2} (x) \rd x 
\le C n^{\sigma}, \quad |x_0| \le 1 - \kappa
\ee
for some positive  $\sigma, \delta, \bar\e$ with $\sigma<1$.
We will take take  $\delta= 1/4$, same as in \cite{PS}.  
Let 
$$
   \bar \psi = \frac{1}{2\ell}\int_{|x-x_0|\le \ell}
 \psi(x) \rd x
$$
be the average of $\psi$ in the interval 
$|x-x_0|\le \ell $ with some $x_0$, $|x_0|\le 1-\kappa$
and $\ell \le \kappa/2$. 
We have
$$
 |\psi(x_0)|\le   |\bar \psi|  +  \| \psi' \|_{L^{2}} \ell^{1/2}.
$$
Using \eqref{22} to estimate $|\bar\psi|\le
C\ell^{-1/2}n^{(\sigma-\delta)/2}$  (under the assumption that $\ell < n^{-\delta}$)
and using \eqref{21}, we obtain
\[
|\psi(x_0)| \le C\ell^{-1/2}n^{(\sigma-\delta)/2} + Cn^{1+\bar\e/2}\ell^{1/2}.
\]
Choosing $\ell= n^{-1+ (\sigma-\delta -\bar\e)/2}$ 
we have 
\be
|\psi(x_0)|  \le n^{\frac{1}{2}+ \frac{1}{4}(\sigma +\bar\e-
\delta)}.
\label{psi0}
\ee
Note that $|\psi(x_0)| =O(n^{\frac{1}{2}- \e'})$ with some $\e'>0$
provided that $\sigma +\bar\e < \delta$.
Suppose  we can also prove that 
\be 
|\varrho'(x)| \le  C  n^{\e''}    ( \psi_{n-1}^2(x) +  \psi_{n}^2(x) )
\label{vr}
\ee
with some small power $\e''$,
then it will follow that $ |\varrho'(x)| \le   o(n)$ and this 
proves the regularity of the density over a distance of order $1/n$. 
Together with the fact that the density is well approximated
with the semicircle law on scales bigger than $1/n$  this will
show that the density is close to the semicirle law
pointwise. In \cite{PS} 
the regularity of the density on larger scales
followed from the smoothness of the potential (Theorem 2.2
of \cite{PS}). In our case
this follows from \eqref{gg2} which
is  a consequence of the fact \cite{ESY3} that the semicircle law
is precise on scales slightly larger than $1/N$ 
that corresponds to scales bigger than $1/n$ after
rescaling.

\bigskip

In proving \eqref{21}, \eqref{22} and \eqref{vr},
one  basic assumption in \cite{PS} requires the potential 
to be in $C^{2+ \nu}$ for some $\nu>0$. 
The potential for our probability measure \eqref{U},  parametrized
by the boundary 
conditions $\by$,   is singular near the boundary points
$\{ \pm 1\}$.  In order to control these singularities,  
besides using  some special properties of orthogonal polynomials,
we  rely on  \cite{ESY3} via \eqref{poteq}
to provide essential estimates such as level repulsions. 
It turns out that we can only establish 
\eqref{21} and \eqref{22} for $\psi_j, j \le n-1$ following this idea. 
The case of $j=n$ has to be treated 
completely differently.  
We now start to prove (\ref{21}) for $\psi_j, j = n-1, n-2$.

\begin{lemma}\label{9.1} Suppose  that $L\in \cG$, $\by\in \cY_L$ and,
after rescaling that sets $y_{-1}=-1$, $y_1=1$, let  the $\by$-configuration
satisfy  
\be\label{a2}
\sup_{|x|\le 1} \sum_{1<|k|<n^B}\frac 1 {|x-y_k|} \le  
 \sum_{1<|k|<n^B} \Big[ \frac{1}{|y_k-y_1|} +
\frac{1}{|y_k-y_{-1}|}\Big]
\le Cn^{1+3\gamma}
\ee
(note that the boundary terms $k=\pm 1$ are not included in the summations).
Furthermore,  assume that the
density $\varrho_n$ satisfies
\be\label{a3}
\int_{-1+n^{-A}}^{1-n^{-A}} [ (x+1)^{-2}+ (1-x)^{-2} ] 
  \varrho_n(x) \rd x  \le C n^{4\gamma}
\ee
for some $A\ge 60 B$.
Then for  the orthonormal functions $\psi_j$ from \eqref{psi}
we have
\be\label{73} 
\int_{-1}^1   \psi_j^{2}(x) 
 \left [ \frac 1 n  \sum_{|k|< n^B } \frac 1 {|x-y_k|} \right ]^2 
 \rd x   \le  n^{6 \gamma}  \qquad j\le n-1.
\ee
and
\be\label{51}
\int_{-1}^1 (\psi'_j(x))^{2} \rd x 
\le C n^{2+6\gamma}  \qquad j\le n-1.
\ee
\end{lemma}

Notice that  the assumptions
\eqref{a2} and \eqref{a3}  follow from \eqref{def:goody} and
\eqref{poteq}.

\medskip

In this section and in the subsequent Sections \ref{sec:smeared}
and \ref{sec:regdens} we work with orthogonal polynomials on
$[-1,1]$ with respect to the potential $U_{\tby} (x)$ (see \eqref{U}).
For brevity, we set $V(x) =  U_{\tby} (x)$ in these three
sections and we make the convention that the summation over the index $k$
that labels the elements of the external configuration $\by$ will
always run over integers with for $1\le |k| < n^B$ unless otherwise
indicated.
\medskip

{\it Proof.}
For simplicity, let $p(x)=p_{j}(x)$
and $\psi(x)= \psi_{j}(x)$. Then
\[
\int_{-1}^1   (p'(x) )^2 e^{-nV(x)} \rd x   
= \int_{-1}^1  -p^{\prime \prime}(x)p (x)  e^{-nV(x)} \rd x 
+ n\int_{-1}^1   p'(x) p (x) V'(x)   e^{-nV(x)} \rd x .
\]
Note that $e^{-nV(x)}$ is zero at the boundary $x=\pm 1$
so the boundary term vanishes in the integration by parts.
Since $p(x)$ is an orthogonal polynomial, it is orthogonal
to all polynomials of lower degree, thus the first integral vanishes. 
By Schwarz inequality, the second integral is bounded by 
\[
n\int_{-1}^1  p'(x) p(x)  V'(x)   e^{-nV(x)} \rd x  \le 
\frac 1 2 \int_{-1}^1
  (p'(x))^{2}  e^{-nV(x)} \rd x +  \frac 1 2  \int   p^2(x)
(n V'(x))^2   e^{-nV(x)} \rd x
\]
We have thus proved that 
\be
\int_{-1}^1    (p'(x) )^2   e^{-nV(x)} \rd x  \le  
2 \int_{-1}^1   p^2(x) (n V'(x))^2   e^{-nV(x)} \rd x.
\label{derp}
\ee

The last integral  is bounded by 
\be 
\int_{-1}^1   p^{2}(x) (nV'(x))^{2} e^{-nV(x)}
\rd x   \le  I_1 + I_2
\label{a1a2}
\ee
with
\be
I_1=   2\int_{-1}^1 \left [ \frac 1 { (x-1)^2} + \frac 1 { (x+1)^2} 
\right ]  
  \psi^{2}(x) \rd x,     \quad I_2 =   2
\int_{-1}^1     \left [  \sum_{k\not =  \pm 1} \frac 1 {x-y_k} \right ]^2
 \psi^{2}(x) \rd x .
\ee
From \eqref{a2}, and the normalization of $\psi$ we have 
\be
  I_2 \le C n^{2+6\gamma}.
\label{a22}
\ee

To control the term $I_1$, 
we separate the integration regimes
$|x\pm 1|\le n^{-A}$ and $-1+n^{-A}\le x \le 1-n^{-A}$
for some big constant $A$.
In the inside regime, we can use
 $|\psi(x)|^2=|\psi_j (x)|^2 \le n \varrho_n(x)$ since $j\le n-1$.
{F}rom 
\eqref{a3} we obtain
\be
  \int_{-1+n^{-A}}^{1-n^{-A}}
   \left [ \frac 1 { (x-1)^2} + \frac 1 { (x+1)^2}  \right ]  
  \psi^{2}(x) \rd x \le Cn^{1+4\gamma}.
\label{middle}
\ee

To estimate the singular part of the
integral in $I_1$ near the boundary points, we can
focus in estimating
$$
   \int_{-1}^{-1+n^{-A}} \frac{\psi^2(x)}{(1+x)^2} \rd x 
$$
the other endpoint being similar.
Let 
\[
g(x)=\frac{\psi(x)}{ x+1}.
\]
Notice that  $g(x)$ is a polynomial of degree 
$\mbox{deg} \; g \le 2n^{2B}+n$.  {F}rom the  Nikolskii inequality (see,
e.g.,
Theorem A.4.4 of \cite{BE1})
\be
\| g \|_{4} \le C(\mbox{deg}\; g)^{7.5}
\| g \|_{1/4} \le C n^{15B} \| g \|_{1/4} 
\label{nik}
\ee
with some universal constant $C$. Here $\| g\|_p$ is defined as  $\big(
\int_{-1}^1 |g(x)|^p\rd x\big)^{1/p}$ for any $0<p<\infty$.
Notice that Nikolskii inequality holds between 
$L^p$ spaces even with exponents $p<1$.
By the H\"older inequality, 
\be
\begin{split}
\| g\|_{1/4}^{1/2}=\left ( \int_{-1}^1  |g(x)|^{1/4}  \rd x  \right )^{2}
& \le  \Big(\int_{-1}^1 |g(x)|^{1/2}|x+1|^{1/2}  \rd x  \Big)
\Big( \int_{-1}^1 |x+1|^{-1/2} \rd x \Big) 
\cr
& \le   C \left ( \int |g(x)|^{2}(x+1)^{2}  \rd x  \right )^{1/4} 
\cr &  = C \|\psi\|_2^{1/2} = C.
\nonumber
\end{split}
\ee
Thus from \eqref{nik} we have $\|g\|_{4}\le C n^{15B}$ and
by H\"older inequality we have
\be\label{*}
\int_{-1}^{-1+n^{-A}}\frac{\psi^2(x)}{(x+1)^{2}} \rd x
\le C n^{-A/2}\| g\|_4^2 \le C n^{30B-A/2} \le C
\ee
provided  $A\ge 60B$.  Together
with \eqref{middle}, this proves $I_1\le Cn^{1+4\gamma}$.
Combining this with \eqref{a22} we obtain 
a bound $Cn^{2+6\gamma}$ for \eqref{a1a2} which proves \eqref{73}.

Using this estimate and \eqref{derp} we obtain that
\[
\int_{-1}^1   |p'(x) |^2 e^{-nV(x)} \rd x  \le C  n^{2+6\gamma}.
\]
Since 
\[
|\psi'(x)|^2 \le C [p'(x)^2 + p^2(x) (n V'(x))^2 ] e^{-nV(x)},
\]
we have thus proved that 
\be\label{5}
\int_{-1}^1   |\psi'(x) |^2   \rd x  
\le C n^{2+6\gamma} + C
\int_{-1}^1   p^2(x) (n V'(x))^2   e^{-nV(x)} \rd x 
\le C   n^{2+6\gamma} 
\ee
by using  \eqref{a1a2}. 
This completes the proof. $\Box$

\section{Bound  on  smeared-out orthogonal polynomials}\label{sec:smeared}

\begin{lemma} \label{lm:smeared} Let $\kappa,\delta_0>0$ be arbitrary positive
numbers.
Let $L\in \cG$, $\by\in\cY_L$,  suppose that the $\by$-configuration satisfies 
\eqref{a2}, \eqref{a3} and the density 
$\varrho_n(x) \ge \delta_{0}>0 $ for all $ |x| \le 1 - \kappa$.
Let $\psi=\psi_{n-1}$ or $\psi_{n-2}$  be an orthogonal function. 
Then we have
\be\label{P2}
n^{1/4} \int_{|x-x_0|\le n^{-1/4}}  \psi^{2} (x) \rd x 
\le C n^{3\gamma}, \quad |x_0| \le 1 - \kappa
\ee
with a constant $C$ depending on $\kappa$ and $\delta_0$.
\end{lemma}

{\it Proof.}
For any $z=u+i\eta\in \CC$ with $\eta>0$, let
$$
  m_n(z) = \int_{-1}^1 \frac{\varrho_n(x)}{x-z} \, \rd x
$$
denote the Stieltjes transform of the density and 
denote by 
\be
G_n(x, y) = \wt p^{(2)}_n (x, y) - \wt \varrho_n(x) \wt \varrho_n(y)
= -\frac {K_n(x, y)^2}{n(n-1)}
 + \frac {\wt\varrho_n(x) \wt \varrho_n(y) } {n-1}
\ee
the truncated correlation function, where $\wt p^{(2)}_n$ 
was defined in \eqref{corr} and computed from \eqref{determ}.
We will again drop the tilde in this proof.

We have the identity 
\be
\int  \frac {V'(x) \varrho_n(x)}{x-z} \rd x  = 
- \frac {n-1} n  m_n^2(z)  - \frac 1 n  \int  
\frac { \varrho_n(x)}{(x-z)^2} \rd x  
- \frac {n-1} n   \int  \frac { G_n(x, y)}{(x-z)(y-z)} \rd x  \rd y .
\ee
This identity follows from expressing  $\varrho_n$ by an integral 
over $n-1$ variables of the equilibrium measure
and then integrating by parts (see also (2.81) of \cite{PS}).
Hence, by using \eqref{repro}, we have 
\be\label{8}
 m_n^2(z) + \int  \frac {V'(x) \varrho_n(x)} {x-z} \rd x  = 
- \frac 1 {2 n^2}  \int    K^2_n(x, y) \left (\frac 1 {x-z}-
\frac 1 {y-z} \right )^2\rd x  \rd y .
\ee
The last integral can be bounded by 
\begin{align*}
\left | \int    K^2_n(x, y) \left (\frac 1 {x-z}- 
\frac 1 {y-z} \right )^2\rd x  \rd y  \right | 
\le & \left | \int    K^2_n(x, y) \frac {(x-y)^2 }  
{(x-z)^2(y-z)^2} \rd x  \rd y  \right | \\
\le & \eta^{-4} \int    K^2_n(x, y)(x-y)^2\rd x  \rd y  \le C  \eta^{-4}, 
\end{align*}
where, to estimate the last integral,  we have used the
Christoffel-Darboux formula
\be\label{cd}
K_n(x, y) = J_{n-1}  \frac { \psi_n(x) \psi_{n-1} (y) - 
\psi_n(y) \psi_{n-1} (x) } {x-y}, \quad J_{n-1} = \int_{-1}^1 x 
\psi_{n-1} (x) \psi_{n} (x) \rd x.
\ee
We have thus proved that   
\be\label{7}
m_n^2(z) + \int  \frac {V'(x) \varrho_n(x)}{x-z} \rd x  
= O(n^{-2} \eta^{-4}),  \quad z =  u + i \eta
\ee

We define a new measure $\mu^{-}_{\by}$ on $[-1,1]^{n-1}$ as
$$
  \mu^{-}_{\tby}(\rd x_1, \ldots, \rd x_{n-1}) = \frac{1}{Z^-_{\tby,n}}
  \exp\Big[ - n \sum_{i=1}^{n-1} V(x_i) + 2\sum_{1\le i<j< n-1}
  \log |x_i-x_j|\Big]
$$
where we already omitted the tildes and recall that $V(x) = U_\by(x)$.
Note that this measure differs from \eqref{mutby} written  
in $n-1$ variables in that we kept the prefactor $n$ in front
of the potential. Define
$$
  \varrho_n^{-}(x) = \frac{n-1}{n}
  \int \mu_\by(x, \rd x_2, \rd x_3, \ldots \rd x_{n-1})
$$
and note that
$$
  \varrho_n^{-}(x) = \frac{1}{n} \sum_{j=0}^{n-2} \psi_j^2(x)
$$
where $\psi_j$'s are defined in \eqref{psi}. This latter formula 
follows from the recursive relation of the correlation functions
for GUE-like ensembles, therefore
\[
\psi_{n-1}^2(x) = n (\varrho_n(x) - \varrho_{n}^-(x)).
\]
Let 
$$
  m_n^-(z) =  \int_{-1}^1 \frac{\varrho_n^-(x)}{x-z} \, \rd x
$$
be the Stieltjes transform of $\varrho_n^-$; then we have the analogue
of \eqref{7}
$$
[m_n^-(z)]^2 + \int  \frac {V'(x) \varrho_n^-(x)}{x-z} \rd x  
= O(n^{-2} \eta^{-4}).
$$
Subtracting this from \eqref{7}, we have 
\be
n  (m_n^2(z) - [m_{n}^-(z)]^2) = -   
\int  \frac {V'(x) \psi_{n-1}^2(x)}{x-z} \rd x  + O(n^{-1} \eta^{-4}).
\label{VV}
\ee
Assume that $u=\mbox{Re}\, z$ satisfies  $|u-x_0| \le n^{-1/4}$.
By adding $n (m_n(z) - m_n^-(z))  V'(u)$ to the both sides 
of \eqref{VV}, we obtain 
\[
n (m_n(z) - m_{n}^-(z)) (m_n(z) + m_{n}^-(z) + V'(u)) = -   
\int  \frac {(V'(x) - V'(u) )\psi_{n-1}^2(x)}{x-z} \rd x 
+ O(n^{-1} \eta^{-4}).
\]
We divide the integral into $ |x-x_0| \le \nu/2$ and
$  |x-x_0| \ge \nu/2$. 
In the first integration regime, since $|x_0|\le 1- \nu$, we have 
\be
\begin{split}
\int_{|x-x_0| \le \nu/2}    \Big|  &
 \frac {V'(x) - V'(u) }{x-z} \Big|  \psi_{n-1}^2(x) \rd x  \\
&\le  \sup _{ |x-x_0| \le \nu/2}  \frac 1 n  \sum_k 
\frac 1 {|x-u|}  \left |    \frac  1 {y_k-x}  -   \frac  1 {y_k-u}
 \right |     \int_{|x-x_0| \le \nu/2}     \psi_{n-1}^2(x) \rd x  . 
\label{supx}
\end{split}
\ee
Since $|x|\le 1-\nu/2$, $|u|\le 1-\nu/2$, 
we have $|y_k-u|\ge 2\nu^{-1}$
for any $k$.
Thus, by \eqref{a2},  the prefactor in \eqref{supx}
is bounded, uniformly in $|x|\le 1-\nu/2$, by
\[
\frac 1 n  \sum_k  \frac 1 {|x-u|}  \left |   
\frac  1 {y_k-x}  -   \frac  1 {y_k-u}  \right |  
\le  \frac 1 n  \sum_k      \frac  1 {|y_k-x| |y_k-u|}  
 \le \frac{2} {\nu n}  \sum_k      \frac  1 {|y_k-x|}
\] 
\be
 \le \frac{C}{n}\Big[ \frac{1}{|1-x|}+\frac{1}{|1+x|}+
  \sum_{k\ne \pm 1}      \frac  1 {|y_k-y_1|}+
 \sum_{k\ne \pm 1}      \frac  1 {|y_k-y_{-1}|} \Big]\le Cn^{3\gamma},
\label{yyu}
\ee
where the constant $C$ depends on $\nu$ and we recall that
$y_{-1}=-1$, $y_1=1$ in the rescaled variables.

In the second integration regime we use $|x-u| \ge |x-x_0| - |x_0-u|
\ge \nu/4$ and obtain
\begin{align*}
\int_{ |x-x_0| \ge \nu/2}  &  \left |   
\frac {V'(x) - V'(u) }{x-z} \right |  \psi_{n-1}^2(x) \rd x   \\
& \le   \frac C n   \int_{ |x-x_0| \ge \nu/2}  \sum_k  
\left |  
 \frac  1 {y_k-x}  -   \frac  1 {y_k-u}  \right |   
  \psi_{n-1}^2(x) \rd x  
\\
&  \le    \frac {C} {n}   \int \sum_k 
  \left |    \frac  1 {y_k-x}   \right |      \psi_{n-1}^2(x) \rd x    
 + \frac {C} {n}  \sum_k  \frac  1 {|y_k-u|} 
\le C n^{3 \gamma}
\end{align*}
where we have used \eqref{73} and H\"older inequality
to estimate the first term in the last line
and using \eqref{a2} for the second term.

We have thus proved that 
\[
n \left | (m_n(z) - m_n^-(z)) (m_n(z) + m_n^-(z) + V'(u)) \right |  \le 
n^{3\gamma} + Cn^{-1} \eta^{-4}.
\]
Hence 
\[
n\left | m_n(z) - m_n^-(z)  \right |  \le 
\frac {n^{3 \gamma} + Cn^{-1} \eta^{-4}} {  \im \, m_n(z) }
\]
using that $\im\, m_n^-(z)>0$.
Since $\varrho_n(x) \ge \delta_{0} >0$ by assumption,  
$\im \, m_n(z)$ is bounded from below. Thus, choosing $\eta = n^{-1/4}$,
we obtain
$$
 \Big| \int \frac{\psi_{n-1}^2(x)}{x-z}\rd x\Big| \le  Cn^{3\gamma}.
$$
with $C$ depending on $\nu$ and $\delta_0$. Taking imaginary
part, we have
$$
   \int \frac{\eta}{(x-u)^2 + \eta^2}\, \psi_{n-1}^2(x)\rd x
\le  Cn^{3\gamma}.
$$
for any $u$ with $|u-x_0|\le \eta=n^{-1/4}$.
Integrating over $|u-x_0| \le \eta$ and using
$$
  \int_{|u-x_0|\le \eta}\frac{\eta}{(x-u)^2 + \eta^2} \ge
 c \cdot {\bf 1}(|x-x_0|\le \eta)
$$
with some positive constant $c$, we have proved \eqref{P2}
for $\psi=\psi_{n-1}$. The case $\psi=\psi_{n-2}$  can be done in a
similar
way. This completes the proof of Lemma \ref{lm:smeared}. $\Box$

\begin{corollary}
Suppose  that the $\by$-configuration satisfies
\eqref{a2}, \eqref{a3} and the density satisfies 
$\varrho_n(x) \ge \delta_{0} $ 
for all $ |x| \le 1 - \kappa$ for some $\delta_0, \kappa>0$.
Let $\psi=\psi_{j}$ with $j= n-2, n-1, n$  be an orthogonal function. 
Then
\be\label{52}
\sup_{|x| \le 1 -  \kappa} | \psi (x) |^2
\le C n^{1 -  \frac{1}{8} +11\gamma} 
\ee
with a constant $C$ depending on $\kappa$ and $\delta_0$.
\end{corollary}

{\it Proof.} For the case $j=n-2, n-1$, the estimate \eqref{52}, even with
a better exponent, follows 
from the argument leading to  \eqref{psi0} from
the two assumptions \eqref{21} and \eqref{22} with $\delta=1/4$,
$\bar\e=6\gamma$ and $\sigma=3\gamma$:
\be\label{52j}
\sup_{|x| \le 1 -  \kappa} | \psi_j (x) |^2
\le C n^{1 -  \frac{1}{8} +\frac{9}{2}\gamma} \qquad j=n-2, n-1.
\ee
 The estimate
\eqref{21} was proven in Lemma \ref{9.1}, the estimate 
\eqref{22} follows from Lemma \ref{lm:smeared}. 

The proof of \eqref{52} for $\psi =\psi_n$ requires a different
argument. Let $a_j$ be the leading coefficient of the
(normalized) $j$-th orthogonal polynomial, i.e. $p_j(x) = a_j x^j +
\ldots$.
Observe that $p_n'(x) = na_n x^{n-1} + \ldots = n(a_n/a_{n-1})p_{n-1}(x)
+\ldots$, where dots mean a polinomial of degree less than $n-1$. Thus
\be
\begin{split}
\frac {n a_{n}} {a_{n-1}} = & \int  p'_n(x) p_{n-1}(x)  e^{-nV(x)} \rd x 
\cr
= & \int  p_n(x) p'_{n-1}(x)  e^{-nV(x)} \rd x + 
\int  p_n(x) p_{n-1}(x) n V'(x)  e^{-nV(x)} \rd x.
\nonumber
\end{split}
\ee
The first integral on the right hand side vanishes. 
By the Schwarz inequality, we have 
\be
\begin{split}
\frac {n |a_{n}|} {|a_{n-1}|}  & \le  \int  | p_n(x) p_{n-1}(x) n V'(x)| 
e^{-nV(x)} \rd x \cr
& \le \left  [  \int   p_n^2(x)   e^{-nV(x)} \rd x 
\right ]^{1/2}
\left [ \int  |  p_{n-1}(x) n V'(x)|^2  e^{-nV(x)} \rd x \right ]^{1/2}
\le C n^{1+3\gamma} ,
\label{aan}
\end{split}
\ee
where the second integral was estimated in \eqref{73}.

Recall the standard three-term recursion relation for orthogonal
polynomials 
\be
x p_{n-1} = a p_n + b p_{n-1} + c p_{n-2}
\label{3term}
\ee
with some real numbers $a,b,c$ depending on $n$.
By comparing the leading coefficients, we have $a_{n-1}=aa_n$
and by orthonormality, we get
$$  
  a^2+b^2+c^2 = \int_{-1}^1 x^2 p_{n-1}^2(x) e^{-nV(x)}\rd x  \le 1.
$$
In particular
$$ 
   \frac{1}{|a|} =   \Big|\frac{a_{n}}{a_{n-1}}\Big| \le Cn^{3\gamma}
$$
from \eqref{aan}.
Hence, from \eqref{3term},  
\[
|p_n(x)| \le \big| a^{-1} [ (x-b)  p_{n-1}(x) 
-  c p_{n-2}(x)  ]\big| \le Cn^{3\gamma}  [ |p_{n-1}(x)|  + |p_{n-2}(x) |
] .
\]
Using the bound \eqref{52j}, 
we obtain \eqref{52} for $\psi=\psi_n$ as well.  $\Box$

\section{Regularity of Density}\label{sec:regdens}

\begin{lemma}\label{de} Let $L\in \cG$, $\by\in \cY_L$.
Suppose  that the external $\by$-configuration satisfies 
\eqref{a2} and \eqref{a3} and assume that $\gamma < \frac{1}{150}$.
 Then  for any $\kappa>0$ we have
\be \label{23} 
\sup_{|x| \le 1 - \kappa} |\varrho_n'(x)| \le 
C n^{3\gamma}    ( \psi_{n-1}^2(x) +  \psi_{n}^2(x)+1 ) \le C  n^{1-
\frac{1}{8} +14\gamma}
\ee
and 
\be
 \sup_{|x|\le 1-2\kappa} |\varrho_n(x) - \varrho_{sc}(\bar y)| 
\le Cn^{-\gamma/12}
\label{tv}
\ee
where the constant $C$ depends on $\kappa$.
\end{lemma}

{\it Proof.}
The derivative of the density can be computed explicitly 
(see, e.g., (3.63) of \cite{PS}) as 
\be\label{1}
\varrho_n'(x) = \int_{-1}^1 [V'(z) - V'(x)] K^2_n (x, z) \rd z.
\ee
In our case
$$
 V'(z) - V'(x) = 
\frac 1 n\sum_k   \left [ \frac 1 {x-y_k} - \frac 1 {z-y_k} \right ] 
$$
and 
\be\label{91}
\frac 1 {x-y_k} - \frac 1 {z-y_k} = 
 -\frac {x-z} {(x-y_k)^2}- \frac {(x-z)^2} {(z-y_k) (x-y_k)^2 }.
\ee
From the Christoffel-Darboux formula, we have 
\[
\Bigg| \int_{-1}^1 (x-z)^{\alpha} K^2_n (x, z) \rd z\Bigg|
\le C ( \psi_{n-1}^2(x) +  \psi_{n}^2(x) ), \quad \alpha = 1, 2.
\]
Since $|x| \le 1 - \kappa$, we can estimate the contribution to \eqref{1}
from the first term in \eqref{91} by 
\[
 \sum_k \frac {1} { n (x-y_k)^2} \Bigg| \int_{-1}^1
(x-z)  K^2_n (x, z) \rd z\Bigg|
\le  Cn^{3\gamma} ( \psi_{n-1}^2(x) +  \psi_{n}^2(x) ),
\]
where we have used  \eqref{a2} 
to bound the factor in front of the  integral 
\be
  \sup_{|x|\le 1-\kappa} \sum_k \frac {1} { n (x-y_k)^2}
 \le  \sum_{k\ne \pm 1} \frac {1} { n \kappa |y_{-1}-y_k|} 
 + \sum_{k\ne \pm 1} \frac {1} { n  \kappa |y_{1}-y_k|} +  
\frac{C}{\kappa^2 n}
\le Cn^{3\gamma}.
\label{sq}
\ee

The contribution from the second term in \eqref{91} is bounded by 
\begin{align*}
  \Bigg| \frac 1 n  \sum_{k}  \frac {C} { (x-y_k)^2} &
\int_{-1}^1 \frac {(x-z)^2} {(z-y_k)} 
K^2_n (x, z) \rd z\Bigg| 
\\ & \le   C n^{3\gamma}\kappa^{-2}
   \int_{-1}^1 \frac{1}{n}\sum_k \frac {1} {|z-y_k|}  
\left [ \psi_{n}(x)\psi_{n-1}(z) - \psi_{n}(z)\psi_{n-1}(x) \right ]^2 \rd
z
\\
&\le  C n^{3\gamma} \sum_{j=0, 1}  \psi_{n-j}^2(x)  
\int_{-1}^1 \frac{1}{n}\sum_k
\frac {1} {|z-y_k|} \psi_{n+j-1}^2(z)   \rd z .
\end{align*}
The integral is estimated as
\be
\begin{split}
  \int_{-1}^1 \frac{1}{n}\sum_k
\frac {1} {|z-y_k|} \psi_{n+j-1}^2(z)   \rd z 
 \le & \int_{-1}^1 \frac{1}{n}\sum_{k \ne \pm 1}
\Big[\frac {1} {|y_{-1}-y_k|}+\frac {1} {|y_{1}-y_k|}\Big]
  \psi_{n+j-1}^2(z)   \rd z  \\
& +  \int_{-1}^1 \frac{1}{n}\sum_{k \pm 1}
\Big[\frac {1} {|1-z|}+\frac {1} {|1+z|}\Big]
  \psi_{n+j-1}^2(z)   \rd z.
\end{split}
\ee
The first term on the right hand side
is bounded by $Cn^{3\gamma}$ using \eqref{a2}.
In the second term, we split the integration into two regimes:
 $|z|\le 1- n^{-A}$ and $1-n^{-A}\le |z|\le 1$ with
some $A\ge 60 B$. In the first regime,
we use the bound \eqref{52} to obtain $ CAn^{-1/8 + 11\gamma}\log n 
\le C$ if $\gamma < \frac{1}{88}$. In the second regime we use
the bound \eqref{*}. This proves \eqref{23}.

\bigskip

For the proof of \eqref{tv} we use the derivative estimate and
the fact that the density is close to the semicircle law
on scale $n^{-1+\gamma}$ as given in \eqref{gg2}. For any $x, y\in
[-1+2\kappa, 1-2\kappa]$ we have
$$
  \varrho(x) = \varrho(y) + \int_x^y \varrho'(u)\rd u
$$
Taking the average on the interval $I=[x- \frac{1}{2} n^{-1+\gamma}, x+ 
\frac{1}{2} n^{-1+\gamma}]$, we get
\be
   \Big| \varrho(x) -  n^{1-\gamma}\int_I \varrho(y)\rd y
 \Big| \le n^{-1+\gamma} \| \varrho'\|_\infty  
 \le Cn^{-1/8+15\gamma}.
\label{varx}
\ee
Using \eqref{gg2}, we have
$$
   n^{1-\gamma}\int_I \varrho(y)\rd y  
  =\EE_{\mu_\by} \frac{\N(I^*)}{N|I^*|} = \varrho_{sc}( T^{-1}(x))
 + O(n^{-\gamma/12}) = \varrho_{sc}(\bar y) + O(n^{-\gamma/12})
$$
with $I^*:= T^{-1}(I)$, where we also used that
$$
 | \varrho_{sc}( T^{-1}(x)) - \varrho_{sc}(\bar y)| 
\le |I_\by| \sup_{|x|\le 2-\kappa} |\varrho'(x)| \le CnN^{-1}.
$$
Combining these inequalities, we arrive at  \eqref{tv} and this
completes the
proof of Lemma \ref{de}. $\Box$

\section{Proof of the Main Theorem \ref{mainthm}}\label{sec:proofmain}

Let $V(x) = U_{\tby}(x)$ be the external potential on $I=[-1,1]$
given by \eqref{U} after rescaling. Notice that $V$ is
continuous on $(-1,1)$ and $\lim_{|x|\to 1}V(x)=\infty$.
Let $\nu(\rd x)$ be
the equilibrium measure, defined as the unique
solution to the variational problem
\be
   \inf_{\nu \in \cM^1} \Big\{ \int_{-1}^1\int_{-1}^1
  \log |s-t|^{-1}\nu(\rd s) 
 \nu(\rd t) + \int_{-1}^1 V(s)\nu(\rd s)\Big\},
\label{varpr}
\ee
where $\cM^1$ is the space of probability measures on $[-1,1]$.
For general properties of the equilibrium measure, see, e.g.
Chapter 2 of \cite{LL1}  (and references therein) that
specifically discusses the case of compact interval $I$
and continuous potential going to infinity at the endpoints.
We point out however, that we follow the convention of \cite{D} and
\cite{PS} in what we call external potential; the potential in
\cite{LL1} and \cite{LL}, denoted by $q(x)$ and $Q(x)$, respectively,
differs by a factor of two from our convention: 
$q(x) = Q(x)= \frac{1}{2}V(x)$.

The equilibrium  measure $\nu$ with
support $S(\nu)$ satisfies the Euler-Lagrange equations
$$
    \int \log|s-t|^{-1} \nu(\rd s) + \frac{1}{2} V(t) = C  \qquad t\in
 S(\nu)
$$ 
$$
   \int \log|s-t|^{-1} \nu(\rd s) + \frac{1}{2} V(t) \ge C  \qquad t\in
 I\setminus S(\nu)
$$ 
and $S(\nu)\subset (-1,1)$
(Theorem 2.1 of \cite{LL1}). Moreover, since $V$ is convex in $(-1,1)$
such that $\lim_{|x|\to 1} V(x) =\infty$,
the support $S(\nu)$ is an interval, $S(\nu) = [a,b]$,
whose endpoints satisfy $-1<a<b<1$ and they are uniquely determined
by the equations
\be
  \int_a^b  \frac{V'(s)\,\rd s}{ \sqrt{(s-a)(b-s)}}
    =0, \qquad
  \frac{1}{2\pi}\int_a^b  \frac{V'(s)\, s\,\rd s}{ \sqrt{(s-a)(b-s)}}
   =1.
\label{ab}
\ee
according to Theorem 2.4 \cite{LL1} (after adjusting a factor of 2).

In our case, the potential $V$ and thus the equilibrium measure $\nu$
depend on 
$n$ and the external configuration $\by$ in a non-trivial
way. The main result of the recent work of Levin and Lubinsky \cite{LL}
proves the
universal sine-kernel behavior for the correlation
function of the orthogonal polynomials with respect to a general
$n$-dependent potential. This result fits exactly our situation, after
the conditions of \cite{LL} are verified.

We  recall the main result of \cite{LL} in a special form we will need.

\begin{theorem} \label{thm:L}  For
each $n\ge 1$, consider a positive Borel measure $\mu_n$ on the real line
whose $2n+1$ moment is finite. Let $I=[-1,1]$ and assume that
each $\mu_n$ is absolutely continuous on $I$ and they
can be written as 
$$
  \mu_n(\rd x) = W_n^{2n}(x)\rd x 
$$
where the non-negative functions $W_n$ are continuous on $I$. 
We define the potential $Q_n=-\log W_n : I\to (-\infty, +\infty]$
and let $\nu_n$ be the solution of the variational problem \eqref{varpr}
with $V=V_n=2Q_n$.
Let $J$ be a compact subinterval of $(-1,1)$. Assume 
the following conditions
\begin{itemize}
\item[(a)] The  equilibrium measure is absolutely continuous
with $\nu_n(\rd x) = g_{n}(x) \rd x  $, where $g_{n}$ 
is  positive and uniformly bounded
in some open interval containing $J$;

\item[(b)]  The family $\{Q'_{n}\}_{n=1, 2,\ldots}$ is
equicontinuous and uniformly bounded in some
open interval containing $J$;

\item[(c)] The density $\varrho_{n}(x)$ of the first $n$ orthogonal
polynomials
with respect to $\mu_n$ on $I$ (defined in \eqref{def:dens}) 
satisfies $C^{-1}\le \varrho_{n}(x) \le C$
in some open interval containing $J$;

\item[(d)]
The following limit holds uniformly for $E \in J$ and $a$ in any fixed
compact 
subset of $\RR$:
\[
\lim_{n \to \infty}  \frac {\varrho_{n}(E ) } {\varrho_{n}(E + \frac x n)
} =1.
\]
\end{itemize}
Then for the $n$-th reproducing  kernel of the measure $\mu_n$ on
$I$ (defined in \eqref{def:Kn})
we have
\be\label{Llim}
\lim_{n \to \infty}   \frac{1}{n\varrho_n(E)} K_n\left( E+ \frac a { n
\varrho_n(E)},
E+ \frac b { n \varrho_n(E)} \right ) 
=   \frac{\sin \pi (a-b)}{\pi (a-b)}
\ee
uniformly for $E\in J$ and for $a, b$ in compact subsets of $\RR$.
\end{theorem}

\medskip
First we verify the conditions of this theorem for our case. 
We consider the sequence of measures $\mu_n$ on $\RR$
that vanish outside of $I=[-1,1]$ and that are given
by $\mu_n(\rd x) = e^{-nU_\by(x)}\rd x$ on $I$, where $\by \in \cY_L$ is a
sequence
of good external configurations after rescaling for some $L\in \cG$.
Recall that the concept of good external configurations depends on $N$,
i.e. $\cG=\cG_N$ 
and we recall the relation \eqref{nchoice} between $n$ and $N$.
We set $J=[-1+\sigma, 1-\sigma]$ for some $\sigma>0$. 
The measure $\mu_n$ is clearly absolutely continuous (actually it
has a polynomial density), and since it
is compactly supported, all moments are finite. 
Conditions (a) and (b) will be verified separately in Appendix
\ref{sec:supp}.
Conditions (c) and (d) follow directly from \eqref{tv} in Lemma
\ref{de}.

\bigskip 

Now we start the proof of the main Theorem \ref{mainthm}.
Throughout this proof, 
$\EE$ is the expectation for the Wigner ensemble
with a small Gaussian component, i.e.
$\EE = \EE_{f_{t}}$ with the earlier notation. All
constants in this proof may depend on $\kappa$.
We will use the results obtained in Sections \ref{sec:goodglobal}--\ref{sec:regdens}.
In these sections, various small exponents $\al, \beta, \gamma, \e$,
and various large exponents $A, B$ need to be specified. The exponent $\beta$ is
given in the theorem and it can be an arbitrary positive constant. The other 
exponents are determined in terms of $\beta$ subject to the following requirements:
$\beta\ge 10\e+\al$  \eqref{L1choice}, $\beta\ge (4A+8)\e +\al$ (Lemma
\ref{lm:equipot}), $B\e<1/2$ (Section \ref{sec:cutoff}), $B\ge 20$ (Lemma
\ref{lm:mumu}) and $A\ge 60B$ (Lemma \ref{9.1}). Finally, $\gamma\le \frac{1}{10}$ can
be an arbitrary positive number, independent of $\beta$. Obviously, these conditions
can be simultaneously satisfied for any $\beta>0$ if $\al, \gamma, \e$
are chosen sufficiently small and $A, B$ sufficiently large. All constants
in the proof depend on this choice.

\medskip

  Let $O(a,b)$ be a bounded  function and $\delta<\kappa/2$.
In \eqref{eq:mainthm} we have to compute the limit of
\be\label{sin}
\begin{split}
\frac{1}{2\delta}\int_{E_0-\delta}^{E_0+\delta} \rd E & \int  \rd a \rd b
\;  
\varrho_{sc}(E)^{-2}  p_N^{(2)}\Big(E+ \frac{ a } {   N \varrho_{sc}(E)},
E+
\frac{ b} {  N \varrho_{sc}(E) }
  \Big)  O\big( a,  b \big) \\
& =   \frac{N^2}{2\delta}\int_{E_0-\delta}^{E_0+\delta} \rd E
\int  \rd u \rd v \; p^{(2)}_N (u, v)  O\Big( (u-E)N\varrho_{sc}(E), 
(v-E)N\varrho_{sc}(E) \Big) \\ 
& = \frac{N}{N-1}\frac{1}{2\delta}\int_{E_0-\delta}^{E_0+\delta} \rd E
\; \EE   \sum_{j\neq  k}^N  O\Big( (\lambda_j-E)N\varrho_{sc}(E), 
(\lambda_k-E)N\varrho_{sc}(E) \Big),  \\
& = :\frac{N}{N-1}  T(N,\delta),
\end{split}
\ee
where we have changed variables. Using the form of $O$ given in
\eqref{def:O}, we have
\be
 T(N,\delta)=  \EE   \sum_{j\neq  k}^N  
 \frac{1}{2\delta}\int_{E_0-\delta}^{E_0+\delta}\rd E\;
g\big( (\lambda_j-\lambda_k)N \varrho_{sc}(E) ) 
h\Big( \big( \frac{\lambda_j+\lambda_k}{2}-E\big)N\varrho_{sc}(E) \Big) .
\label{def:TN}
\ee
We first show that 
\be\label{63}
\sup_{\delta \le \kappa/2}
\sup_{N\in \NN} T(N,\delta)\le C
\ee
with a constant depending on $\kappa$.
To see this, let $R$ be a large number so that  $g(x)=h(x)=0$ for $|x| \ge
R$,
then
\be
\begin{split}
T(N,\delta) & \le
C \EE   \sum_{j\neq  k}^N  
\frac{1}{2\delta}\int_{E_0-\delta}^{E_0+\delta}\rd E\;
  \prod_{\ell= j, k}  {\bf 1} \big [ |\lambda_\ell- E| \le C R/N  \big]
\\
& \le C  \frac{1}{2\delta}\int_{E_0-\delta}^{E_0+\delta}\rd E\;
 \EE \; \N^2[E- C R/N, E+ C R/N] \le C,
\end{split}
\ee
where we have used that $\inf\{ \varrho_{sc}(E) \; : \; | E-E_0|\le
\delta\}
\ge c >0$  and that 
\be
 \EE \, \N_I^k \le C_k(N|I|)^k
\label{Nmom}
\ee
for any interval $I$ of length $|I|\ge 1/N$. The bound \eqref{Nmom} 
follows from Eq. (3.11) in \cite{ESY3} after cutting
the interval $I$ into subintervals of size $1/(2N)$.

The estimate \eqref{63} and similar ideas allow
us to perform many cutoffs and approximations. 
For example, we can replace $ \varrho_{sc}(E)$ in $g$ and $h$ by
$\varrho:=
\varrho_{sc}(E_0)$ in the definition of $T(N,\delta)$, see \eqref{def:TN},
at the expense of
an error that vanishes in 
the limit $\delta \to 0$. 
 We shall give a proof in case we perform the change for, say, $g$:
\begin{align*}
\EE   \sum_{j\neq  k}^N  
\frac{1}{2\delta}\int_{E_0-\delta}^{E_0+\delta}\rd E
\, & \Big | g\big( (\lambda_j-\lambda_k)N \varrho_{sc}(E) ) -
g\big( (\lambda_j-\lambda_k)N \varrho_{sc}(E_0) ) \Big | 
h\Big( \big( \frac{\lambda_j+\lambda_k}{2}-E\big)N\varrho_{sc}(E) \Big) 
\\
&\le C \delta 
\EE   \sum_{j\neq  k}^N  
\frac{1}{2\delta}\int_{E_0-\delta}^{E_0+\delta}\rd E
  \prod_{\ell= j, k}  {\bf 1} \big [ |\lambda_\ell- E| \le C R/N  \big]
\le  C\delta,
\end{align*}
where we used that $\varrho_{sc}'(E)$ is uniformly bounded 
on $[E_0-\delta, E_0+\delta]\subset [-2+\kappa/2, 2-\kappa/2]$.
We will not repeat this type of
simple argument in this proof. 

After this replacement, we can perform the $\rd E$ integration
using that $\int h =1$:
\be
\begin{split}
  T(N,\delta) & = \EE   \sum_{j\neq  k}^N   
g\big( (\lambda_j-\lambda_k)N \varrho ) 
 \frac{1}{2\delta}\int_{E_0-\delta}^{E_0+\delta}\rd E\;
h\Big( \big( \frac{\lambda_j+\lambda_k}{2}-E\big)N\varrho \Big)  +
O(\delta) \\
& =   \frac{1}{2N\varrho \delta} \, \EE
\sum_{j\neq  k}^N   g\big( (\lambda_j-\lambda_k)N\varrho\big) 
\prod_{\ell=j, k}  {\bf 1}\Big( \Big| \lambda_\ell- E_0\Big|
 \leq \delta\Big)  + O(\delta) + O(\delta^{-1}N^{-1}),
\end{split}
\ee
where the last error comes from the contribution of
eigenvalues within $CR/N$ distance to $E_0\pm \delta$.
With the notation
$$
 T^*(N,\delta): = \frac{1}{2N\varrho \delta} \, \EE
\sum_{j\neq  k}^N   g\big( (\lambda_j-\lambda_k)N\varrho\big) 
\prod_{\ell=j, k}  {\bf 1}\Big( \Big| \lambda_\ell- E_0\Big|
 \leq \delta\Big), 
$$
and using \eqref{sin},
we thus need to prove that 
$$
\lim_{\delta\to0}\lim_{N\to\infty}  T^*(N,\delta)
=   \int g(a-b) \Big [1 -
\Big( \frac{\sin \pi(a-b)}{\pi(a-b)}\Big)^2  \Big ] \rd a\rd b.
$$
Recall the definition of $\fN_{sc}(E)$ from \eqref{def:Nsc}
and its inverse function $\fN_{sc}^{-1}(E)$.
Note that 
\be
(\fN_{sc}^{-1})'(E) \le C\kappa^{-1/2} \quad \mbox{if}\quad 
  -2+\kappa\le E \le 2-\kappa.
\label{Nscder}
\ee
We define
\[
\chi_{N, E_0, \delta} (j): =  {\bf 1}   ( M_- \le  j \le  M_+),  
\quad M_\pm =N\cdot  \fN_{sc} (E_0\pm\delta),
\]
and write
\be
 {\bf 1}\Big( \big|  \lambda_j- E_0\big|
 \leq \delta\Big) 
=    \chi_{N, E_0, \delta} (j)   +  U_{j},
\ee
where $U_j$ is the error term, defined as the difference of 
${\bf 1}\big( \Big|  \lambda_j- E_0\Big|
 \leq \delta\big)$ and $\chi_{N, E_0, \delta} (j)$.
We thus have 
\be
\begin{split}\label{Tsplit}
T^*(N,\delta)  =  &   \frac{1}{2N\varrho \delta}
\EE\sum_{j\neq  k}^N   g\big( (\lambda_j-\lambda_k)N\varrho\big) 
\chi_{N, E_0, \delta} (j) {\bf 1}\Big( \big|  \lambda_k- E_0\big|
 \leq \delta\Big)  \\
& + \frac{1}{2N\varrho \delta} \EE
\sum_{j\neq  k}^N   g\big( (\lambda_j-\lambda_k)N\varrho\big)
U_{j} {\bf 1} \Big( \big|  \lambda_k- E_0\big|
 \leq \delta\Big).
\end{split}
\ee
The last term is bounded by 
\be
\left ( \EE \; \frac{1}{2N\varrho \delta}  \sum_j 
\left [ \sum_{k: k\neq  j}^N  {\bf 1} \Big( \big|  \lambda_k- E_0\big|
 \leq \delta\Big)
 g\big( (\lambda_j-\lambda_k)N\varrho\big) \right ]^2  \; \EE
\left [  \frac{1}{2N\varrho \delta} \sum_j   
U_{j}^2   \right ] \right )^{1/2}.
\label{lastt}
\ee
The first expectation is bounded by 
$$ 
  \frac{1}{2N\varrho \delta}\EE \sum_{k,k',j} 
{\bf 1} \Big( \big|  \lambda_k- E_0\big|
 \leq \delta\Big)  {\bf 1} \Big( \big|  \lambda_{k'}- E_0\big|
 \leq \delta\Big)  {\bf 1} \Big( \big| \la_j-\la_k\big|\le C/N\Big)
 {\bf 1} \Big( \big| \la_j-\la_{k'}\big|\le C/N\Big)  
$$
Splitting the interval $[E_0 -\delta - C/N, E_0+\delta + C/N]$
into overlapping subintervals $I_\ell$ of length $4C/N$ with an overlap
at least $2C/N$, we get that this last expectation is bounded by
$$ 
 \frac{1}{2N\varrho \delta}\sum_\ell \EE \; \N_{I_\ell}^3 \le C,
$$
where we used  the  moment bound \eqref{Nmom}   with $k=3$
and the fact that the number of subintervals is $CN\delta$.

Since $\fN_{sc}$ is monotonic, the second expectation in \eqref{lastt} is
bounded by 
\[
 \frac{1}{2N\varrho \delta} \sum_j   \EE 
 \Big [ {\bf 1}\Big( \Big|  \lambda_j- E_0\Big|
 \leq \delta\Big) 
-  {\bf 1}   (   | \fN_{sc}^{-1} ( j/N ) - E_0 | \le \delta  ) \Big ]^2.
\]
On the set $\Omega^c$ we estimate the difference of the
two characteristic functions by 2, and we get from \eqref{NI1}
that the contribution is subexponentially small in $n$.
On the set $\Omega$ we can use \eqref{y} and we see that 
the difference of the two characteristic functions 
can be nonzero only if 
$$
\delta - Cn^{-\gamma/6} \le |\fN_{sc}^{-1} ( j/N ) - E_0| \le \delta +
Cn^{-\gamma/6}
$$
i.e. the number of $j$'s this can happen is bounded by $CNn^{-\gamma/6}$.
Recalling  \eqref{nchoice}, we get
$$
\lim_{N\to \infty}\EE
\left [  \frac{1}{2N\varrho \delta} \sum_j   
U_{j}^2   \right ] =0,
$$
therefore the second term in \eqref{Tsplit} vanishes in the $N\to\infty$
limit.

This shows that we can replace $ {\bf 1}\Big( \Big|  \lambda_j- E_0\Big| 
\leq \delta\Big) $ by $  \chi_{N, E_0, \delta} (j)$ 
in the definition of $T^*$ 
a with negligible error and we can do similarly for $k$ instead of $j$.
Therefore, we need to 
prove that  
$$
 \lim_{\delta\to 0}\lim_{N\to\infty}
 \frac{1}{2N\varrho \delta} \EE
\sum_{M_-\le j, k\le M_+, \atop j\neq  k}   g\big(
(\lambda_j-\lambda_k)N\varrho\big)  
=   \int g(a-b) \Big [1 - \Big( \frac{\sin \pi(a-b)}{\pi(a-b)}\Big)^2 
\Big ] \rd a\rd b.
$$
and without loss of generality, we can assume that $g\ge 0$.

We define
$$
 X_{L}:=
  n^{-1} \, \sum_{L  \le j, k\le L+ n\atop j\ne k} 
 g\big( (\lambda_j-\lambda_k)N\varrho\big).   
$$
and let
$$
   Q_{L}: =  \EE \, X_{L}.
$$

We claim that
\be
\frac{1}{2N\varrho \delta} \EE
\sum_{M_-\le j, k\le M_+, \atop j\neq  k}   g\big(
(\lambda_j-\lambda_k)N\varrho\big)  
 = \frac{1+ O(n^{\gamma-1})}{2N\varrho \delta} 
\sum_{M_-\le L\le M_+}  Q_{L} + O\big(N^2 e^{-cn^{\gamma/6}}\big).
\label{QLL}
\ee
To see this, we consider the expectation value separately 
on $\Omega$ and $\Omega^c$. Since the double sum contains
at most $N^2$ terms and $\PP(\Omega^c)$ is subexponentially small
\eqref{NI1}, it is sufficient to check \eqref{QLL}
when the expectations are restricted to
the set $\Omega$. On the set $\Omega$ we have
\be
(1-Cn^{\gamma-1}) \sum_{M_-\le L \le M_+}
  X_{L} \le  
\sum_{M_-\le j, k\le M_+, \atop j\neq  k}   g\big(
(\lambda_j-\lambda_k)N\varrho\big) 
\le  (1+ Cn^{\gamma+1})\sum_{M_-\le L \le M_+}
  X_{L},
\label{comp}
\ee
where $C$ depends on $\|g\|_\infty$.
This follows from the fact that, by the support of $g$,
only those $(j,k)$ index pairs give nonzero contribution for which
$|\lambda_j-\la_k| \le C/N$,
and thus $|j-k|\le Cn^\gamma$  by \eqref{yy}.  Therefore the sum $\sum_L
X_L$ contains 
each pair $(j,k)$ at least $[ n - Cn^\gamma]$-times  and at most
$[n + Cn^\gamma]$-times.  Taking the expectation of \eqref{comp}
on $\Omega$, we obtain \eqref{QLL}.

Since $Q_{L}$ is bounded by using \eqref{Nmom}, and 
\be
\lim_{\delta \to 0} \lim_{N \to \infty}   \frac{1}{2N\varrho \delta} 
\sum_{M_-\le L \le M_+} 1 = 1, 
\label{numb}
\ee
we only have to estimate $Q_{L}$  for a typical $L$. Additionally
to $L\in \{ M_-, M_-+1, \ldots , M_+\}$, we can thus assume that 
$L \in \cG$, since the relative proportion of good indices approaches
one within any index set with cardinality proportional with $N$
and which is away from the boundary
(see \eqref{fractioncG}). 
More precisely, we fix two sequences $L_-(N)$ and $L_+(N)$
such that $L_\pm(N)\in\cG=\cG_N$ 
$$
   Q_{L_-(N)} = \min \{ Q_L, L\in \cG_N\}, \qquad 
Q_{L_+(N)} = \max \{ Q_L, L\in \cG_N\},
$$
 then it follows from \eqref{numb} that
$$
   (1- \e_{N,\delta}) Q_{L_-(N)}\le     \frac{1}{2N\varrho \delta} 
\sum_{M_-\le L \le M_+} Q_L \le (1+\e_{N,\delta})Q_{L_+(N)}
$$
where $\lim_{\delta\to0}\lim_{N\to\infty}\e_{N,\delta}=0$.
We thus have to show that $Q_{L_\pm(N)}$ converges
to the sine kernel. We will actually prove that
$Q_L$ converges to the sine-kernel for any 
sequence $L=L(N)\in \cG =\cG_N$. The dependence on $N$
will be omitted from the notation.

For $L\in \cG$, we can compute the expectation as
\[
Q_{L} 
=     \EE_{f_t}    \EE_{f_{\by}} X_{L} = \EE\, \EE_{f_{\by}} X_{L} 
\]
according to the convention that $\EE = \EE_{f_t}$.
Recall that definition of the sets $\Omega_1=\Omega_1(L)$,
$\Omega_2 =\Omega_2(L)$ and $\Omega_3(L)$
from \eqref{def:om1},\eqref{def:om2} and \eqref{def:om4}. 
Setting $\wt\Omega:= \Omega_1\cap\Omega_2\cap \Omega_3$,
we see that the probability of its
complement is $\PP (\wt\Omega^c) \le Cn^{-2}$
(see \eqref{om1}, \eqref{om2} and \eqref{om4}).
Since $X_{L}\le C n$, we only have to consider external configurations
such that
$\by \in \wt\Omega$. Thus  
\be
\begin{split}
Q_{L}= & \EE \, {\bf 1} (\by \in \wt\Omega)  \EE_{f_{\by}} X_{L}
  + O(n^{-1})\\
= & \EE  \,  {\bf 1} (\by \in \wt\Omega)  \left [   \EE_{\mu_{\by}} X_{L}
+ 
\int (f_{\by} - 1) X_{L} \rd \mu_{\by} \right ]+O(n^{-1}).
\end{split}
\ee
The second  term in the square bracket will be an error term since it  is
bounded by 
\[
\EE \,  {\bf 1} (\by \in \wt\Omega)   \int |f_{\by} - 1|  |X_{L}| \rd
\mu_{\by} 
\le   C n  \EE {\bf 1} (\by \in \wt\Omega)   \int |f_{\by} - 1| \rd
\mu_{\by}.
\]
Since $\by \in \wt\Omega$ and $L \in \cG$, we have 
\[
\int |f_{\by} - 1| \rd \mu_{\by} \le n^{-2}
\]
from \eqref{L1} and \eqref{L1choice} and we thus
obtain 
$$
Q_{L}
= \EE  \,  {\bf 1} (\by \in \wt\Omega)   \EE_{\mu_{\by}} X_{L} 
+O(n^{-1}).
$$

For the main term, by using \eqref{mumu} and assuming that $B$ is large
enough,   
we can also replace the measure $\mu_{\by}$ by its cutoff version
$\mu_{\by}^{(1)}$ with a negligible error. Let $\varrho_\by
=p^{(1)}_\by:=
p^{(1)}_{\mu_{\by}^{(1)}}$
denote the density and $p^{(2)}_\by:=
p^{(2)}_{\mu_{\by}^{(1)}}$
denote the two point marginal of this measure.
Thus we have
\be  Q_{L}=    (n-1)\EE \, 
{\bf 1} (\by \in \wt\Omega)  \int_{y_{-1}}^{y_1} \rd \alpha  
   \int_{y_{-1}}^{y_1} \rd  \beta \,  p^{(2)}_\by(\alpha, \beta) 
  g\big( (\alpha-\beta)N\varrho\big) 
   +O(n^{-1}).
\label{QL}
\ee
Since $\mu_{\by}^{(1)}$ is an equilibrium measure, its correlation
functions can be obtained as determinants of the appropriate $K$
kernels, see \eqref{determ}. In particular
\be
 0\le p^{(2)}_\by(u,v) = \frac{n-1}{n} p^{(1)}_\by(u)p^{(1)}_\by(v) -
\frac{1}{n(n-1)} K^2(u,v)
 \le  \varrho_\by(u)\varrho_\by(v)
\label{p2}
\ee
holds for the marginals of the measure $\mu_{\by}^{(1)}$.
The lower bound on $p^{(2)}$ follows from the fact that
$K$ is the kernel of a positive operator, i.e. $|K(u,v)|^2\le
K(u,u)K(v,v)$.

Let $0<\kappa\le 1/10$. We now show that, up to an error of order
$\kappa$,
the $\rd\al$ integration in \eqref{QL} can be
restricted from $I_\by = [y_{-1}, y_1]$ onto
$$
 I^*_\by = [y_-^*, y_+^*]:=\Big[ \bar y - \frac{1-2\kappa}{2} |I_\by|, 
\bar y + \frac{1-2\kappa}{2} |I_\by|\Big], \qquad \bar y =
\frac{y_{-1}+y_1}{2},
$$
i.e. onto an interval in the middle of
$I_\by$ with length $(1-4\kappa)|I_\by|$. 
Similarly, the $\rd\beta$ integration will be restricted to
$$
 I^{**}_\by = [y_-^{**}, y_+^{**}]:=\Big[ \bar y - \frac{1-\kappa}{2}
|I_\by|, 
\bar y + \frac{1-\kappa}{2} |I_\by|\Big], \qquad \bar y =
\frac{y_{-1}+y_1}{2},
$$
i.e. onto an interval in the middle of
$I_\by$ with length $(1-2\kappa)|I_\by|$. 
We show how to restrict the $\rd \al$ integration, the other one is
analogous.

The difference  between the full $\rd \al$ integral and the restricted one
is given by
\be 
 C n \EE \,  {\bf 1} (\by \in \wt\Omega)  \int_{I_\by\setminus I^*_\by}
  \rd \alpha  
   \int_{y_{-1}}^{y_1} \rd  \beta \,  p^{(2)}_\by(\al, \beta) 
g\big( (\alpha-\beta)N\varrho\big).
\label{toest}
\ee
To do this estimate, we go back from the equilibrium measure
$\mu^{(1)}_\by$ to $f_\by$
and we also remove the constraint $\wt\Omega$.
As above, all these changes result in negligible errors. Moreover, we can
insert $\Omega$ at the expense of a negligible error since 
$\PP (\Omega)$ is subexponentially small.
Thus \eqref{toest} can be estimated by
\be
 \frac{C}{n}\EE \, {\bf 1}_\Omega  
  \sum_{L\le j,k\le L+n\atop j\neq k}
g\big((\lambda_j-\lambda_k)N\varrho)
 \Bigg[ {\bf 1}\Big( \lambda_j -\lambda_L \le 2\kappa(\lambda_{L+n}-
\lambda_L)\Big)
 +{\bf 1}\Big( \lambda_j -\lambda_L
  \ge (1-2\kappa)(\lambda_{L+n}- \lambda_L)\Big)\Bigg]
\label{toest1}
\ee
up to negligible errors. 
On the set $\Omega$ we know from \eqref{yy} that 
$$
  N\varrho(\lambda_{L+n}- \lambda_L) = n + O(n^{4/5}), \qquad
  N\varrho(\lambda_{j}- \lambda_L) = (j-L) + O(n^{4/5})
$$
assuming that $\gamma \le 1/20$.
Thus the first term in the square bracket of \eqref{toest} can be
estimated by
\be
 Cn^{-1} \EE \, {\bf 1}_\Omega  \sum_{L\le k\le L+n} 
  \sum_{j\ne k} {\bf 1}\Big( L\le     j \le L+2\kappa n + Cn^{4/5}\Big)
 g\big((\lambda_j-\lambda_k)N\varrho) \le C\kappa
\label{toest2}
\ee
taking into account  \eqref{Nmom} as before. Similar estimate holds for
the
second term in \eqref{toest1}. Thus, restricting the 
$\rd \al$-integration to $I_\by^*$ results in an error of order $O(\kappa)$.

Doing the same restriction for the $\rd\beta$ integral, we can 
from now on assume that both  integrations in \eqref{QL} are restricted
to $I_\by^*$, i.e. it is separated away from
the boundary. In particular, from \eqref{tv} and after rescaling,
 we know that $\varrho_\by(\al)$ and
$\varrho_\by(\beta)$ are
essentially constant and equal to $|I_\by|^{-1}(1+O(n^{-\gamma/12})$.
Moreover, on the set $\wt \Omega$,
we know from \eqref{Il1} that $|I_\by|^{-1}= \frac{N\varrho}{n}(1+
O(n^{\gamma-1/4}))$, i.e.
\be
 \varrho_\by(\beta) = \frac{N\varrho}{n}\big(1+  O(n^{-\gamma/12})\big),
\qquad
\mbox{for any}\;\; \beta\in I_\by^{**}
\label{varrho}
\ee
Since $I_\by^*\subset I_\by^{**}$, the same formula holds for
$\varrho_\by(\al)$ for all $\al\in I_\by^*$.

We now compute the restricted integrals in \eqref{QL}. Changing
variables from $\beta$ to $b$ with   $  \beta = \alpha+ b(n
\varrho_\by(\alpha) )^{-1}$, 
we have 
\be \label{64}
\begin{split}
Q_{L}= &    \EE  {\bf 1} (\by \in \wt\Omega)   \int_{y_-^*}^{y_+^*} \rd
\alpha  
 \frac {n-1} {n  \varrho_\by(\alpha)} 
   \int_{(y_-^{**} - \alpha)n \varrho_\by(\alpha)}^{(y_+^{**} - \alpha)
n \varrho_\by(\alpha) }
\rd  b \,  p^{(2)}_\by\Big ( \alpha, 
\alpha+ \frac b {n \varrho_\by(\alpha)}  \Big )    
g\Big ( \frac {- N\varrho   b} {n \varrho_\by(\alpha) }  \Big )
\\ & +O(n^{-1})+O(\kappa).    
\end{split}
\ee
Since $g$ is smooth and has compact support, we have 
\be
g\Big ( \frac {- N\varrho   b} {n \varrho_\by(\alpha) }  \Big ) 
=   g( -b)  + \xi , \quad 
|\xi| \le  C \left | \frac { N\varrho  } {n  \varrho_\by(\alpha) } - 1
\right |
 \le Cn^{-\gamma/12}
\label{xi}
\ee
from \eqref{varrho}.
Therefore,  when we insert \eqref{xi} into \eqref{64} and use \eqref{p2},
the error term involving $\xi$ is bounded by 
\be
\begin{split}
 C  \EE   {\bf 1} (\by \in \wt\Omega)  & \int_{I_\by^*} \rd \alpha   
\frac {1} { \varrho_\by(\alpha)}   
\int_{(y_-^{**} - \alpha)n \varrho_\by(\alpha)}^{(y_+^{**} -
\alpha)n\varrho_\by(\alpha) }  \rd  b \,  
  p^{(2)}_\by\Big ( \alpha, 
\alpha+ \frac b {n  \varrho_\by(\alpha)} \Big )  
    \left | \frac { N\varrho  } {n  \varrho_\by(\alpha) } - 1 \right | 
\\ 
\le &  C n^{-\gamma/12} \EE   {\bf 1} (\by \in \wt\Omega)   \int_{I_\by^*}
\rd \alpha  
   \int_{(y_-^{**} - \alpha)n \varrho_\by(\alpha)}^{(y_+^{**} -
\alpha)n\varrho_\by(\alpha) }  \rd  b \,  
\varrho_\by \Big ( 
\alpha+ \frac b {n  \varrho_\by(\alpha)} \Big )      
\\
\leq & C n^{-\gamma/12} \EE   {\bf 1} (\by \in \wt\Omega) 
 \int_{I_\by^*} \varrho_\by(\al)\rd \alpha  \int_{I_\by^{**}}
\varrho_\by(\beta) \rd \beta 
\\ \le &   C n^{-\gamma/12},
\end{split} 
\ee 
using that, by definition,
$$
  \int_{I_\by^*} \varrho_\by(\al)\rd \alpha \le
\int_{I_\by} \varrho_\by(\al)\rd \alpha =1
$$
and similar bound holds for the $\beta$-integral.

Thus we can replace the variable of $g$ in \eqref{64} by $-b$ 
with negligible errors. 
Now Theorem \ref{thm:L} states that  
\[
\frac {1} {{ \varrho_\by(\alpha)}^2} p^{(2)}_\by\Big ( \alpha, 
\alpha+ \frac b {n \varrho_\by(\alpha)}  \Big )   
\to  \left [ 1-  \Big(  \frac{\sin \pi b}{\pi b}\Big)^2  \right ]   
\]
Clearly, as $n \to \infty$, 
\[
(y_{\pm}^{**} - \alpha)n \varrho_\by(\alpha) \to \pm \infty
\]
for all $ \alpha \in I_\by^*$, i.e. the integration limits can be
extended to infinity, noting that $g$ is compactly supported. 
Finally, from \eqref{varrho} we have
$$
  \int_{I_\by^*} \varrho_\by(\al)\rd\al \ge 1- O(\kappa) -
O(n^{\gamma-1}).
$$
Combining all these estimates with Theorem \ref{thm:L}, we obtain
\[    
Q_{L} 
=  \int_{-\infty}^{\infty} \rd  b \,    \left [ 1-  
\Big(  \frac{\sin \pi b}{\pi b}\Big)^2  \right ]   
g ( -b )   + O(n^{-\gamma/12}) + O(\kappa) +o(1),
\]
where the last term error term is from Theorem \ref{thm:L} that goes
to zero as $N\to\infty$.
Taking  the $N\to\infty$, $\delta\to0$ and $\kappa\to 0$ limits
in this order, we arrive at the
proof of Theorem \ref{mainthm}.  \qed

\appendix

\section{Proof of Theorem \ref{thm:semi}}\label{sec:B}

We start with the proof of \eqref{eq:mup} and \eqref{eq:omup}. 
{F}rom Theorem 4.6 of \cite{ESY3}, we have 
\[ \P \left( |m (x+iy)| \geq K \right) \leq C e^{-c \sqrt{K N |y|}}
\] 
for all $K>0$ sufficiently large, and $|y| \geq (\log N)^4/N$. 
Since moreover $|m (x+iy) | \leq |y|^{-1}$ with probability one, 
we obtain, under the assumption $N |y| \geq (\log N)^4$, 
\[ 
\EE |m (x+i y)|^q \leq K^q + C |y|^{-q} e^{-c \sqrt{KN|y|}} \leq C_q 
\]
uniformly in $N,x$. The bound (\ref{eq:omup}) follows because $\omega_y
(x) = \pi^{-1}
\text{Im } m (x+iy)$.

\bigskip

To prove the results about the closeness of  $m(z)$ or $\EE\, m(z)$ to
$m_{sc}(z)$, we 
first recall  the key identity about the trace of a resolvent in
terms of resolvents of minors (see, e.g., (4.5) of \cite{ESY1}):
\begin{equation}\label{eq:mz}
m(z) = \frac{1}{N} \, \Tr \, \frac{1}{H-z} 
=  \frac{1}{N}
\sum_{k=1}^N \frac{1}{-m(z) - z + \delta_k (z)}
\end{equation}
with
\begin{equation}\label{eq:deltadef}
 \delta_k (z)  = h_{kk} + m(z) - \left(1-\frac{1}{N} \right)m^{(k)} (z) -
X^{(k)} (z) , 
\end{equation}
and
\[
m^{(k)} (z) = \frac{1}{N-1} \Tr \, \frac{1}{B^{(k)} - z}, \qquad X^{(k)}
(z) 
= \frac{1}{N} \sum_{\alpha} \frac{\xi_{\alpha}^{(k)}
-1}{\lambda_\alpha^{(k)} - z},
\qquad \xi^{(k)}_{\alpha} = N| \ba^{(k)}\cdot \bv_\alpha^{(k)}|^2 \, .
\] 
Here $B^{(k)}$ is the $(kk)$-minor of $H$ (the $(N-1)\times (N-1)$ matrix
obtained
by removing the $k$-th row and the $k$-th column from $H$),
$\lambda^{(k)}_{\alpha},
\bv_{\alpha}^{(k)}$ are the eigenvalues and the eigenvectors of
$B^{(k)}$, and 
$\ba^{(k)} = (h_{k1}, \dots, h_{k,k-1}, h_{k,k+1}, \dots h_{kN})$. 
Throughout the proof we let $x, y$ denote the real and imaginary parts of
$z=x+iy$. Moreover, we will restrict our attention to $y >0$. 
The case $y <0$ can be handled similarly. 

\bigskip

{\it Step 1. Lower bound on $|m(z) +z|$.} There exist constants $C,c>0$
such that 
\begin{equation}\label{eq:lowm} 
\P \left( |m(x+iy) + (x+iy)|  \leq c \right) 
\leq e^{-C \sqrt{Ny}} 
\end{equation}
for all $x \in \bR$, $y \geq (\log N)^4/N$,
and for all $N$ large enough (depending only on the choice of $C,c$).

\medskip

To show (\ref{eq:lowm}), we use a continuity argument. 
We claim that there exist positive
constants $C_1,C_2, C_3, c >0$ 
such that the following four conditions are satisfied: 
\begin{equation}\label{eq:cond} \begin{split}  &\inf_{z \in \bC\setminus
[-2,2]}
|z + m_{\text{sc}} (z)| \geq 2c , \\  &\P \left( |m(x+iy)| \geq
\frac{1}{2c} \right) \leq \frac{e^{-C_1 \sqrt{Ny}}}{3} \qquad
\text{for all } x \in \bR, y \geq (\log N)^4/N \\ &\P \left( 
\sup_{1 \leq k \leq N}  |\delta_k (x+iy)| \geq \frac{c}{16} \right)
\leq \frac{e^{- C_2 \sqrt{Ny}}}{3} \qquad  \text{for all } x \in \bR, y
\geq (\log N)^4/N \\
&\P \left( |m(x+iy) - m_{\text{sc}} (x+i y)| \geq c 
\right) \leq \frac{e^{-C_3 \sqrt{Ny}}}{3} \qquad \text{for all } |x| \leq
1,
y \geq (\log N)^4 /N \, .
\end{split} \end{equation}
The first condition can be checked explicitly from \eqref{def:msc}.
The second condition follows from the
upper bound in Theorem 4.6 of \cite{ESY3}. The third condition can be
satisfied
because of Lemma 4.2 in \cite{ESY3}, combined with the fact that 
$\P (\max_k \, |h_{kk}| \leq (c/48)) \leq e^{-CN}$ and with the
observation that
\be
  \Big|m (z) - \big(1- \frac{1}{N}\big) m^{(k)} (z)\Big| \leq
\frac{C}{Ny}
\label{mmm}
\ee
with probability one 
(see, for example (2.7) in \cite{ESY2}). Finally, the last condition can
be verified by Theorem 4.1 of \cite{ESY3}. Note that the last three 
conditions only need to hold for all $N > N_0 (c,C_1, C_2, C_3)$ large 
enough. Fix $C = \min (C_1, C_2, C_3)$.

\medskip

For $|x|\leq 1, y \geq (\log N)^4/N$ we have (using the first and the
last equation in (\ref{eq:cond})) 
\[ \begin{split} \P \Big( &|m(x+iy) + (x+iy)| \leq c \Big) \leq \P  
\left( |m(x+iy) -m_{\text{sc}} (x+iy)| \geq c \right) \leq e^{-C
\sqrt{Ny}} \, .
\end{split} \] Hence (\ref{eq:lowm}) holds true (with the defined
constants $c,C$)
for every $|x| \leq 1$, $y \geq (\log N)^4 /N$ . Suppose now that
(\ref{eq:lowm})
holds for a given $z = x+iy \in \bC$. Then we claim that (\ref{eq:lowm})
holds
true for all \[ z' = x' + i y'  \in B_z = \{ z' \in \bC : |z- z'| \leq D
N^{-2}, 
\quad \text{Im } z' \geq (\log N)^4 /N \}\] for a constant $D$ depending
only on $c$,
and for all $N > N_0$; this implies immediately that (\ref{eq:lowm})
holds true
for all $x \in \bR$, $y \geq (\log N)^4/N$ and $N > N_0$. 

\medskip

To prove (\ref{eq:lowm}) for $z' \in B_z$, notice that $|m' (z)| \leq N^2$
for all
$z \in \bC$ with $\text{Im } z \geq (\log N)^4 /N$ with probability one.
Therefore, using (\ref{eq:lowm}) for $z$, we find that 
\begin{equation}\label{eq:lowmz'} \P \left( \, |m(z') + z'| \leq
\frac{c}{2}\right)
\leq e^{-C \sqrt{Ny'}} \end{equation} for all $z' \in B_z$. Expanding
(\ref{eq:mz}),
we obtain that
\[ 
m(z') + \frac{1}{m(z') +z'} = -\frac{1}{N} \sum_{k=1}^N \frac{1}{m(z')
+z'} 
\delta_k (z')\frac{1}{m(z')+z' -\delta_k (z')}\, .
\]
Therefore, 
\[ 
\begin{split} \P \Big( \frac{1}{|m(z') +z'|} \geq \frac{1}{c} \Big)
&\leq \P \Big(|m(z')| \geq \frac{1}{2c}\Big) + \P \left( \frac{ 1  }{N} 
\sum_{k=1}^N \frac{ |\delta_k(z')|}{|m(z') +z'| |m(z') +z' -\delta_k(z')|}
\geq \frac{1}{2c} \right) \\ & \leq  \P \Big( |m(z')| 
\geq \frac{1}{2c} \Big) + \P \left( |m(z') +z'| \leq \frac{c}{2} \right)
 + \P \left( \sup_{1 \leq k \leq N}  |\delta_k (z')| \geq \frac{c}{16}
\right)
\\ &\leq e^{-C \sqrt{Ny}} \end{split} \]
where we used (\ref{eq:cond}) and (\ref{eq:lowmz'}). This implies
(\ref{eq:lowm}) 
for $z' \in B_z$, and completes the proof of Step 1.

\bigskip

{\it Step 2. Convergence to the semicircle in probability.} Suppose that
$|x| \leq K$, $(\log N)^4/N \leq y \leq 1$. Then there exist constants
$c,C,\delta_0$, only depending on $K$, such that 
\begin{equation}\label{eq:step2} 
\P \left( |m (x+iy) -  m_{\text{sc}} (x+i y)| 
\geq \delta \right) \leq C\,
e^{-c \delta \sqrt{Ny \, |2-|x||}} 
\end{equation} 
for all $\delta < \delta_0$, and all $N \geq 2$.

\medskip

To show (\ref{eq:step2}), we first observe that, by increasing the
constant $C$, we can assume $N$ to be sufficiently large. Then we expand
(\ref{eq:mz}) into 
\begin{equation}\label{eq:exp1}  
m(z) + \frac{1}{m(z) +z} = -\frac{1}{N} 
\sum_{k=1}^N \frac{1}{m(z) +z} \delta_k (z)\frac{1}{m(z)+z -\delta_k
(z)}\, .
\end{equation}
We define the complex random variable 
\[ 
Y(z) =  \frac{1}{N} \sum_{k=1}^N \frac{1}{m(z) +z} 
\delta_k (z)\frac{1}{m(z)+z -\delta_k (z)}\,. 
\]
{F}rom (\ref{eq:lowm}) and since, by Theorem 4.2 of \cite{ESY3}, 
\begin{equation}\label{eq:deltaP} 
\P \left( \sup_{1 \leq k \leq N} |\delta_k (z)|
\geq \delta\right) \leq C e^{-c \min (\delta \sqrt{Ny}, \delta^2 N y)}
\end{equation} 
for all $y \geq (\log N)^4/N$ and $\delta >0$, we find 
\begin{equation}
\label{eq:PYz} \P \, (|Y(z)| \geq \delta) \leq \P \, (|m(z) + z| \leq c)
+
\P \left( \sup_{k \leq N} |\delta_k (z)| \geq \min \, 
\left( \frac{c^2\delta}{2}, \frac{c}{2} \right) \right) \leq C e^{-c\delta
\sqrt{Ny}}
\end{equation}
for $\delta$ small enough, $y \geq (\log N)^4 /N$, and $N$ large enough
(independently of $\delta$). 

\medskip

To prove (\ref{eq:step2}) for $|x| < 2$, we use that, 
from (6.14) in \cite{ESY3},
\[ \left| m + \frac{1}{m+z} \right| \leq \delta \quad \Rightarrow \quad
|m-m_{sc}| \leq \frac{C\delta}{(2-|x|)^{1/2}} \]
for all $z = x+iy$ with $|x| < 2$ and $0<y<1$. This implies, using
(\ref{eq:PYz}), that
\[ \P \left(|m(z) - m_{sc} (z)| \geq \delta \right) \leq \P \left( |Y(z)|
\geq c \delta (2-|x|)^{1/2} \right) \leq C e^{-c\delta \sqrt{Ny
(2-|x|)}} \] for all $\delta$ small enough, $N$ large enough, $|x| \leq
2$, $(\log N)^4/N \leq y \leq 1$.

\medskip

It remains to show (\ref{eq:step2}) for $2 \leq |x| \leq K$. 
To this end, for $(\log N)^4/N \leq y \leq 1$ and  $2 \leq |x| \leq K$,
 we consider the event 
\[ 
\Omega^*
= \Big\{  |m(z)+z|\ge c, \;\sup_{1 \leq k \leq N} |\delta_k (z)|
\leq \delta, \; 
   |Y(z)| \leq \delta \sqrt{|2-|x||} \Big\} \,, \quad z=x+iy \, . 
\]
{F}rom \eqref{eq:lowm}, \eqref{eq:deltaP}
and (\ref{eq:PYz}), we have $\P \, ( [\Omega^*]^c) \leq e^{-c \delta 
\sqrt{Ny |2-|x||}}$ for all $\delta$ small and $N$ large enough. 
Solving (\ref{eq:exp1}) for $m$ on the set $\Omega^*$, we get 
\[ \begin{split} m(z) &
= - \frac{z}{2} - \frac{Y(z)}{2} + \sqrt{\frac{z^2}{4} -1 
-\frac{zY(z)}{2} +\frac{Y(z)^2}{4}} \, .
\end{split} \] 
Since $m(z)$ is the Stieltjes transform of an empirical measure with
finite support, it is analytic away from a compact subset of the real
axis. Similarly,
on the set $\Omega^*$,  $Y(z)$ is  bounded and 
analytic away from a compact subset of the real axis. 
The square root in the above formula is therefore uniquely 
defined as the branch analytic on $\bC \backslash (-\infty,0]$, 
characterized by the property that the real part of
the square root is non-negative.
Hence,  on $\Omega^*$,
\[ \begin{split}
m(z) - m_{sc} (z) = & -\frac{Y(z)}{2} + 
\sqrt{\frac{z^2}{4} -1 - \frac{z Y(z)}{2} + \frac{Y(z)^2}{4}}
 - \sqrt{\frac{z^2}{4} - 1} \\ = & 
-\frac{Y(z)}{2} - \frac{1}{4}
 \frac{ 2zY(z) - Y(z)^2}{\sqrt{\frac{z^2}{4} -1 - 
\frac{z Y(z)}{2} + \frac{Y(z)^2}{4}} + \sqrt{\frac{z^2}{4} - 1}}
\end{split}
\]
using the explicit formula \eqref{def:sc} for $m_{sc}(z)$,
and therefore 
\begin{equation}\label{eq:mbd}
| m(z) - m_{sc} (z) | \leq  \frac{|Y(z)|}{2} + 
\frac{ 2|z| |Y(z)| + |Y(z)|^2}{4 \text{Re} \, \sqrt{\frac{z^2}{4} - 1}}
 \leq C \frac{|Y(z)| + |Y(z)|^2}{\sqrt{|2-|x||}}
\end{equation}
using the fact that
\[ \text{Re } \sqrt{\frac{z^2}{4} - 1}  
\geq C |2-|x|| \qquad \text{for all } 2 \leq |x| \leq K, \; |y| \leq 1 \,. \]
{F}rom (\ref{eq:mbd}), we obtain that
\[ 
\begin{split} 
\P \left( \left| m (z) - m_{sc} (z) \right| 
\geq \delta \right) \leq \P \left( |Y (z)| + |Y(z)|^2 \geq 
c \delta \sqrt{|2-|x||} \right) \leq e^{-c \delta \sqrt{Ny |2-|x||}}
\end{split}
\]
for all $\delta$ small enough, $2 \leq |x| \leq K$, 
$(\log N)^4/N \leq y \leq 1$, and $N$
large enough. 

\bigskip

{\it Step 3. Fluctuations of $m(z)$.} Suppose that $|x| \leq K$,
$(\log N)^4 /N \leq y \leq 1$ and 
$Ny |2-|x|| \geq (\log N)^4$. Then there exist constants $C, c >0$ such
that 
\begin{equation}\label{eq:step3}
\P \left( \left| m(z) - \EE m(z) \right| 
\geq \delta \right) \leq C\, e^{-c \delta \sqrt{Ny |2-|x||}} 
\end{equation} 
for all $0<\delta\le \delta_0$, with $\delta_0$ small enough and all $N$
large enough.

\medskip

To show (\ref{eq:step3}), we observe first that 
\[ 
\left| \EE m(z) - m_{sc} (z) \right| \leq \EE |m(z) - m_{sc} (z)| \leq 
\int_0^{1/y} \rd t \, \P \left( |m(z) - m_{sc} (z)| \geq t \right),
\]
where we used that $|m(z)| \leq y^{-1}$. 
Using (\ref{eq:step2}), we obtain 
\begin{equation} \label{eq:Em-sc}
\begin{split} \left| \EE m(z) - m_{sc} (z) \right| 
&\leq \frac{C}{\sqrt{Ny |2-|x||}}+ \frac{1}{y} e^{-c \sqrt{Ny|2-|x||}} 
\leq  \frac{2C}{\sqrt{Ny |2-|x||}}
\end{split} \end{equation}
for $N$ large enough. For $\delta \sqrt{Ny|2-|x||} \ge 4C$ we thus obtain
\[ \P \left( |m(z) - \EE m(z)| \geq \delta\right) \leq 
\P \Big(|m(z) - m_{sc} (z)| \geq \frac{\delta}{2}\Big)
\leq C\, e^{-c \delta \sqrt{Ny |2-|x||}}\, 
\]  
where we used (\ref{eq:step2})  again. For $\delta \sqrt{Ny|2-|x||} \le
4C$
the bound \eqref{eq:step3} is trivial.

As a consequence of \eqref{eq:step3}, we immediately obtain
\eqref{eq:E|m|}.
If $Ny|2-|x||\le (\log N)^4$, we directly use \eqref{eq:mup}. Otherwise,
we
use
$$
\EE\, |m(z)- \EE \, m(z)|^q  \le C_q\int_0^{\delta_0} t^{p-1}
 \PP (|m(z)- \EE \, m(z)|\ge t) \rd t + Cy^{-q} 
e^{-c\delta_0\sqrt{Ny|2-|x||}}
$$
from \eqref{eq:step3}, and we obtain the first term on the r.h.s. of
\eqref{eq:E|m|}.

\bigskip

{\it Step 4. Convergence to the semicircle in expectation.}
Assume that $|x| \leq K$, $(\log N)^4 /N \leq y \leq 1$ and 
$Ny |2-|x|| \geq (\log N)^4$. Then 
\be
\left| \EE m(z) - m_{sc} (z) \right| \leq \frac{C}{N y |2-|x||^{3/2}} 
\label{eq:step4}
\ee
for a universal constant $C$.  Note that this bound
gains an additional $(N\eta)^{-1/2}$
factor on the precision of the estimates compared with  Step 2 and Step 3,
but the negative power of $|2-|x||$ has increased.

\medskip

To prove \eqref{eq:step4}, with $c_0: = \inf_z |m_{sc} (z) + z| >0$, we
have 
\begin{equation}\label{eq:Em+z}
|\EE m(z) + z| \geq |m_{sc} (z) + z|
- |\EE m(z) - m_{sc} (z) | \geq c_0 - \frac{C}{\sqrt{Ny |2-|x||}}
\geq \frac{c_0}{2}
\end{equation}
for $N$ large enough (here we used (\ref{eq:Em-sc})). 
Expanding the denominator in the r.h.s. of (\ref{eq:mz}) 
around $\EE m(z) + z$, we find
\begin{equation}
\begin{split} m(z) = \; &-\frac{1}{\EE m(z) + z} - 
\frac{1}{N} \sum_{k=1}^N \frac{1}{(\EE m(z) + z)^2}
\left( m(z) - \EE m(z) + \delta_k (z) \right) \\ 
&+ \frac{1}{N} \sum_{k=1}^N \frac{1}{(\EE m(z) +z)^2} 
\left( m(z) - \EE m(z) + \delta_k(z) \right)^2 \frac{1}{m(z) + z -\delta_k
(z)} \, . 
\end{split}
\end{equation}
Taking expectation, we find 
\begin{equation}
\begin{split}
\EE m(z) +\frac{1}{\EE m(z) + z} = &\;  -\frac{1}{(\EE m(z) + z)^2} \, 
\EE \delta_1 (z) \\ &+ \frac{1}{(\EE m(z) + z)^2} \EE 
\left[ (m(z) - \EE m(z) + \delta_1 (z))^2 \frac{1}{m(z) + z -\delta_1 (z)}
\right] .
\end{split}
\end{equation}
With a Schwarz inequality, we get 
\begin{equation}\label{eq:Schwarz} 
\begin{split} 
\left| \EE m(z) +\frac{1}{\EE m(z) + z} \right| \leq \;
&\frac{1}{|\EE m(z) + z|^2} |\EE \delta_1 (z)|  \\ 
&+ 2\frac{ \left(\EE |m(z) - \EE m(z)|^4 + \EE |\delta_1 (z)|^4
\right)^{1/2}}{|\EE m(z) + z|^2} \left(\EE \frac{1}{|m(z) + z -\delta_1
(z)|^2}
\right)^{1/2} \, .
\end{split} 
\end{equation} 
{F}rom (\ref{eq:step3}), we find
\be
\EE \left| m (z) - \EE m(z) \right|^q \leq \frac{C_q}{(Ny |2-|x||)^{q/2}}
\label{EMM}
\ee
for arbitrary $q \geq 1$. Moreover, with $c$ fixed in (\ref{eq:lowm}), 
we have 
\[
\P \left( |m(z) + z -\delta_1 (z)| \leq \frac{c}{2} \right) \leq 
\P (|m(z) +z| \leq c) +   \P \left( |\delta_1 (z)| \geq \frac{c}{2}
\right)
\leq e^{-C \sqrt{Ny}} \leq e^{-C (\log N)^2} 
\]
using \eqref{eq:deltaP},
and hence 
\begin{equation}\label{eq:m+z-delta}
\EE \, \frac{1}{|m(z) + z - \delta_1 (z)|^q} \leq 
\frac{1}{y^q} e^{-C (\log N)^2} + \frac{2^q}{c^q} 
\leq \frac{2^{q+1}}{c^q} 
\end{equation} 
if $N$ is large enough. Here we used the fact that 
$\text{Im } m(z) + z - \delta_1 (z) \geq \text{Im } z = y$.
{F}rom (\ref{eq:deltaP}), we also have 
\begin{equation}\label{eq:delta1qE} 
\EE |\delta_1 (z)|^q \leq \frac{C^q}{(Ny)^{q/2}}\,. 
\end{equation}
{F}rom the definition of $\delta_1 (z)$
in  (\ref{eq:deltadef}), from $\EE X^{(k)}=0$ and from \eqref{mmm}, we get
\begin{equation}\label{eq:delta1z}
\left| \EE \, \delta_1 (z) \right| \leq \frac{1}{Ny}
\end{equation} 
Combining this bound with \eqref{EMM},  we find, from (\ref{eq:Schwarz}),
that 
\begin{equation}\label{eq:EE-stab} 
\left| \EE m(z) + \frac{1}{\EE m(z) + z} \right| \leq
 \frac{C}{Ny |2- |x||}.
\end{equation}
Recall that $m_{sc}(z)$ solves the equation
$$
   m_{sc}(z) + \frac{1}{m_{sc}(z) + z} =0.
$$
This equation is stable in a sense that the
inverse of the function $m\to m+(m+z)^{-1}$ near zero
is Lipschitz continuous with a constant proportional 
to $|2-|x||^{1/2}$. Thus we obtain
\[
|\EE m(z) - m_{sc} (z)| \leq \frac{C}{Ny |2- |x||^{3/2}} \,
\]
and this completes Step 4.

\bigskip

{\it Step 5. Alternative bound on $|\EE m(z) - m_{sc} (z)|$.} 
Assuming $|x|\le K$ and $(\log N)^4/N\le y \le 1$, 
there exists a constant $C>0$ such that 
\begin{equation}\label{eq:step5} 
\left| \EE m(z) - m_{sc} (z) \right| \leq \frac{C}{N y^{3/2}
|2-|x||^{1/2}}\, 
\end{equation}
for all $N$ large enough (independently of $z=x+iy$).

\medskip

To prove (\ref{eq:step5}), we use the bound 
\begin{equation}\label{eq:EE-LSI} 
\EE \left| m(z) - \EE m(z) \right|^q 
\leq \frac{C^q}{(Ny^{3/2})^q} 
\end{equation} 
which is valid for all $q \geq 1$ and it follows from Theorem 3.1 in
\cite{ESY1}. 
Expanding again the denominator in the r.h.s. of (\ref{eq:mz}) 
around $\EE m(z) + z$, we get
\begin{equation}\begin{split}
m(z) = \; &-\frac{1}{\EE m(z) + z} - \frac{1}{N} \sum_{k=1}^N 
\frac{m(z) - \EE m(z) + \delta_k (z)}{\big[\EE m(z) + z\big]\; 
\big[m(z) +z -\delta_k(z)\big]} \\
= \; &-\frac{1}{\EE m(z) + z} - \frac{1}{N} \sum_{k=1}^N 
\frac{m(z) - \EE m(z)}{\big[\EE m(z) + z\big]\; 
\big[m(z) +z -\delta_k(z)\big]}
- \frac{1}{N} 
\sum_{k=1}^N \frac{ \delta_k (z)}{(\EE m(z) + z)^2} \\ 
&- \frac{1}{N}  \sum_{k=1}^N
\frac{\delta_k (z)\,(\EE m(z) - m(z) + \delta_k (z))}{\big[\EE m(z) +
z\big]^2\;
\big[ m(z) +z -\delta_k(z)\big]} \,.
\end{split}
\end{equation}
Taking the expectation, we find
\begin{equation}\begin{split}
\Big| \EE m(z) + &\frac{1}{\EE m(z) + z} \Big| \\ \leq &\;  
\frac{1}{|\EE m(z) + z|} \left(\EE \left| m(z) - \EE m(z) \right|^2
\right)^{1/2} 
\left( \EE \frac{1}{|m(z) +z -\delta_1 (z)|^2} \right)^{1/2}  \\
&+ \frac{ \left| \EE \, \delta_1 (z) \right|}{|\EE m(z) + z|^2} +
\frac{ \left(\EE |\delta_1 (z)|^4\right)^{1/2}}{|\EE m(z) + z|^2} 
\left( \EE \frac{1}{|m(z) +z -\delta_1 (z)|^2}  \right)^{1/2} \\ 
&+ \frac{ \left(\EE |\delta_1 (z)|^4\right)^{1/4}
\left( \EE |m(z) - \EE m(z)|^4 \right)^{1/4} }{|\EE m(z) + z|^2}
\left( \EE \frac{1}{|m(z) +z -\delta_1 (z)|^2} \right)^{1/2}.
\end{split}
\end{equation}
Using (\ref{eq:m+z-delta}) with $q=2$, (\ref{eq:EE-LSI}) with $q=2$ and
$q=4$, 
(\ref{eq:delta1qE}) with $q=4$ and (\ref{eq:delta1z}), we find, by the
stability
argument, that 
\[
\Big| \EE m(z) + \frac{1}{\EE m(z) + z} \Big| \leq \frac{C}{Ny^{3/2}} 
\]
which implies (\ref{eq:step5}). This completes the proof of
Theorem \ref{thm:semi}. \qed

\section{ Proof of Proposition \ref{prop:distr}}\label{sec:C}

We start with the proof of \eqref{NNint}.
{F}rom the moment method, we know that
if $\la_{\text{min}} (H)$ and $\la_{\text{max}} (H)$ denote the smallest
and 
the largest eigenvalues of the hermitian Wigner matrix $H$, and if 
$K$ is large enough, then
\[ 
\P \left( \la_{\text{min}} (H) \leq -K \right) = 
\P \left( \la_{\text{max}} (H) \geq K \right) \leq K^{-cN^{2/3}}
\] 
(for example, one can use the bound $\EE \mbox{Tr}\, H^{N^{2/3}} \le C$
from \cite{Sosh}; the symmetry condition on the distribution can
be removed by symmetrization).
This implies that $\fN(E) \leq N K^{-cN^{2/3}}$ for $E < -K$ and 
$1-\fN(E) \leq N K^{-cN^{2/3}}$ for all $E >K$. Therefore
\be
\int_{-\infty}^{\infty} |\fN(E) - \fN_{sc} (E)| \rd E \leq 
\int_{-K}^K \big| [\fN(E)-\fN(-K)] - \fN_{sc} (E)\big| \rd E + 2 N K^{-cN^{2/3}}
\label{err1}
\ee for $K >0$ large enough.  The last term is negligible.
The main estimate is contained  in
the following lemma whose proof is given at the end of this
section.

\begin{lemma}\label{lm:HS} Let $\varrho^*=\varrho_+-\varrho_-$ be a
difference of two
finite measures  with support in $[-K,K]$
for some $K>0$. Let
$$
    m^*(z) = \int_\RR \frac{\varrho^*(\rd x)}{x-z}, \qquad
  \fN^*(E): = \int_{-K}^E \varrho^*(\rd x)
$$
be the Stieltjes transform
and the distribution function of $\varrho^*$,
respectively. Denote moreover by $m_\pm^* (z)$
the Stieltjes transforms of $\varrho^*_\pm$.
We assume that $m^*, m_+^*, m_-^*$ satisfy the following bounds for
$|x|\le K+1$:
\begin{align} \label{mcond1} &| m_+^*(x+iy) | + |m_-^* (x+i y)|  \leq L_1
\quad  &\text{for all }  \quad  (\log N)^4/N\le |y| \le 1 \\
\label{mcond2}
 &|m^*(x+iy) |  \leq \frac{L_2}{N |y|^{3/2} |2-|x||^{1/2}}
\qquad &\text{for all } \quad (\log N)^4/N\le |y| \le 1 \\
\label{mcond3}
&| m^*(x+iy)  | \leq \frac{L_3}{N |y| |2-|x||^{3/2}}
\qquad &\text{ for all} \quad N |y| |2-|x|| \geq (\log N)^4,
\end{align}
with some constants $L_1, L_2, L_3$. 
Then 
\be
      \int_{-K}^K |\fN^*(E)| \rd E \le \frac{CL}{N^{6/7}}.
\label{NEb}
\ee
with $L=\max\{ L_1, L_2, L_3\}$.
The constant $C$ 
in \eqref{NEb} depends only on $K$.
\end{lemma}

We apply this lemma for the signed measure
$\varrho^*(\rd x)={\bf 1}(|x|\le K)\big[
\varrho(x)-\varrho_{sc}(x)\big]\rd x$. 
The bounds \eqref{mcond1}, \eqref{mcond2}, and \eqref{mcond3}
follow from \eqref{eq:mup}, \eqref{eq:Em-msc-2}  and \eqref{eq:Em-msc-1} 
(choosing $K+1$ instead of $K$), respectively.
{F}rom \eqref{NEb} we obtain 
$$
\int_{-K}^K \big| [ \fN(E) -\fN (-K)] - \fN_{sc} (E)\big| \rd E \le
\frac{C}{N^{6/7}},
$$
which, together with \eqref{err1}, completes the proof of \eqref{NNint}.

\medskip

For the proof of \eqref{NNE}, we fix $|E|\le K$ and we choose
$N^{-3/4}\le\eta\le 1$ to be optimized later. 
Define a function $f=f_E:\bR\to \bR$ such that $f(x)=1$ 
for $x\le E-\eta$, $f(x)=0$ for $x>E+\eta$ with
$|f'|\leq C\eta^{-1}$ and $|f''| \leq C \eta^{-2}$. 
We have
\be
  \Big| \Big(\frac{\N(-\infty, E)}{N}-\fN(E)\Big) - 
 \int_{-\infty}^\infty f_E(\la)[\om(\rd \la)- \varrho( \la)\rd\la] \Big|
\leq  
 \frac{\N[E-\eta, E+\eta]}{N}+ \EE \frac{\N [E-\eta,E+\eta]}{N}  \, .
\label{nf1}
\ee
The second term on the r.h.s is estimated by $C\eta$, using
(\ref{eq:omup}).
For the first term we use Theorem 4.6 of \cite{ESY3}:
\be
 \PP\Big( \frac{\N[E-\eta, E+\eta]}{N} \ge \delta/4\Big) \le
Ce^{-c\sqrt{\delta N}}
\label{nene}
\ee
with some positive $c>0$.

Now we consider the fluctuation of the smoothed distribution function
$$
W:=\int_{-\infty}^\infty f_E(\la)[\om(\rd \la)- \varrho( \la)\rd\la]  
 =\frac{1}{N} \sum_{\al=1}^N \big[ f_E(\lambda_\al) - \EE
f_E(\lambda_\al)\big].
$$
We partition 
$[-K-2, K+2]$ into intervals $I_r$ of length $\eta$. 
For $M\ge M_0$ with a sufficiently large $M_0$, and set
$$
  \Omega_k:=  \Big\{ kMN\eta\le \sup_r \N(I_r)<
  (k+1)M N\eta \Big\} , \qquad k=0, 1,2,
 \ldots ,\eta^{-1},
$$
then from
Theorem 4.6 of \cite{ESY3} we know that
$$
\PP \{ \Omega_k\}
\le  C e^{-c\sqrt{kMN\eta} }.
$$
Analogously to the calculation \eqref{Ader}, 
the size of the variance of $W$ is determined  by the size of $|\nabla
W|$. 
On the event $\Omega_k$, we have
\begin{equation}
\begin{split}
|\nabla W|^2 &= \frac{1}{N} \sum_{1\leq i \leq j \leq N} 
  \Big| \frac{\partial W}{\partial \text{Re }h_{ij}}\Big|^2 +  
\Big|\frac{\partial W}{\partial \text{Im }h_{ij}}\Big|^2 \\ &= 
\frac{1}{N} \sum_{1\leq i \leq j \leq N} 
  \Big| \frac{1}{N} \sum_\alpha f'_E (\lambda_\alpha) \text{Re }
\overline{\bu}_{\alpha} (i) \cdot \bu_{\alpha} (j) \Big|^2 +
\Big|\frac{1}{N}
 \sum_\alpha f'_E (\lambda_\alpha)
\text{Re } \overline{\bu}_{\alpha} (i) \cdot \bu_{\alpha} (j)\Big|^2 \\ 
&=  \frac{1}{N^3} \sum_{\alpha} |f' (\lambda_{\alpha})|^2 \leq \frac{k
M}{N^2 \eta}
\end{split}
\end{equation}
(Note that the derivative in $\nabla W$ is with 
respect to the original random variables $z_{ij} = \sqrt{N} h_{ij}$).
 {F}rom the concentration inequality 
(Theorem 2.1 of \cite{BG}) we obtain that
\be
\begin{split}
 \PP (W \ge \delta/4) \le & \; e^{-T\delta/4} \; \EE \; e^{TW} \cr 
 \le & \; e^{-T\delta/4} \; \EE \exp{ \Big[ S T^2
 |\nabla W|^2\Big]} \cr \le &  C e^{-T\delta/4} \sum_{k=0}^{1/\eta}
\EE {\bf 1}_{\Omega_k}  \, e^{S T^2 kM N^{-2}\eta^{-1}} \\ 
\le &  C e^{-T\delta/4} \sum_{k=0}^{1/\eta}
e^{-c \sqrt{kMN\eta}}  \, e^{S T^2 kM N^{-2}\eta^{-1}} \, .
\end{split}
\ee
Choosing $T = c N^{1/2}$, and $\eta = N^{-3/4}$, it follows that
\[  \PP (W \ge \delta/4) \le e^{-\delta \sqrt{N}}\, . \] Repeating 
the same argument with $W$ replaced by $-W$, we conclude that
\[  \PP (|W| \ge \delta/4) \le e^{-c \delta \sqrt{N}}. \] 
Combining this with \eqref{nf1} and \eqref{nene}, we have
$$
  \PP (|\fN(E)-\fN_{sc}(E)|\ge \delta) \le Ce^{-c\delta\sqrt{ N}} .
$$
which completes the proof of Proposition \ref{prop:distr}. \qed

\bigskip

{\it Proof of Lemma \ref{lm:HS}}. For simplicity, in the proof we omit the
star from the notation.  First notice that \eqref{mcond1} implies that,
after taking imaginary part,
\be
 |\varrho|(I)\le CL_1|I|
\label{rhocon}
\ee
for any interval of length $|I|\ge (\log N)^4/N$, $I\subset [-K-1, K+1]$.

Let $(\log N)^4/N\le  \eta \le 1$  to be chosen later.
Fix $E \in [-K,K]$ and define a function $f=f_E:\bR\to \bR$ such that
$f(x)=1$ 
for $x\in [-K,E-\eta]$, $f(x)=0$ for $x>E+\eta$ and $x<-K-1$ with 
$|f'|\leq C\eta^{-1}$ and $|f''| \leq C \eta^{-2}$. We have
\be
  \Big| \fN(E) - \int_{-\infty}^\infty f_E(\la)\varrho(\rd \la) \Big| \leq 
 |\varrho|(E-\eta, E+\eta)\le
CL_1\eta \,.
\label{nf}
\ee
To express $f(\la)$ in terms of the Stieltjes transform,
we use the Helffer-Sj\"ostrand functional calculus, see, e.g.,  
\cite{Dav}.
Let $\chi(y)$ be a smooth cutoff function with 
 support in $[-1,1]$, with $\chi(y)=1$ for  $|y|\leq 1/2$
and with bounded
derivatives.
Let
$$ 
  \wt f(x+iy)= (f(x) + iyf'(x))\chi(y),
$$
then
\be
   f(\lambda) =\frac{1}{2\pi}\int_{\bR^2}
\frac{\partial_{\bar z} \wt f(x+iy)}{\lambda-x-iy} \rd x  \rd y 
=\frac{1}{2\pi}\int_{\bR^2}
\frac{iy f''(x)\chi(y) +i(f(x) + iyf'(x))\chi'(y) }{\lambda-x-iy} \rd x 
\rd y, 
\label{1*}
\ee
and therefore, since $f$ is real, 
\begin{equation}\label{eq:1+2}
\begin{split} 
\Big| \int_{-\infty}^{\infty} f(\la) \varrho(\rd \la) \Big| = & \Big|
\text{Re } \int_{-\infty}^{\infty} f(\la) \varrho(\rd \la) \Big| \\ \leq &
\left| \frac{1}{2\pi}\int_{\bR^2} \, yf'' (x) \,\chi (y) \, 
\text{Im} \, m(x+iy) \rd x  \rd y \right|  \\ 
&+ C \int_{\RR^2}  \left(|f(x)| + |y| |f'(x)| \right) |\chi'(y)| 
\left|  m (x+iy) \right| \rd x  \rd y \, .
\end{split}\end{equation}
Using (\ref{mcond2}) and the support properties of $\chi'$ and
$f$, the second contribution is bounded by
\be
\begin{split}
L_2\int_{|x| \leq K+1} \rd x\int_{\frac{1}{2}\leq |y| \leq 1} \rd y\;
\frac{\big[|f(x)| + |y| |f'(x)|\big]}{N|y|^{3/2} |2-|x||^{1/2}}
 \leq & \frac{CL_2}{N} +
\frac{CL_2}{N \eta} \int_{E-\eta}^{E+\eta} \frac{\rd x}{|2-|x||^{1/2}} 
\\ \leq & \frac{CL_2}{N|2-|E||^{1/2}}\, .
\label{secc}
\end{split}
\ee

\medskip

For the first term in \eqref{eq:1+2}, we split the integration:
\be\label{2*}\begin{split}
\Big| \int_{\bR^2} \, yf'' (x) \,&\chi (y) \, 
\text{Im}\, m(x+iy)  \rd x  \rd y \Big| \,
\\ \leq \; &C \int_{|y| \leq \eta} \int_{|x| \leq K+1} |y| |f'' (x)| 
\big(|\text{Im} \, m_+ (x+iy)| + |\text{Im} \, m_- (x+iy)|\big) \rd x \rd
y
\\&+ \Big|  \int_{\eta \leq |y| \leq 1} \int_{|x| \leq K+1} 
yf'' (x) \,\chi (y) \,  m(x+iy) \rd x  \rd y \Big| 
\end{split}\ee
where, in the second term, we dropped the imaginary part since $f$ and
$\chi$ are real.
To bound the first 
term we note that, for every fixed $x$, the functions 
\[ 
|y| |\text{Im} \, m_{\pm} (x+iy)| =  \, \int \rho_{\pm} (\rd s) 
\frac{y^2}{(s - x)^2 + y^2} 
\] 
are monotonically increasing in $|y|$. This implies that, 
for all $|y| \leq \eta$, 
\[ 
|y|  |\text{Im} \, m_\pm (x+iy)|
\leq \eta |\text{Im} \, m_\pm (x+i\eta)|
\leq C \eta 
\]
by (\ref{mcond1}). Therefore, we find 
\be\label{logf}
  \int_{|y|\leq \eta} \int_{|x|\leq K+1} |y| |f''(x)| 
 \big(|\text{Im} \, m_+ (x+iy)| + |\text{Im} \, m_- (x+iy)|\big)
\rd x \rd y  \leq C L_1 \eta .
\ee
As for the second term on the r.h.s. of (\ref{2*}), we integrate by parts
first 
in $x$, then in $y$. It is sufficient to consider the regime $\eta\le y
\le 1$,
the case of negative $y$'s is treated identically.
We find 
\be\begin{split}
\int_{\eta \leq y\leq 1} \int_{|x|\leq K+1} iy & f''(x) \chi(y) 
 m(x+iy)  \rd y  \rd x  \\
= &-\int_{\eta \leq y\leq 1} \int_{|x|\leq K+1} iy f'(x) \chi(y)
 m' (x+iy)  \rd y  \rd x   \\
  = &\int_{\eta \leq y\leq 1} \int_{|x|\leq K+1} 
\partial_y (y\chi(y)) f'(x) m(x+iy)   \rd y  \rd x  
\\ &+ \int_{|x|\leq K+1} \eta f'(x) \chi(\eta) m(x+i \eta)  \rd x.  
\label{3*} \end{split}
\ee
Using \eqref{mcond1}, the second term is bounded in absolute value by 
\be
CL_1 \int_{|x|\leq K+1} \eta |f'(x)| \rd x  \leq CL_1\eta \, . 
\label{logff}
\ee
The absolute value of the first term on the r.h.s. of (\ref{3*}) is
estimated by

\begin{equation}
C \eta^{-1}
\int_{\eta}^1 \int_{E-\eta}^{E+\eta} | m(x+iy)| \rd x \rd y. 
\end{equation}

Putting all terms together, we find  from  \eqref{nf}, \eqref{eq:1+2}, 
\eqref{secc},
\eqref{logf}
and \eqref{logff} that
\be
\label{integrate}
\int_{|E| \leq K} | \fN (E) | \rd E 
\leq CL \Big( \eta + N^{-1}+ \int_{\eta}^1 \rd y\int_{|x| \leq K+1} \rd x
\,
 \left|  m(x+iy) \right| \Big).
\ee
We will  use the bounds \eqref{mcond1}--\eqref{mcond3} and we split the
integration into separate regions:
\begin{equation}\begin{split}\label{eq:inte}
\int_{\eta}^1 \rd y &\int_{|x| \leq K+1}  \rd x 
\left|  m(x+iy) \right| \\
\leq \; & C L \int_{\eta}^1 \rd y \int \rd x \; {\bf 1} \Big( |2-|x||
\leq \frac{(\log N)^4}{Ny}\Big) \\
&+ C L \int \rd y \, \int_{|x| \leq K+1}  
\rd x \,
\min \left(1, \frac{1}{N|y|^{3/2} |2-|x||^{1/2}},
\frac{1}{N |y| |2-|x||^{3/2}} \right) \\ =: \; &
C L \left( \text{I} + \text{II} \right) \, .
\end{split}\end{equation}
Clearly
\[
\text{I} \leq  \int_\eta^1 \rd y \int  \rd x\; {\bf 1} 
\left( |2-|x|| \leq \frac{(\log N)^4}{Ny} \right) \leq 
C \frac{(\log N)^5}{N} \, .
\]
As for the second term on the r.h.s. of (\ref{eq:inte}), 
we divide the integral 
into several pieces:
\begin{equation}\begin{split}
\text{II} \leq \; & \int_{\eta}^{N^{-4/7}} \rd y \,
\left[ \int_{|2-|x|| \leq (Ny)^{-2/3}}   \rd x  +
\int_{(Ny)^{-2/3} \leq |2-|x|| \leq K+3} \, \rd x \, 
\frac{1}{N y |2-|x||^{3/2}} \right] \\
&+\int_{N^{-4/7}}^1 \rd y \, \left[ \int_{|2-|x|| \leq y^{1/2}}  \rd x \, 
\frac{1}{N y^{3/2} |2-|x||^{1/2}}  + \int_{y^{1/2} \leq |2-|x|| \leq K+3}
\, \rd x 
\,  \frac{1}{N y |2-|x||^{3/2}} \right]  \\ \leq \;
&\int_{\eta}^{N^{-4/7}} \rd y \frac{1}{(Ny)^{2/3}} + \int_{N^{-4/7}}^1
\rd y 
\frac{1}{N y^{5/4}} \\ \leq \; & C N^{-6/7} 
\end{split}\end{equation}
independently of $\eta$. Inserting in (\ref{integrate}), 
and choosing  $\eta = N^{-6/7}$,  we conclude the proof of \eqref{NEb}.
$\Box$

\section{Proof of Lemma \ref{lm:grid}.} \label{sec:grid}

We partition the interval $[-2+\kappa, 2-\kappa]$
into a disjoint union of intervals 
\be
     I_r = \Big[ n^{\gamma} N^{-1}(r-\frac{1}{2}), n^{\gamma} N^{-1}
  (r+\frac{1}{2})\Big]
\label{Ir}
\ee
of length $n^{\gamma}N^{-1}$ and center $w_r= rn^{\gamma}N^{-1}$,
where $r\in \ZZ$, $|r|\leq r_1:= Nn^{-\gamma}(2-\kappa)$. 
Then, for any $r$, 
\be
   \big| \frac{\N(I_r)}{n^{\gamma}}-\varrho_{sc}(w_r)\big|\leq
n^{-\gamma/6}
\label{ir}
\ee
by \eqref{omm}.
To prove \eqref{y}, first we locate middle eigenvalue.
Let $r_0$ be the index such
that
$$
 \sum_{r< r_0} \N(I_r) < \frac{N}{2} \le \sum_{r\le r_0}\N(I_r)
$$
in other words
\be
    \lambda_{N/2} \in I_{r_0}.
\label{N2}
\ee

For definiteness, we can assume that $r_0\ge 0$.
Using the second event in \eqref{Omstar} we obtain that
\be
  \sum_{r=1}^{r_0-1} \N(I_r) \le N/2- \N[(-\infty, 0)] \le
CNn^{-\gamma/6}.
\label{r01}
\ee
On the other hand, with the notation
 $r_1:= \min \{ (r_0-1)_+, Nn^{-\gamma}\}$, we have
by \eqref{ir} that
\be
   \sum_{r=1}^{r_0-1} \N(I_r) \ge   \sum_{r=1}^{r_1} \N(I_r) \ge
   n^{\gamma} (1- n^{-\gamma/6}) \sum_{r=1}^{r_1} \varrho_{sc}(w_r)
  \ge  cr_1n^{\gamma},
\label{r02}
\ee
where we used that  $w_r\le 1$ for any $r\le r_1\le Nn^{-\gamma}$
and thus  $\varrho_{sc}(w_r)\ge \varrho_{sc}(1)\ge c$.
 {F}rom
\eqref{r01} and \eqref{r02} we conclude that 
 $r_0\le C Nn^{-7\gamma/6}$,
i.e.
$w_{r_0}\le Cn^{-\gamma/6}$. Thus
we proved that
\be
|w_{r_0}|\leq Cn^{-\gamma/6}, \quad |\lambda_{N/2}|\le Cn^{-\gamma/6} .
\label{r0}
\ee

Starting the proof of \eqref{y}, we can 
 assume that $a\ge N/2$ by symmetry.
Suppose first that $\lambda_a \in [-2+\kappa, 2-\kappa]$,
i.e. $\lambda_a\in I_r$ for some $|r|\le r_1$,
i.e. $a\ge N/2$ implies  $r\ge r_0$.
Then we have
$$
     \sum_{u=r_0+1}^{r-1} \N(I_u)\le   a- N/2  \le \sum_{u=r_0}^{r}
\N(I_u)
$$
i.e.
\be
\sum_{u=r_0+1}^{r-1} \N(I_u)\le   a- 
N\int_{-\infty}^{0}\varrho_{sc}(E)\rd E  \le \sum_{u=r_0}^{r} \N(I_u)
\label{a-}
\ee
using  $\int_{-\infty}^{0}\varrho_{sc}(E)\rd 
E = 1/2$.

\medskip

Note that 
\be 
\sum_{u=r_0}^{r} \N(I_u) \le n^{\gamma} (1+ C n^{-\gamma/6}) 
\sum_{u=r_0}^r \varrho_{sc} (w_u) \leq (1+Cn^{-\gamma/6})
  N\int_{0}^{w_r} \varrho_{sc}(E)\rd E
\label{nv}
\ee
using \eqref{ir} \eqref{r0} and that $\gamma $ is small. Similarly
\be
   \sum_{u=r_0+1}^{r-1} \N(I_u) \ge  (1-Cn^{-\gamma/6})N
  \int_{0}^{w_r} \varrho_{sc}(E)\rd E - Cn^{\gamma}.
\label{nv1}
\ee
Thus, combining these estimates with \eqref{a-}, we have
$$
    \Big| aN^{-1} -\int_{-\infty}^{w_r} \varrho_{sc}(E)\rd E\Big|
 \le Cn^{-\gamma/6}
$$
i.e.
$$
  | \fN_{sc}^{-1}(aN^{-1}) - w_r| \le C\kappa^{-1/2}n^{-\gamma/6}
$$
using \eqref{Nscder} and  $\kappa^{3/2}\le aN^{-1}\le 1-\kappa^{3/2}$.
Since $\lambda_a\in I_r$, i.e.
$|\lambda_a-w_r|\le n^{\gamma}N^{-1}$, we obtain \eqref{y}.
Finally, we consider the case when $\lambda_a > 2-\kappa$. 
The lower bound in \eqref{a-} and the estimate  \eqref{nv1}
hold with $r=r_1$ so we get
\be
\begin{split}
  a\ge & \; (1-Cn^{-\gamma/6})N\int_{-\infty}^{w_{r_1}} \varrho_{sc}(E) \rd E
 -Cn^\gamma  \\
& \ge 
 N\int_{-\infty}^{2-\kappa} \varrho_{sc}(E) \rd E
 -Cn^{-\gamma/6}N \ge (1-\pi^{-1}\kappa^{3/2})N -Cn^{-\gamma/6}N,
\end{split}
\ee
which contradicts the assumption  $a\le N(1-\kappa^{3/2})$ for
large $N$.

\bigskip

For the proof of \eqref{yy}, 
suppose that $\lambda_a\in I_r$, $\lambda_b\in I_s$.
Using \eqref{y} and $N\kappa^{3/2}\le a< b\le N(1-\kappa^{3/2})$,
we know that $-2+\kappa/2\le w_r\le  w_s\le 2-\kappa/2$.
By \eqref{y}, we have the apriori bound
$$
 |\lambda_a-\lambda_b| \le |\fN_{sc}^{-1}(aN^{-1})- 
\fN_{sc}^{-1}(bN^{-1})|+ C_\kappa
n^{-\gamma/6} \le C_\kappa N^{-1}|b-a| + C_\kappa n^{-\gamma/6}
 \le C_\kappa n^{-\gamma/6}
$$
by the assumption $|b-a|\le CN n^{\gamma/6}$.
In particular
\be
   |w_r-w_s|\le |\lambda_a-\lambda_b| + C_\kappa n^{-\gamma/6} 
\le C_\kappa n^{-\gamma/6}.
\label{rs}
\ee
 The constants $C_\kappa$
depend on $\kappa$ as $C_\kappa \le C\kappa^{1/2}$.

{F}rom $\lambda_a\in I_r$, $\lambda_b\in I_s$ it also follows that
\be
    (s-r-1) n^{\gamma}N^{-1}\le
   \lambda_b-\lambda_a\le (s-r+1) n^{\gamma}N^{-1}
\label{ll}
\ee
and
\be
 \sum_{u=r+1}^{s-1} \N(I_u)\le    b-a \le \sum_{u=r}^s \N(I_u).
\label{nba}
\ee

Let $s-r+1= \sum_{j =0}^{j_0} 2^{m_j}$, $m_0<m_1<\ldots$ 
 be the binary representation of $s-r+1$
with $j_0= [\log_2(s-r+1)]\le \log N$.
Using this representation, we can 
concatanate the intervals $I_u$, $r\le u \le s$, into
longer intervals  $J_0, J_1, \ldots$ of length
$|J_j|= 2^j  n^\gamma N^{-1}$ such that
$$
   I:= \bigcup_{u=r}^s I_r = \bigcup_{j =0}^{j_0} J_j.
$$
Since $\varrho_{sc}'$ is bounded on $I$, 
we have
$$
      \varrho_{sc}(w) \le (1+ C_\kappa|I|)\varrho_{sc}(\la_a),
\quad \mbox{for any $w\in I$.}
$$
On the set $\Omega$ we thus have (see \eqref{omm})
\be
\begin{split}
  \sum_{u=r}^s \N(I_u) =  \sum_{j=0}^{j_0} \N(J_j)
& \le (1+ C_\kappa|I|)\varrho_{sc}(\la_a) \sum_{j=0}^{j_0}N|J_j|\Big[1 + 
(N|J_j|)^{-1/4}n^{\gamma/12}\Big] \\
& \le (1+ C_\kappa|I|)\varrho_{sc}(\la_a) 
\Big[ N|I| + (N|I|)^{3/4} n^{\gamma/12}\log N\Big].
\end{split}
\ee
Similary, one can get a lower bound on 
$\sum_{u=r+1}^{s-1} \N(I_u)$. Recalling $|I|= (s-r+1)n^\gamma N^{-1}$,
and that $|I|\le Cn^{-\gamma/6}$ from \eqref{rs},
we conclude from \eqref{nba} that
$$
      \Big| (b-a) -  \varrho_{sc}(\lambda_a) n^{\gamma}(s-r)\Big|
  \le C_\kappa n^\gamma|b-a|^{3/4} + C_\kappa N^{-1}|b-a|^2+ C_\kappa 
n^\gamma .
$$
But from \eqref{ll}
$$
 \big|  N\varrho_{sc}(\la_a)(\la_b-\la_a)  - 
\varrho_{sc}(\lambda_a) n^{\gamma}(s-r)\big| \le C_\kappa n^{\gamma}
$$
thus
$$
\big|  N\varrho_{sc}(\la_a)(\la_b-\la_a)  - (b-a)\big| 
\le C_\kappa n^\gamma|b-a|^{3/4} + C_\kappa N^{-1}|b-a|^2
$$
with $C_\kappa\le C\kappa^{1/2}$,
and we have proved \eqref{yy}. $\Box$

\section{Proof of Lemma \ref{lm:detappr}}\label{sec:det}

We start with the outline of the proof and indicate the origin
of the restriction $\al>1/4$.  We will first regularize the logarithmic
interaction on a scale $\eta$ at the expense of an error of
$O(\eta)$ for each pair of eigenvalues, modulo logarithmic corrections (Lemma \ref{lm:detreg}).
By a Schwarz inequality \eqref{DS11}, the fluctuation of the regularized two body interaction 
is split into the product of the fluctuation of the regularized potential $A_x$  \eqref{Axdef}
and the fluctuation of the local semicircle law regularized on scale $\eta$.
The latter is of order $O(N^{-1/2}\eta^{-1/2})$
by the improved fluctuation bound on
the local semicircle law \eqref{eq:E|m|}. The former is
of order $O(N^{-1}\eta^{-1/2})$ using that
the logarithmic Sobolev inequality \eqref{Sobol} on the
single site distribution can be turned into a spectral gap estimate
for $A_x$. Finally, we optimize the regularization error $O(\eta)$
and the fluctuation error $O(N^{-3/2}\eta^{-1})$ per particle pairs,
 which gives a total  error of
order $N^2\cdot N^{-3/4}= N^{1+1/4}$.

\medskip

The proof of the following regularization lemma
is postponed until the end of the section:

\begin{lemma}\label{lm:detreg}
Let $(\log N)^4/N \le\eta\le 1$, then
\be
 \Bigg| \frac{1}{N^2}\EE \Big[\sum_{j<k} \log |\lambda_j-\lambda_k| -
 \frac{1}{N^2}\sum_{j<k} \log |\lambda_j-\lambda_k+ i\eta|\Big] \Bigg|
 \le C \eta \, \log N
\label{detreg}
\ee
with respect to any Wigner ensemble whose single-site
distribution satisfies \eqref{cond1} and \eqref{cond2}.
\end{lemma}

Then Lemma \ref{lm:detappr} directly follows from the
following statement:

\begin{lemma}\label{lm:regdet} Suppose $\eta =N^{-3/4}$,  then 
\be \begin{split} 
\Big| \frac{1}{N^2}\, \EE \; \sum_{i<j} \log |\lambda_i-\lambda_j +i\eta|
-
\frac{1}{2} 
\int\!\int \log|x-y |\,  \varrho_{sc}(x)\varrho_{sc}(y)\,\rd x\,\rd y \;
\Big| 
\leq C \frac{\log N}{N^{3/4}}
\label{loglog}\end{split}
\ee
for a universal constant $C>0$ and all $N$ large enough.
\end{lemma}

{\it Proof of Lemma \ref{lm:regdet}.} Recall that $\om(\rd x)$ denotes
the empirical measure of the eigenvalues \eqref{def:om}.
We have 
\[ \Big| \sum_{i < j} \log \, |\lambda_i - \lambda_j + i\eta|
- \frac{N^2}{2} \int \, \log |x-y +i\eta| \omega (\rd x) \omega (\rd y)
\Big| 
\leq N |\log \eta| \]
because of the contribution of the diagonal terms.

\bigskip

{\it Step 1.} 
Recall the definition of $\om_\eta(x)$ from \eqref{def:ometa},
then
\begin{equation}\label{eq:log1} \Big| \EE \int \log |x-y+i \eta| \,
\omega (\rd x) \omega (\rd y) - \EE \int \rd x \rd y \log |x-y+i\eta|
\, \omega_{\eta} (x)
\omega_{\eta} (y) \Big| \leq C \eta  (\log N)^2 \, . \end{equation}

\medskip

To prove (\ref{eq:log1}), we observe that
\begin{equation*}
\begin{split}
\int \om(\rd x) &\om(\rd y) \, \log |x-y+i\eta|  - \int \rd x \rd y 
\log |x-y+i\eta| \omega_{\eta} (x) \omega_{\eta} (y) \\
&=  \int \om(\rd x )\om(\rd y) \, \int \rd t \rd r \, 
\frac{\eta}{(t-x)^2 + \eta^2} \frac{\eta}{(r-y)^2 + \eta^2} 
\left( \log |x-y +i\eta| - \log |t-r + i \eta| \right) \, .
\end{split}
\end{equation*} 
Clearly
\begin{equation}
\begin{split}
\Big| \EE &\int  \om(\rd x)\om( \rd y) \, \log |x-y+i\eta| - 
\EE \, \int \rd x \rd y \log |x-y+i\eta| \omega_{\eta} (x) \omega_{\eta}
(y) \Big|
\label{eq:Elog1} \\ 
\leq \; & \EE  \int \om(\rd x)\om( \rd y)  \, \int \rd t \rd r \, 
\frac{\eta {\bf 1}(|t-x| \leq 1)}{(t-x)^2 + \eta^2} 
\frac{\eta {\bf 1} (|r-y| \leq 1)}{(r-y)^2 + \eta^2} 
\big|\log |x-y +i\eta| - \log |t-r + i \eta| \big| \\ &+C \eta|\log
\eta|\,.
\end{split}\end{equation}
Here we also used that $\PP \{ \mbox{supp} \, \om \in [-K,K]\} \ge 1-
e^{-CN}$
for some large constant $K$.
Next we observe that
\begin{equation}
\begin{split}
\int \rd t \rd r \, &\frac{\eta {\bf 1} ( |t-x| \leq 1)}{(t-x)^2 + \eta^2}
\frac{\eta {\bf 1} (|r-y| \leq 1)}{(r-y)^2 + \eta^2} \left|
\log |x-y +i\eta| - \log |t-r + i \eta| \right| \\  
\leq \; &\int_0^1 \rd s  \int \rd t \rd r \, 
\frac{\eta{\bf 1} ( |t-x| \leq 1)}{(t-x)^2 + \eta^2}
\frac{\eta{\bf 1} (|r-y| \leq 1)}{(r-y)^2 + \eta^2}
\frac{|(x-y) - (t-r)|}{ | s (t-r) + (1-s) (x-y) + i \eta| } \\ \leq \;
&\eta \int_0^1 \rd s \, \int \rd t \rd r \, (|t| + |r|) \,
\frac{{\bf 1} ( |t| \leq \eta^{-1}) }{t^2 + 1} 
\frac{{\bf 1} (|r| \leq \eta^{-1})}{r^2 + 1} 
\frac{1}{ | s\eta (t-r) + (x-y) + i \eta| } \,.
\end{split}\end{equation}
Inserting this bound back into (\ref{eq:Elog1}), we find
\begin{equation}\label{eq:Elog11}
\begin{split}
\Big| \EE \int &\om(\rd x)\om( \rd y) \, \log |x-y+i\eta|  -
\EE \, \int \rd x \rd y \log |x-y+i\eta| \omega_{\eta} (x) \omega_{\eta}
(y) \Big| \\
\leq \; &C \eta|\log\eta| + C \eta \int_0^1 \rd s \, \int \rd t \rd r
\,(|t| + |r|) 
\frac{{\bf 1} ( |t|, |r| \leq \eta^{-1})}{(t^2 + 1)(r^2+1)} \,  
\EE \frac{1}{N^2} \sum_{i,j} \frac{1}{ | \lambda_i - 
\lambda_j + s\eta (t-r) + i \eta| } \\ 
\leq \; &C \eta (\log N)^2 .
\end{split}\end{equation}
Here we used the bound
\begin{equation}\label{eq:doublesum} 
\frac{1}{N^2} \EE \sum_{i,j} \frac{1}{|\lambda_i - \lambda_j + x\eta + i
\eta |} 
\leq C \log N,
\end{equation}
 which holds uniformly in $x \in \bR$, if $\eta \geq (\log N)^4 /N$. 
To prove (\ref{eq:doublesum}), consider the event 
\be
 \Theta_0 = \{ \max_j |\lambda_j| \leq K_0 \}
\label{Theta0}
\ee
for some $K_0>0$.
Moreover, define the intervals $I_k = [-(k+1) \eta , -k \eta ] 
\cup [ k \eta , (k+1) \eta]$, for all nonnegative integer $k \leq K_0 /
\eta$, 
and consider the event 
\be
\Theta_1 = \{ 
\N_{I_k} \leq K N \eta\; , \; k=0,1,2, \ldots, K_0\eta^{-1} \}.
\label{Theta1}
\ee
For sufficiently large $K_0$ and  $K$ we have
\be
\P \,( \Theta^c_0 )\leq e^{-c N K_0^2},\qquad
\P (\Theta^c_1)
\leq e^{-c \sqrt{KN\eta}}
\label{Ptheta}
\ee
by Lemma 7.4 \cite{ESY1} and by \eqref{NILDE}, after adjusting $c$.
Then 
\begin{equation}
\begin{split}
\frac{1}{N^2} \EE   \sum_{j <\ell}
\frac{1}{|\lambda_\ell-\lambda_j +x\eta+ i\eta|} \leq \; &\eta^{-1} \big[
\P(\Theta_0^c) + \P(\Theta_1^c ) \big]+ 
\EE \Bigg[ \, \frac{C}{N^2}
\sum_{k,m}^{K_0 \eta^{-1}} \frac{\N_{I_k} \N_{I_m}
{\bf 1}_{\Theta_0 \cap \Theta_1}}{(|k-m+x|+1)\eta}\Bigg]
\\ \leq \; & 
\eta^{-1} (e^{-cK_0^2 N} + e^{-c\sqrt{KN\eta}})+ CK^2 \eta
\sum_{k,m}^{K_0 \eta^{-1}} 
\frac{1}{(|k-m+x|+1)}  \\ \leq \; & C |\log \eta|
\end{split}\end{equation}
because $N\eta \geq (\log N)^4$ by assumption. This completes the proof of
Step 1.

\bigskip

{\it Step 2.} Let $\varrho_{\eta} (x) = \EE \, \om_\eta (x)$, and
assume  $(N\eta)\ge (\log N)^{8}$, then
\begin{equation}\label{eq:Elog2}
\Big| \EE   \int \log |x-y+ i\eta| \om_\eta (x ) \om_\eta (y ) \rd x \rd y
- 
\int \log |x-y +i\eta| \varrho_{\eta} (x) \varrho_{\eta} (y) \rd x \rd y
\Big| 
\leq C \left( \frac{1}{N^{3/2} \eta} +\frac{\eta^{1/2}}{N} \right)
\end{equation}

\medskip

We note that
\begin{equation} \label{eq:Elog2-1} \begin{split} \EE  
\int \log |x-y+ i\eta| \om_\eta (x ) &\om_\eta (y ) \rd x \rd y -
\int \log |x-y +i\eta| \varrho_{\eta} (x) \varrho_{\eta} (y) \rd x \rd y
\\
&=  \EE \, \int \log |x-y+ i\eta| (\om_\eta(x) - \varrho_\eta (x))
(\om_\eta (y)-\varrho_\eta (y))\rd x \rd y  \\ 
&= \EE \, \int \rd x \left( A_x - \EE A_x \right)  \; (\om_\eta(x) -
\varrho_\eta (x)),
\end{split} \end{equation}
where we defined the random variable 
\be A_x := \int \rd y \log |x-y +i\eta| \, \om_\eta(y)  = \frac{1}{N} 
\sum_j f_\eta(\lambda_j - x) \, 
\label{Axdef}
\ee
with $f_\eta(\la) = (\log |\cdot|* \theta_\eta)(\la)$.

To estimate the fluctuations of $A_x$ we 
use that  the logarithmic Sobolev inequality \eqref{Sobol} implies the
spectral gap,
i.e.,  we have
\be
  \EE \, |A_x - \EE A_x|^2 \le S \EE \, |\nabla A_x|^2.
\label{spgap}
\ee
Let $\bu_\al$ denote the orthonormal set of  eigenvectors
belonging to the eigenvalues $\la_\al$ of $H$. Taking into account
the scaling \eqref{scaling},  we
have
\be \begin{split}  | \nabla A_x |^2 =\; &\frac{1}{N} \sum_{1\le i\le j\le
N} 
\Bigg[ \Big|\frac{\partial A_x}{\partial \text{Re} h_{ij}} \Big|^2
+\Big|\frac{\partial A_x}{\partial \text{Im} h_{ij}} \Big|^2\Bigg]
\\ = \; & \frac{1}{N}\sum_{ij} \Bigg[ \left| \frac{1}{N}
\sum_{\al} f_\eta'(\lambda_\al - x)\text{Re} \,
\overline{{\bf u}}_\al (i) {\bf u}_\al (j)\right|^2 + 
\left| \frac{1}{N} \sum_{\al} f_\eta'(\lambda_\al - x)
\text{Im} \, \overline{{\bf u}}_\al (i) {\bf u}_\al (j) \right|^2\Bigg]
\\  = \; &\frac{1}{N^3}
\sum_{\al, \beta} f_\eta'(\lambda_\al - x) f_\eta'(\lambda_\beta - x)
|{\bf u}_\alpha \cdot {\bf u}_\beta|^2 \\ =
\; &\frac{1}{N^3} \sum_{\al} |f_\eta'(\lambda_\al - x)|^2 \\ 
 \le \; &  \frac{C}{N^2\eta} \; \om_\eta(x),
  \label{Ader} \end{split} \ee
using that $|f_\eta'(\lambda)|^2\le C(\lambda^2+\eta^2)^{-1}$.
We have from
\eqref{spgap}, \eqref{Ader} and \eqref{eq:omup} that
\be
\EE \, |A_x - \EE A_x|^2 \le \frac{C}{N^2\eta}.
\label{AA}
\ee
On the other hand, from \eqref{eq:E|m|} and
$\om_\eta(x)=\pi^{-1}\mbox{Im}\;
m(x+i\eta)$ we have
\[ \EE \left| \omega_{\eta} (x) - \varrho_{\eta} (x) \right|^q \leq 
\frac{C_q}{(N \eta |2- |x||)^{q/2}} +
C_q \, {\bf 1} \big( N\eta |2 -|x|| \le (\log N)^4\big)
\] 
for all $q \geq 1$ and for $|x|\le K$
with some large constant $K$.

In order to insert this estimate into \eqref{eq:Elog2-1}, we need
to extract the necessary decay for large $x$ from
$\omega_\eta(x)-\varrho_\eta(x)$.
For $|x| \geq 2K_0$ sufficiently large and for any $q\ge 1$
we can estimate
\[ \begin{split} \EE \, \omega^q_{\eta} (x) &\leq \eta^{-q} \, 
\EE \, {\bf 1} (|\lambda_{\text{max}}| \geq |x|/2) + \EE \,
{\bf 1} (|\lambda_{\text{max}}| \leq |x|/2) \left( \frac{1}{N}
\sum_\al \frac{\eta}{(\lambda_\al - x)^2 + \eta^2} \right)^q
\\ &\leq \eta^{-q} e^{-c |x|^2 N} + \frac{\eta^q}{|x|^{2q}}  
\leq \frac{C_q \eta^q}{|x|^{2q}} \, . \end{split} \]
Inserting the last three equations into (\ref{eq:Elog2-1}) with $q=2$, we
find
\be \begin{split}
\Big| \EE   \int \log |x-y+ i\eta| & \om_\eta (x )\om_\eta (y ) \rd x \rd
y 
- \int \log |x-y +i\eta| \varrho_{\eta} (x) \varrho_{\eta} (y) \rd x \rd y
\Big| 
\\ &\leq \int \rd x  \left(\EE \left| A_x - \EE A_x \right|^2
\right)^{1/2}
\left(\EE \left|\om_\eta(x) - \varrho_\eta (x) \right|^2 \right)^{1/2} 
\\ &\leq \frac{CK_0 }{N^{3/2} \eta} + \frac{C(\log N)^4}{N^2\eta^{3/2}}
+\frac{C}{N\eta^{1/2}} \int_{|x| \geq 2 K_0} \rd x 
\sqrt{\EE \, \omega_{\eta}^2 (x)} \\  &\leq \frac{C}{N^{3/2} \eta} +
\frac{C\eta }{N\eta^{1/2}} \int_{|x| \geq 2 K_0} \frac{\rd x}{|x|^2} \,.
\end{split}  
\label{DS11}
\ee
This completes the proof of Step 2. 

\bigskip

{\it Step 3.}  We have 
\begin{equation}\label{eq:Elog3} \left| \int \rd x \rd y \, \log
|x-y+i\eta| 
\varrho_\eta (x) \varrho_\eta (y) - \int \rd x \rd y \, \log |x-y+i\eta|
\,
\varrho_{sc} (x) \varrho_{sc} (y) \right| \leq C N^{-6/7} +
C\eta.\end{equation}

\medskip

To prove (\ref{eq:Elog3}), we write
\begin{equation}\label{eq:Elog3-1}\begin{split}
\int \rd x \rd y \, \log |x-y+i\eta| 
\varrho_\eta (x) \varrho_\eta (y) &- \int \rd x \rd y \, \log |x-y+i\eta|
\,
\varrho_{sc} (x) \varrho_{sc} (y) \\ = \; & \int \rd x \rd y \, 
\log |x-y+i\eta| (\varrho_\eta (x) - \varrho_{sc} (x) )
\varrho_\eta (y) \\ &+ \int \rd x \rd y \, \log |x-y+i\eta| \varrho_{sc}
(x) 
(\varrho_\eta (y) - \varrho_{sc} (y) )\,. 
\end{split}\end{equation}
To control the first term on the r.h.s. of the last equation, we recall
that
\[
\fN(x) =  \frac{1}{N}\, \EE \, \N (-\infty ; x), \qquad 
\fN_{sc} (x) = \int_{-\infty}^x \varrho_{sc} (t) \rd t \,
\]
denote the expected number of eigenvalues up to $x$ normalized by $N$
(integrated
density of states) and the distribution function of the semicircle law. 
Note that $\fN(x)-\fN_{sc}(x)$ vanishes at $x=\pm\infty$.
Introducing $\fN_\eta(x) : = \int_{-\infty}^x \varrho_\eta$
and integrating by parts we find
\begin{equation}
\begin{split}
\Big| \int \rd x \rd y \, \log |x-y+i\eta| & (\varrho_\eta (x) -
\varrho_{sc} (x) )
\varrho (y)\Big| \\ = \; & \Big| \int \rd x \rd y \, \log |x-y+i\eta| 
\frac{\rd}{\rd x} \left( \fN_{\eta} (x) - \fN_{sc} (x) \right) \varrho (y)\Big|
\\ =
\; & \Big|\int \rd x \rd y \frac{(x-y)}{(x-y)^2 + \eta^2} \,
\left(\fN_{\eta} (x) -
\fN_{sc} (x)\right) \varrho (y) \Big|\\ = \; & \Big|\int \rd x \, \text{Re
}
\EE \, m(x+i\eta) \, \left(\fN_{\eta} (x) - \fN_{sc} (x)\right)\Big| \\ 
\leq \; & \sup_x \, \EE \, |m(x+i\eta)| \int \rd x \, 
\Big| \fN_{\eta} (x) - \fN_{sc} (x)\Big| \,.
\end{split}
\end{equation}
{F}rom the upper bound  \eqref{eq:mup} on  $|\EE \, m(x+i\eta)|$ and
from 
\[ \int \rd x \, \Big| \fN_{sc} (x) - ( \fN_{sc} * \theta_{\eta}) (x)\Big|
\leq C \eta
\] 
we find, by \eqref{NNint}, 
\[  
\begin{split} 
\Big| \int \rd x \rd y \, \log |x-y+i\eta|  (\varrho_\eta (x)
- \varrho_{sc} (x) ) \varrho (y)\Big| & \leq C\eta + C \int \rd x \Big| 
\fN_{\eta} (x) - (\fN_{sc} \ast \theta_{\eta}) (x)| \\ &\leq C\eta + C \int 
\rd x |\fN(x) - \fN_{sc} (x) | \cr  & \leq C \eta + C N^{-6/7}\,.
\end{split}
\]
The second term on the r.h.s. of (\ref{eq:Elog3-1}) can be bounded
similarly.
This completes the proof of Step 3. Combining the estimates in
Step 1--3 and choosing $\eta = N^{-3/4}$,
we finish the proof of the Lemma \ref{lm:regdet}. \qed

\bigskip

{\it Proof  of  Lemma \ref{lm:detreg}.}
We split the summation into three parts:
$$
\sum_{j<k} \log |\lambda_j-\lambda_k+i\delta| = Y_1(\delta)
+ Y_2(\delta) + Y_3(\delta)
$$
for any $0\le \delta \le \eta$ 
with
\be
\begin{split}
  Y_1(\delta) & = \sum_{j<k} {\bf 1}(|\lambda_j-\lambda_k|\ge \eta) 
 \log |\lambda_j-\lambda_k+i\delta|
\cr
Y_2(\delta) & = \sum_{j<k} {\bf 1}(N^{-10} \le |\lambda_j-\lambda_k|
\le \eta) 
 \log |\lambda_j-\lambda_k+i\delta|
\cr
Y_3(\delta) & = \sum_{j<k} {\bf 1}(|\lambda_j-\lambda_k|\le N^{-10}) 
 \log |\lambda_j-\lambda_k+i\delta|.
\end{split}
\ee
We have 
\be\label{Y1} \begin{split}
 \EE \, |Y_1(\eta) - Y_1(0)| \le \; &\EE \, \sum_{j<k} {\bf 1}
(|\lambda_j - \lambda_k| \geq \eta) \int_0^1 \rd s \left|\frac{\rd}{\rd
s} 
\log |\lambda_j - \lambda_k + i s \eta| \right| \\
   \leq \;& \EE \, \sum_{j<k} {\bf 1} (|\lambda_j - \lambda_k|\geq \eta)
\int_0^1 \rd s \frac{\eta}{|\lambda_j - \lambda_k + i s \eta|} \\ 
\leq \; & \EE \, C \eta \sum_{j<k} \frac{1}{|\lambda_j - \lambda_k + i
\eta|} 
\\ \leq \;&  C N^2 \eta |\log N|
\end{split} \ee
by (\ref{eq:doublesum}). For the $Y_2$ term, we remark that, 
for arbitrary $0 \leq \delta \leq \eta$, 
\be
    |Y_2(\delta)| \le C \log N \, \sum_{j<k} {\bf 1}
(|\lambda_j - \lambda_k| \leq \eta)\,.
\ee
To bound the r.h.s. we consider the events $\Theta_0, \Theta_1$ from
\eqref{Theta0}, \eqref{Theta1} with sufficiently large $K$ and $K_0$ so
that \eqref{Ptheta} holds.
Then 
\be\label{eq:Y2} \begin{split} \EE \, |Y_2 (\delta)| \leq 
\; &C N^2 (\log N) \, \left( \P(\Theta_0^c) + 
\P(\Theta_1^c) \right) \\ &+ C (\log N)
\EE \Bigg[ {\bf 1} (\Theta_0 \cap \Theta_1)
\sum_{k=0}^{K_0 \eta^{-1}} \N_{I_k} (\N_{I_{k-1}} + \N_k + \N_{I_{k+1}})
\Bigg]
\\ \leq \; & CN^2 (\log N) \,e^{-c(\log N)^2} + 
C (\log N) \eta^{-1} (N\eta)^2 \\ \leq \; &C N^2 (\log N) \eta
\end{split}\ee
for every $0 \leq \delta\leq \eta$. Finally, for the $Y_3$ 
term we use the level repulsion estimate \eqref{eq:lev}
from Theorem \ref{thm:rep}, 
which implies that for any 
interval $I=[E-\e/N, E+\e/N]$ with $E\in \RR$ and $0<\e \le 1$
$$ 
  \PP (\N_I \ge 2)
\leq C \eps^4 N^{18}.
$$
Let
$$
   J_r = \Big[ \frac{r-1}{N^{10}}, \frac{r+1}{N^{10}} \Big], \qquad r\in
\ZZ
$$
be overlapping intervals covering $\RR$. 
We can then write
\be
  |Y_3(\delta)|  \le  \sum_{j<k} \sum_{r\in \ZZ}  \sum_{m=0}^\infty
  {\bf 1}\Big\{\lambda_j\in J_r,  \frac{2^{-m-1}}{ N^{10}}\le
|\la_j-\la_k|
 \le \frac{2^{-m}}{N^{10}}\Big\} |\log (2^{m}N^{10})|.
\label{Y3}
\ee
We  split the interval $J_r$ into overlapping subintervals
of length $2^{-m+1}N^{-10}$ by defining
$$
  J_{r,s}:= \Big[\frac{r-1}{N^{10}} + \frac{s}{2^{m}N^{10}}, 
\frac{r-1}{N^{10}} + \frac{s+2}{ 2^{m}N^{10}}\Big], 
\qquad 0\le s \le 2^{m+1}-2.
$$
Then
\be
\begin{split}
 \PP \Big\{\lambda_j\in I_r,  \frac{2^{-m-1}}{ N^{10}}\le |\la_j-\la_k|
 \le \frac{2^{-m}}{N^{10}}\Big\}
& \le \sum_{s=0}^{2^m} \PP 
  \Big\{\lambda_j\in I_r,  \N_{J_{r,s}}\ge 2 \Big\}
\cr
& \le \frac{2^m CN^{18}}{ (2^{m-1} N^9)^4}\le C \, 2^{-3m} \, N^{-{18}}.
\end{split}
\ee
For large $|r|\ge KN^{10}$, we can also use the bound 
$$
  \PP \{ \lambda_j\in I_r \} \le C\exp{\big( - c N (N^{-10}r)^2\big)},
$$
that follows from the 
trivial large deviation estimate for the largest eigenvalue
(Lemma 7.4 \cite{ESY1}).
Inserting these last two estimates into \eqref{Y3}, we have for  
every $0 \leq \delta \leq \eta$
\be \begin{split}
 \EE \,  |Y_3 (\delta)| \le \; &CN^2\sum_{m=0}^\infty (m+\log N ) \Big[ 
  \sum_{|r|\le r^*} 2^{-3m} N^{-18} +
  \sum_{|r|> r^*} \exp{\big( - cN (N^{-10}r)^2\big)} \Big] \\
\le \; & CN^{-2} (\log N),
\label{Y31}
\end{split}
\ee
where $r^*=  K N^{10} \log (m+2)$ for brevity. Combining \eqref{Y1},
\eqref{eq:Y2}, 
and \eqref{Y31}, we obtain \eqref{detreg}. $\Box$

\section{Level repulsion near the spectral edge}\label{sec:D}

We need to establish a Wegner-type inequality, and bounds on the level
repulsion in the same spirit as in Theorem 3.4 and Theorem 3.5 of
\cite{ESY3},
for energy intervals close to the spectral edges. Since we only
need these bounds for very small values of $\e \simeq N^{-\alpha}$, 
we are not aiming at the most general result here.
The statements we present can be proven by simply replacing,
in the proof of Theorems 3.4 and Theorem 3.5 of \cite{ESY3}, 
the convergence to the semicircle law stated in Theorem 3.1 of
\cite{ESY3} with  Theorem \ref{thm:semi}. 
Recall that  Theorem 3.1 of
\cite{ESY3} is 
valid up to the smallest possible scale $\eta > K/ N$
but only away from the spectral edges, while
Theorem \ref{thm:semi} holds all the way to the spectral edges, but only
up to the
logarithmic scale $\eta > (\log N)^4 /N$. A better $N$-dependence
of the bounds in the following theorem (but a worse $\kappa$-dependence)
can be achieved by following the dependence on $\kappa$ of the
constants in Theorem 3.1 of \cite{ESY3}. 

\medskip

All statements assume the conditions
\eqref{Sobol}--\eqref{cond1}. We introduce
the notation that $[x]_+$ denotes the positive part of a real number $x$.

\begin{theorem}[Gap distribution] \label{thm:gap}
Let $H$ be an $N \times N$ hermitian Wigner matrix and let $|E|<2$. Denote
by 
$\lambda_\al$ the largest eigenvalue below $E$ and assume 
that $\al\leq N-1$. Then there are positive constants $C,D,c,d$
such that
\be
   \P \Big(\lambda_{\al+1} -E\ge \frac{K}{N}, \; \al\leq N-1\Big)
\leq C\; e^{-c  [2-|E|]^{3/2}  \sqrt{K}}
\ee
for any $N\ge 1$ and any $D (\log N)^4/ (2-|E|) \leq K \leq \kappa N d$.
\end{theorem}

{\it Proof.} The proof of this theorem can be obtained following the proof
of 
Theorem 3.3 in \cite{ESY3}, making use of Theorem \ref{thm:semi} instead
of 
Theorem 3.1 of \cite{ESY3} (in order to follow the $|2-|E||$ dependence of
the probability).  More precisely, we observe that the event
$\lambda_{\al+1} - E \ge K/N$ implies that there is a gap of size $K/N$
about the energy $E'=E+K/(2N)$. Choosing $M= D^{1/2} \kappa^{-1/2}$ for a
sufficiently large constant $D>0$, and $\eta = K/(N M^2) \geq (\log N)^4$,
we find, similarly to (7.3)-(7.4) in \cite{ESY3}, that, apart from a set
$\Omega^c$ of measure $\P (\Omega^c) \leq C e^{-c\sqrt{K}}$, \[ \text{Im}
\, m (E' + i\eta) \leq \frac{16}{M} \leq \frac{16\sqrt{\kappa}}{D}, \]
which implies, for sufficiently large $D$, that \be\label{eq:dev} |m
(E'+i\eta) - m_{sc} (E' + i\eta)| \geq \frac{c_0}{5}\ee where $c_0 = \pi
\varrho (E') \geq c \sqrt{\kappa}$. The theorem then follows because, by
Theorem \ref{thm:semi}, the event (\ref{eq:dev}) has probability 
\[ 
\P
\left( |m(E'+i\eta) - m_{sc} (E' + i\eta)| \geq c \kappa^{1/2} \right)
\leq C e^{-c \kappa^{1/2} \sqrt{N \eta \kappa}} \leq C e^{-c \kappa^{3/2}
\sqrt{K}} \,. 
\]

\qed

\begin{theorem}[Wegner Estimate] \label{thm:wegner} 
Let $E\in \bR$ and set $\kappa: = [2-|E|]_+$. 
There exists a  constant $C>0$ such that for the number of eigenvalues
$\N_I$ in the interval
$I =[E- \eps/(2N) ; E+\eps/(2N)]$, we have
\begin{equation}\label{eq:weg2}
\EE \; \N_I \leq
\frac{C\e (\log N)^4\,}{\left( \kappa + N^{-1} \right)^9} 
\end{equation}
for every $E \in \bR$ and $\e\le 1$. Moreover  
\begin{equation}\label{eq:weg1}
y \; \EE |m (x+iy)|^2 \leq  \frac{C (\log N)^4\,}{\left( \kappa + N^{-1}
\right)^9} 
\end{equation}
for all $x\in \bR$, $y>0$. 
\end{theorem}

\begin{theorem}[Level Repulsion]\label{thm:rep}
Let $E\in \bR$ and set $\kappa: = [2-|E|]_+$.
There exists a universal constant $C$ such that
 for the number of eigenvalues
$\N_I$ in the interval
$I =[E- \eps/(2N) ; E+\eps/(2N)]$ we have
\begin{equation}\label{eq:lev}
\P \left( \N_I 
\geq 2 \right) \leq  \frac{C\e^4 (\log N)^4\,}{\left( \kappa + N^{-1}
\right)^{18}} 
\end{equation}
for all $E \in \bR$, all $0<\e <1$, and all $N$ large enough.
\end{theorem}

{\it Proof.} The proof of Theorem \ref{thm:wegner} and Theorem
\ref{thm:rep} 
follows exactly the proof of Theorem 3.4 and, respectively, Theorem 3.5 in
\cite{ESY3}, after replacing Theorem 3.3 of \cite{ESY3} by Theorem
\ref{thm:gap} 
above (in order to follow the dependence on the distance from the edges). 
\qed

\bigskip

Note that the results of the last three theorems are 
only useful in the regime of very small $\e = N |I| \ll (\log N)^{-4}$.

\section{Properties of the equilibrium measure}\label{sec:supp}

Here we check the conditions (a) and (b) in Theorem \ref{thm:L}. The main
ingredient is the following:

\begin{lemma} \label{lm:endpoint} Let $L\in \cG$ and $\by\in \cY_L$. 
After rescaling, then for any fixed $\sigma>0$ with $J'=[-1+\sigma/2, 1-\sigma/2]$,
the first and second derivatives of the potential are uniformly
bounded on $J'$, i.e.
\be
    \sup_{x\in J'} |U_{\tby}^{(\ell)}(x)|  \le C_\ell, \qquad \ell =1,2,
\label{Ubound}
\ee
where the constant is independent of $\by$.
Furthermore, 
the  endpoints $a, b$ of the support of the equilibrium measure $\nu=
\nu_\by$ satisfy
\be
     |a+1|, |b-1|\le C n^{-\gamma/3}\log n .
\label{ends}
\ee
\end{lemma}

Condition (b) of  Theorem \ref{thm:L} is given now by \eqref{Ubound}.
To see condition (a) of Theorem \ref{thm:L}, 
let $[a_n, b_n]$ denote the support of the equilibrium measure $\nu_n$,
then $a_n\to -1$ and $b_n\to 1$ as $n\to\infty$,
thus $g_n$ is positive on $J=[-1+\sigma, 1-\sigma]$ for any fixed
$\sigma>0$
and any sufficiently large $n$. 

For the uniform boundedness of $g_n(x)$ on $J$, 
we use the explicit formula (see, e.g. Theorem 2.5. of \cite{LL1}):
\be
  g_n(x) = \frac{1}{2\pi^2} \sqrt{(x-a_n)(b_n-x)}\; \mbox{P.V.}
   \int_{a_n}^{b_n} \frac{V'_n(s)}{s-x} \frac{1}{\sqrt{(s-a_n)(b_n-s)}}
\, \rd s,
\label{gn}
\ee
where P.V. denoted principal value. For sufficiently large $n$ and for any
$x\in J$
the singularity of $(s-x)^{-1}$ is uniformly separated away from $a_n$ and
$b_n$,
i.e. from the singularity of the square roots.
Moreover,
$V'_n(x)$ is a smooth function inside $(-1,1)$
with 
$$
 \sup_n \sup_{x\in J'} |V'_n(x)| + |V''_n(x)| \le C.
$$
according to \eqref{Ubound}. 
Thus the uniform boundedness of $g_n$ on $J$ follows immediately
from \eqref{gn} with standard estimates on the principal value. 

\bigskip

{\it Proof of Lemma \ref{lm:endpoint}.}
Recall the definition $E_L=\fN_{sc}^{-1}(LN^{-1})$ from  \eqref{ELdef}.
For $\by\in \cY_L$
we know from the first bound in
\eqref{Il1} that $\mbox{dist}(I_\by, E_L)\le Cn^{-\gamma/6}$,
and from \eqref{yy} that 
$$
    y_k =  y_{-1} +  \frac{k + O(k^{4/5})}{N\varrho_0},
$$
with $\varrho_0:=\varrho_{sc}(E_L)$, assuming $\gamma\le 1/20$ and
$Cn\le |k|\le n^B \le N^{1/2}$.
After rescaling, this corresponds to
\be
 \wt y_k = \frac{k + O(k^{4/5})}{n\varrho_0},
\qquad \varrho_0:=\varrho_{sc}(E_L),
\label{yky1}
\ee
and  we will drop the tilde for the rest of this proof.
This bound on the location of $y_k$'s will be used to estimate the
derivatives of $U_\by$. For $\ell=1,2$ and $x\in J'$ we have
\be
\begin{split}
  |U_\by^{(\ell)}(x)| &
  \le \frac{2}{n} \sum_{|k|< Cn} \frac{1}{|x-y_k|^\ell} 
 + \frac{2}{n} \Bigg| \; \sum_{Cn\le |k| < n^B} \frac{1}{(x-y_k)^\ell} \Bigg| \\
& \le C\sigma^{-\ell} + \frac{C}{n}\sum_{Cn\le k <n^B}
  \Bigg| \frac{1}{(x-y_k)^\ell} +\frac{1}{(x-y_{-k})^\ell}\Bigg| .
\end{split}
\ee
For $\ell=2$ we can use in the second sum that
$|x-y_{\pm k}|\ge Ck [n\varrho_0]^{-1}$ for $k\ge Cn$ by \eqref{yky1},
thus  $ |U_\by''(x)|\le C(\sigma)$. For $\ell=1$ we estimate
$$
    \Bigg| \frac{1}{x-y_k} +\frac{1}{x-y_{-k}}\Bigg| = \Bigg|
\frac{2x - y_k-y_{-k}}{(x-y_k)(x-y_{-k})}\Bigg|
 \le \frac{2|x| n\varrho_0 +Ck^{4/5}}{k^2} n\varrho_0
$$
where we used \eqref{yky1} and $k\ge Cn$. After summation we  conclude
that $ |U_\by'(x)|\le C(\sigma)$ and thus \eqref{Ubound} is proven.

To estimate the location of the endpoints, we
substitute $V(x)= U_\by(x)$ into the equations \eqref{ab}. We have
\be
  \frac{2}{n}\sum_{|k|< n^B} \int_a^b  \frac{1}{ \sqrt{(s-a)(b-s)}}
    \frac{ \rd s }{s-y_k}=0
\label{V1}
\ee
\be
   \frac{1}{n\pi}\sum_{|k|< n^B} \int_a^b  \frac{s}{ \sqrt{(s-a)(b-s)}}
    \frac{ \rd s }{s-y_k}=-1.
\label{V2}
\ee
We will need the following
explicit integration formulae for $a<b$
(see, e.g. Formula 2.266 in \cite{GR})
\be
\begin{split}
  \int_a^b \frac{1}{\sqrt{(s-a)(b-s)}}\frac{\rd s }{s-y}
 & = \frac{\pi}{\sqrt{(a-y)(b-y)}}  \quad \mbox{if}\quad y<a<b,
\cr   \int_a^b \frac{1}{\sqrt{(s-a)(b-s)}}\frac{\rd s }{s-y}
 &  = -\frac{\pi}{\sqrt{(a-y)(b-y)}}  \quad \mbox{if}\quad a<b<y.
\label{expl}
\end{split}
\ee
\be
 \int_a^b \frac{\rd s}{\sqrt{(s-a)(b-s)}}=\pi.
\label{pii}
\ee
With these formulae, \eqref{V1} and \eqref{V2} can be written as
\be
  \frac{1}{n}\sum_{-n^B < k \le -1} \frac{1}{ \sqrt{(a-y_k)(b-y_k)}}
- \frac{1}{n}\sum_{1\le  k < n^B} \frac{1}{ \sqrt{(a-y_k)(b-y_k)}}=0,
\label{V1a}
\ee
\be
   \frac{1}{n}\sum_{-n^B < k \le -1} \Big[ 
  \frac{y_k}{ \sqrt{(a-y_k)(b-y_k)}}+1\Big]
-\frac{1}{n}\sum_{1\le  k < n^B}\Big[ \frac{y_k}{ \sqrt{(a-y_k)(b-y_k)}}
 -1\Big]=-1.
\label{V2a}
\ee
Using the bound \eqref{yky1} on the location of $y_k$'s,
we replace the limit $-n^B < k$ with $-Y\le y_k$
and the limit $k< n^B$ with $y_k\le Y$
in the summations in \eqref{V1a} and
\eqref{V2a}, where $Y:= n^{B-1}\varrho_0^{-1}$.
We have, for example, for the first sum \eqref{V1a},
\be
   \frac{1}{n}\sum_{-n^B < k \le -1} \frac{1}{ \sqrt{(a-y_k)(b-y_k)}}
=\frac{1}{n}\sum_{-Y < y_k \le -1} \frac{1}{ \sqrt{(a-y_k)(b-y_k)}}
+O(n^{-1/5} Y^{-1}),
\label{cut}
\ee
and the estimate for the other three  sums in \eqref{V1a}, \eqref{V2a}
is identical.

With similar argument, we can remove the $y_k$'s that are too
close to $[-1,1]$. Let $X= n^{\gamma-1}$, then
$$
 \frac{C}{n\sqrt{a+1}}\le \frac{1}{n}\sum_{-1-X < y_k \le -1} 
\frac{1}{ \sqrt{(a-y_k)(b-y_k)}}
\le \frac{Cn^{\gamma-1}}{\sqrt{(a+1)(b+1)}},
$$
where, for the lower bound, we used that $y_{-1}=-1$, while
for the upper bound we used 
that the number of $y_k$'s in $[-1-X, -1]$ is at most
$Cn^{\gamma}$ (see the third set in the
definition of \eqref{Omstar}) . Similarly we have
$$
\frac{C}{n\sqrt{1-b}}\le \frac{1}{n}\sum_{1\le y_k < 1+X} 
\frac{1}{ \sqrt{(a-y_k)(b-y_k)}}
\le \frac{Cn^{\gamma-1}}{\sqrt{(1-b)(1-a)}}
$$
and for the sums in \eqref{V2a}
$$
 - \frac{Cn^{\gamma-1}}{\sqrt{(a+1)(b+1)}} 
\le \frac{1}{n}\sum_{-1-X < y_k \le -1} \Big[ 
  \frac{y_k}{ \sqrt{(a-y_k)(b-y_k)}}+1\Big] \le Cn^{\gamma-1}
-\frac{C}{n\sqrt{a+1}}
$$
$$
 -Cn^{\gamma-1}+\frac{C}{n\sqrt{1-b}} \le
 \frac{1}{n}\sum_{1\le  y_k < 1+X}\Big[ \frac{y_k}{
\sqrt{(a-y_k)(b-y_k)}}
 -1\Big]\le \frac{Cn^{\gamma-1}}{\sqrt{(1-b)(1-a)}}.
$$
Define
\be
  W_1:= \frac{1}{n}\sum_{ -Y < y_k<-1-X}
   \frac{1}{ \sqrt{(a-y_k)(b-y_k)}} - \frac{1}{n}\sum_{1+X < y_k < Y}
   \frac{1}{ \sqrt{(a-y_k)(b-y_k)}} 
\label{V13}
\ee
\be
   W_2:= \frac{1}{n}\sum_{ -Y < y_k<-1-X}
   \Big[\frac{y_k}{ \sqrt{(a-y_k)(b-y_k)}} +1\Big]
  - \frac{1}{n}\sum_{1+X < y_k < Y}
   \Big[\frac{y_k}{ \sqrt{(a-y_k)(b-y_k)}}-1\Big] +1
\label{V23}
\ee
to be the truncated summations.
Combining the above estimates with estimates of type \eqref{cut}
and using $B\ge 2$ so that $n^\gamma Y^{-1}\le n^{\gamma-1}$,
we get from \eqref{V1a}, \eqref{V2a} that
\be
\frac{C}{n\sqrt{1-b}} - \frac{Cn^{\gamma-1}}{\sqrt{(a+1)(b+1)}}\le W_1\le
 \frac{Cn^{\gamma-1}}{\sqrt{(1-b)(1-a)}} - \frac{C}{n\sqrt{a+1}}
\label{w1bound}
\ee
and 
\be
 \frac{C}{n\sqrt{a+1}} +\frac{C}{n\sqrt{1-b}}  -Cn^{\gamma-1}
\le  W_2 \le\frac{Cn^{\gamma-1}}{\sqrt{(a+1)(b+1)}}+ \frac{Cn^{\gamma-1}}{\sqrt{(1-b)(1-a)}}.
\label{w2bound}
\ee

Using that for $\by\in \cY_L$ the number of eigenvalues
in any interval of size at least $n^\gamma N^{-1}$
(before rescaling) is approximated
by the semicircle law with a precision $n^{-\gamma/3}$ 
(see \eqref{omm}), we get
\be
  W_1 = \varrho_0 
\int_{-Y}^{-1-X} \frac{\rd y}{ \sqrt{(a-y)(b-y)}}
  -  \varrho_0 
\int_{1+X}^{Y} \frac{\rd y}{ \sqrt{(a-y)(b-y)}} + O(n^{-\gamma/3}\log n)
\label{w1}
\ee
and
\be
 W_2 =  \varrho_0 
\int_{-Y}^{-1-X} \Big[\frac{y}{ \sqrt{(a-y)(b-y)}}+1\Big]\rd y
  -  \varrho_0 
\int_{1+X}^{Y} \Big[\frac{y}{ \sqrt{(a-y)(b-y)}} -1\Big]\rd y 
+1+ O(n^{-\gamma/3}\log n).
\label{w2}
\ee
Here we also used that
$$
 \sup_{a,b\in [-1,1]}
\int_{1+X}^{Y} \frac{\rd y}{ \sqrt{(a-y)(b-y)}} \le C\log n
$$
and
$$
 \sup_{a,b\in [-1,1]}
\int_{1+X}^{Y} \Big[\frac{y}{ \sqrt{(a-y)(b-y)}} -1\Big]\le C\log n.
$$

Let $u=\frac{1}{2}(a+b)$ and $v = \frac{1}{2}(b-a)$
and we can assume, by symmetry, that $u\ge0$.
Then we can change
variables in the integrals
in \eqref{w1}
$$
 W_1 = \varrho_0 
\int_{-Y-u}^{-1-X-u} \frac{\rd y}{ \sqrt{y^2-v^2}}
  -  \varrho_0 
\int_{1+X-u}^{Y-u} \frac{\rd y}{ \sqrt{y^2-v^2}} + O(n^{-\gamma/3}\log n)
$$
$$
  =  \varrho_0 
\int_{Y-u}^{Y+u} \frac{\rd y}{ \sqrt{y^2-v^2}}
  -  \varrho_0 
\int_{1+X-u}^{1+X+u} \frac{\rd y}{ \sqrt{y^2-v^2}} + O(n^{-\gamma/3}\log n).
$$
The first term is of order $Y^{-1}$ and thus negligible. Thus,
from the lower bound in \eqref{w1bound}, we have
$$
\varrho_0 
\int_{1+X-u}^{1+X+u} \frac{\rd y}{ \sqrt{y^2-v^2}} 
\le \frac{Cn^{\gamma-1}}{\sqrt{(a+1)(b+1)}}-
\frac{C}{n\sqrt{1-b}} +Cn^{-\gamma/3}\log n.
$$
Estimating  $y^2-v^2\le (1+X+a)(1+X+b)$ on the integration domain,
we get
$$
\frac{Cu\varrho_0}{\sqrt{ (1+X+a)(1+X+b)}  }  
\le \frac{Cn^{\gamma-1}}{\sqrt{(a+1)(b+1)}}+Cn^{-\gamma/3}\log n,
$$
$$
  Cu\varrho_0 \leq Cn^{\gamma-1}\Big( 1+ \frac{X}{a+1}\Big) 
 + Cn^{-\gamma/3}\log n.
$$
Clearly $a+1\ge 2u$, thus
$$
  Cu\varrho_0 \leq Cn^{\gamma-1}\Big( 1+ \frac{X}{u}\Big) 
 + Cn^{-\gamma/3}\log n,
$$
from which it follows that
$$
   u \le Cn^{-\gamma/3}\log n
$$
if $\gamma\le 1/2$. The case $u\le 0$ is treated similarly,
thus we have shown that
\be
  |a+b|\le Cn^{-\gamma/3}\log n.
\label{a+b}
\ee

Now we consider the $W_2$ and assume again that $u\ge0$.
With the same change of variables
as above, we have
\be\begin{split}
  W_2= & \varrho_0 
\int_{-Y-u}^{-1-X-u} \Big[\frac{y+u}{ \sqrt{y^2-v^2}}+1\Big]\rd y
  -  \varrho_0 
\int_{1+X-u}^{Y-u} \Big[\frac{y+u}{ \sqrt{y^2-v^2}}-1\Big] \rd y 
+1+ O(n^{-\gamma/3}\log n)
\\
= & 
\varrho_0 
\int_{-Y-u}^{-1-X-u} \Big[\frac{y}{ \sqrt{y^2-v^2}}+1\Big]\rd y
  -  \varrho_0 
\int_{1+X-u}^{Y-u} \Big[\frac{y}{ \sqrt{y^2-v^2}}-1\Big] \rd y 
+1 + O(n^{-\gamma/3}(\log n)^2),
\end{split}\label{w22}
\ee
where we used   \eqref{a+b} and 
$$
\int_{-Y-u}^{-1-X-u} \frac{1}{ \sqrt{y^2-v^2}} \rd y \le C\log n
\qquad\mbox{and}\qquad \int_{1+X-u}^{Y-u} \frac{1}{ \sqrt{y^2-v^2}} \rd y \le C\log n.
$$
The integrals on the r.h.s of \eqref{w22} can be explicitly computed:
$$
   \int_{1+X-u}^{Y-u} \Big[\frac{y}{ \sqrt{y^2-v^2}}-1\Big] \rd y
  = \frac{-v^2}{\sqrt{y^2-v^2}+ y}\Bigg|_{1+X-u}^{Y-u} = 
 \frac{v^2}{\sqrt{(1+X-u)^2-v^2}+ 1+X-u} + O(Y^{-1}),
$$
$$
\int_{-Y-u}^{-1-X-u} \Big[\frac{y}{ \sqrt{y^2-v^2}}+1\Big]\rd y
  = \frac{-v^2}{\sqrt{(1+X+u)^2-v^2}+ 1+X+u} + O(Y^{-1}),
$$
thus we have
$$
   W_2 \ge 1- 2\varrho_0v^2 - Cn^{-\gamma/3}(\log n)^2 - CY^{-1} \ge 1- \frac{2}{\pi} -  
Cn^{-\gamma/3}(\log n)^2
$$
by using that $v^2\le 1$ and $\varrho_0\le \pi^{-1}$ (see \eqref{def:sc}).
Combining this estimate 
with the upper bound in \eqref{w2bound}, we have
$$   
    1 -\frac{2}{\pi}-  Cn^{-\gamma/3}(\log n)^2 \le
\frac{Cn^{\gamma-1}}{\sqrt{(a+1)(b+1)}}+ \frac{Cn^{\gamma-1}}{\sqrt{(1-b)(1-a)}}
 \le \frac{Cn^{\gamma-1}}{a+1}+ \frac{Cn^{\gamma-1}}{1-b}
$$
by using $a<b$. Therefore either $a+1$ or $1-b$ is smaller than $Cn^{\gamma-1}$,
but then by using \eqref{a+b} we obtain that both of them are smaller
then  $Cn^{-\gamma/3}\log n $.
This completes the proof of Lemma \ref{lm:endpoint}. $\Box$

\end{document}